\documentclass[aps,prl,showpacs,twocolumn,superscriptaddress]{revtex4-2}%

\usepackage{graphicx}% Include figure files
\usepackage{dcolumn}% Align table columns on decimal point
\usepackage{bm}% bold math

\usepackage{float}  %%%%%%%%%%%%%%%%%%%%%%%%%%%%%%%%%%%%%%%%%%%%%%%%%%%%%%%

\usepackage{graphicx}
\usepackage{graphics}
\usepackage[caption=false]{subfig}
\usepackage{color}
\usepackage[english]{babel}
\usepackage[utf8]{inputenc}
\usepackage{ae}
\usepackage{latexsym}
\usepackage{dcolumn}
\usepackage{bm}
\usepackage[table]{xcolor}
\usepackage{array}
\usepackage{url}
\usepackage{epsf}
\usepackage{epsfig}
\usepackage{pstricks,pst-grad}
\usepackage[pdftex,colorlinks]{hyperref}
\usepackage{amsmath}
\usepackage{amssymb}
\usepackage{ragged2e}
\usepackage{comment}
\usepackage{adjustbox}

\begin{document}

\title{End-Functionalized Ions Promote Stability of Highly Frustrated Phases in Diblock Copolymers
%Electrostatic Fluctuations Drive Microphase Separation and Hierarchical Superlattices in Charged Diblock Copolymers
}

\author{Chao Duan}
\affiliation{Division of Chemistry and Chemical Engineering, California Institute of Technology, Pasadena, California 91125, United States}

\author{Zhen-Gang Wang}
\email {zgw@caltech.edu}\affiliation{Division of Chemistry and Chemical Engineering, California Institute of Technology, Pasadena, California 91125, United States}

\date{\today}% It is always \today, today,

\begin{abstract}
Block copolymers self-assemble into ordered nanostructures whose geometry 
is governed by a competition between interfacial energy and chain conformational entropy. While this competition produces a rich sequence of  morphologies, topologically complex ``frustrated'' phases, such as the primitive cubic network and single networks, incur severe packing penalties and are difficult to access in neutral systems. Here we show that ions functionalized at the termini of one block in an AB diblock copolymer melt introduce a qualitatively new stabilization mechanism.
Strong ion correlations drive chain-end association and generate a curvature preference 
toward the charged domain;
the resulting tendency of end-localized ion clusters to adopt compact, curved geometries selectively favors the highly frustrated single-primitive-cubic network ($Pm\bar{3}m$) and single-gyroid network ($I$4$_1$32) over the classical phases,
in a region of parameter space with a segregation strength lying below the order-disorder transition of the neutral system. Free energy decomposition reveals that the electrostatic energy, arising almost entirely from beyond-mean-field ion correlations, becomes increasingly negative with increasing interfacial curvature. In the primitive cubic network, pronounced local segregation of ions into the cylindrical struts generates compact curved clusters whose correlation energy gain more than offsets the enhanced packing frustration, so the very geometry that is the source of packing frustration in neutral systems becomes the source of its stability here. 
Increasing ion size weakens correlations and suppresses the network phases, consistent with experimental observations. Our results establish curvature-selective end-group association as a general principle for accessing frustrated topologies in block copolymer systems.
\end{abstract}

\maketitle

%\section{1. Introduction}

The capacity of block copolymers (BCPs) to spontaneously organize into periodic nanostructures with precisely tunable geometry has made them one of the most versatile platforms in soft materials science \cite{Hamley_Book1998,Bates_1999,Bates_2012,Bates_2016}. This self-assembly emerges from a competition between two opposing thermodynamic forces: the chemical incompatibility of the distinct blocks, which drives segregation, and the entropic penalty of stretching covalently connected chains away from their Gaussian conformations to fill space uniformly \cite{Matsen_2002,Matsen_2012}. For simple AB diblock copolymers, the interplay of these forces, parameterized by the segregation strength $\chi N$ and the block volume fraction $f$, generates the now-classical phase sequence---body-centered cubic spheres, hexagonally packed cylinders, lamellaes, double-gyroid networks, and orthorhombic $Fddd$ (O$^{70}$) networks; their boundaries have been mapped quantitatively by theory, simulation, and experiment \cite{Leibler_1980,Bates_1990,Foerster_1994,Matsen_1996,Matsen_2023,Gavrilov_2013,Tyler_2005,Takenaka_2007}. The ordered morphologies of BCPs facilitate a broad range of potential applications, including nanolithography \cite{Stoykovich_2005,Ruiz_2008,Tang_2008}, photonic materials \cite{Kim_2009,Sveinbj_rnsson_2012,Stefik_2015}, membrane separations \cite{Yoo_2015,Rangou_2022}, and as structural templates for functional nanomaterials \cite{Robbins_2016,Yang_2022}. Self-consistent field theory (SCFT) has proven the preeminent quantitative framework for predicting BCP phase behavior, achieving remarkable agreement with experiment across a wide range of compositions and architectures \cite{Matsen_1994,Drolet_1999,Bohbot-Raviv:2000aa,Fredrickson_Book2006,Xu_2024}.

The full topological richness of what BCPs can in principle form extends well beyond these classical structures. A broader class of ordered phases---including Frank–Kasper A15, $\sigma$ and Laves phases \cite{Grason_2003,Lee_2010,Xie_2014,Liu_2016,Zhao_2019,Kim_2017,Kim_2018,Bates_2019,Cheong_2020,Dorfman_2021,Zhuang_2024,Zhao_2025,Geng_2025}, quasicrystalline tilings \cite{Gillard_2016,Schulze_2017,Lindsay_2020,Mueller_2021,Mueller_2024,Li_2025}, as well as diamond networks and primitive cubic networks \cite{Lin_2018,Takagi_2019,Takagi_2021,Lai_2021,Wang_2025,Mart_nez_Veracoechea_2007,Martinez_Veracoechea_2009,Qiang_2020,Chang_2021,LiQY_2022,HChen_2025,Lee_2024a,Chang_2024,Chang_2025}---have been identified both computationally and experimentally over the past two decades.
What these complex phases share is geometric frustration: an incompatibility between the locally preferred packing of polymer chains and the global requirement for space-filling periodicity \cite{Reddy_2018,Reddy_2021,Reddy_2022,Grason:2023aa,Dimitriyev_2023}. In the 6-connected primitive cubic (P) network, for instance, the mean curvature of interdomain interface varies more sharply between the cylindrical struts and the highly branched nodes than in the 3-connected gyroid networks, forcing chains that occupy the nodes to stretch substantially beyond their equilibrium dimensions \cite{Matsen_1997,Martinez_Veracoechea_2009,Grason:2023aa,Dimitriyev_2023,Hou_2024,Chang_2024}. In neutral diblock copolymers, this frustration-induced penalty raises both the interfacial and conformational free energies of the P phase above those of the neighboring classical structures, rendering it only metastable for neutral systems \cite{Dimitriyev_2023,Hou_2024,Chang_2024}. The fundamental challenge---and opportunity---is therefore to identify driving forces that can preferentially stabilize frustrated phases by providing a free-energy gain large enough to compensate the packing penalty.

Electrostatic interactions introduce an additional driving force for microphase separation \cite{Weber_2011,Nakamura_2011,Scalfani_2012,Qin_2016,Hou_2018,Kong_2021,Kim_2017PNAS,Brown_2018,Zhang_2021,ZhangZ_2021,Min_2021b,Lee_2024b,Huo_2023,Zhang_2023,Tsai_2025,Hou_2026}.
When charges are distributed along the polymer backbone, e.g., as in sulfonated or single-ion conducting BCPs, ion correlations in the low-dielectric environment of polymer melts enhance the effective immiscibility between blocks, shift order–disorder and order–order transition boundaries, and can give rise to morphologies with no neutral analog \cite{Park_2008,Shim_2019,Min_2021a}. Yet uniformly charged BCPs still overwhelmingly favor the same sequence of classical morphologies as their neutral counterparts \cite{Yan_2020,Park_2021AN,Park_2021,Brown_2025}.

A qualitatively different scenario emerges when charges are sparse and located only at the chain ends, rather than being uniformly distributed along the polymer backbone. In this end-localized regime, ions and their counterions can associate into multiplets (hereafter called clusters), which tend to adopt compact, curved morphologies instead of flat or diffuse ones. The propensity of sparse ionic groups to aggregate within a low-dielectric polymer matrix is well documented in the ionomer literature, where clustering has been shown to depend on ion concentration, local chain architecture, and molecular weight  \cite{Williams_1986,Eisenberg_1990,Hall_2011}.
%[https://pubs.acs.org/doi/pdf/10.1021/ma00165a036, Eisenberg, Hird and Moore: A. Eisenberg, B. Hird, and R. B. Moore, "A New Multiplet-Cluster Model for the Morphology of Random Ionomers," Macromolecules 23, 4098–4107 (1990). DOI: 10.1021/ma00220a012, Hall et al., https://pubs.acs.org/doi/pdf/10.1021/ja209142b ].
Such clustering introduces an additional contribution beyond the interfacial tension and conformational entropy that govern neutral systems, and suggests that ion–ion correlations may preferentially stabilize frustrated topologies.

%Indeed, charged block copolymers (BCPs) show phase behavior that is qualitatively distinct from that of neutral BCPs: ion correlations can enhance the effective immiscibility between blocks [41–45], shift phase boundaries, and generate morphologies with no counterparts in neutral systems [46]. %These properties, combined with the mechanical robustness of the BCP matrix and the high ion mobility achievable in ordered conducting domains, have established ionic BCPs as leading candidates for solid-state battery electrolytes [33–35] — applications where the nanostructure geometry directly controls Li⁺ transport and where accessing low-symmetry, bicontinuous morphologies could dramatically improve device performance [36–40].

Park et al. recently demonstrated that adding trace lithium salt to polystyrene‑b‑poly(ethylene oxide) (PS‑b‑PEO) stabilizes the frustrated double-primitive-cubic network ($Im\bar{3}m$) phase \cite{Lee_2025}. Li$^+$ ions were shown to preferentially coordinate with the PEO hydroxyl end groups, causing clustering of the PEO chain ends. The resulting $Im\bar{3}m$ networks are not only stable but display higher ionic conductivity than other morphologies, because Li$^+$ transport along localized end‑chain pathways is more efficient than along the PEO backbone, thus providing a direct link of the frustrated topology to functional benefit. Subsequent studies from the same group showed that decreasing anion size or replacing the PEO hydroxyl end with $-$PO$_3$H$_2$ groups facilitates ion localization and significantly expands the stability window of the $Im\bar{3}m$ phase \cite{Lee_2025}. These results suggest end‑localized interactions as a general route to access topologically complex phases. In this work, we use an end‑charged diblock copolymer system to elucidate the underlying self‑assembly mechanisms.

%Why the primitive cubic topology in particular? What is the microscopic free-energy mechanism by which end-localized ion correlations overcome packing frustration? And can theory predict and ultimately guide the rational design of such phases?

The presence of strong electrostatic interactions requires a theoretical framework capable of treating strong correlation effects such as ion clustering in dense, inhomogeneous systems. Standard SCFT, extended to incorporate electrostatics within the mean-field Poisson--Boltzmann approximation \cite{Shi_1999,Wang_2004}, predicts that including charge weakens microphase separation \cite{Kumar_2007,Yang_2011,Liu_2011}, in direct contradiction with experiment \cite{Park_2008,Young_2009,Loo_2018}.
Sing et al. developed a hybrid approach combining liquid-state integral equation theory for ionic correlations with SCFT to treat neutral--charged diblock copolymers and predicted a highly skewed, chimney-like phase diagram \cite{Sing_2014a,Sing_2014b,Sing_2015}. But the method assumes local charge neutrality and treats polymer-bound charges as a smooth local density field, making it unsuitable for polymers with discrete or patterned charge sequences such as end-functionalized BCPs. Fredrickson and coworkers incorporated fluctuation corrections through a second-order (Gaussian-level) perturbative treatment suitable for weak electrostatic interactions \cite{Audus_2015}, but the method becomes inadequate in the regime of strong ion correlations and binding that is most relevant to low-dielectric polymer melts. %No existing theoretical framework simultaneously accounts for nonperturbative ion correlations, respects polymer backbone connectivity, and accommodates arbitrary charge sequences — the combination of capabilities essential for treating end-functionalized charged BCPs.

In this work, we apply a recently developed ion-correlation augmented self-consistent field theory \cite{Duan2025PEBrush} to investigate the microphase separation of AB diblock copolymer melts with end-functionalized ions.
%\textcolor{red}{This framework incorporates a nonperturbative variational Gaussian renormalized fluctuation theory for electrostatics into the polymer SCFT. Ion-correlation effects, including the enrichment of opposite-charges and the depletion of like-charges around a tagged ion, can be well captured, as shown from the pair correlation function (See the Materials and Methods). Additionally, our theory treats each charged segment explicitly, faithfully representing backbone connectivity and end-localized charge patterns.}
This framework incorporates a nonperturbative variational Gaussian renormalized fluctuation theory for electrostatics \cite{Wang:2010wk,Agrawal:2022ux} into the polymer SCFT \cite{Fredrickson_Book2006}, and treats each charged segment explicitly, faithfully representing backbone connectivity and end-localized charge patterns. We show that end-functionalized ion correlations stabilize the single-primitive-cubic network ($Pm\bar{3}m$) as the thermodynamically preferred phase over the classical morphologies, in a region of parameter space lying {\emph{below}} the order–disorder transition (in terms of $\chi N$) of the neutral system.
A stable single-gyroid network ($I$4$_1$32) also appears adjacent to the $Pm\bar{3}m$ phase.
Through free energy decomposition, we identify the microscopic mechanism: ions in the P network segregate locally into the cylindrical struts, forming compact curved clusters whose correlation energy gain more than compensates the interfacial and chain-stretching penalties that make this phase metastable in neutral systems.
Smaller ions expand the stability window of the primitive network phase, consistent with the experimental trends of Park and coworkers \cite{Lee_2025}.
Our results establish curvature-selective ion clustering as a general design principle for accessing frustrated topologies in charged BCPs.

\begin{comment}
Ion-containing polymers and soft matters are ubiquitous in commercial products, functional materials, and biological systems \cite{Decher_1997,Galvanetto_2023}.
In particular, charged block copolymers (BCPs) have been widely used in surfactants, sewage treatment, coatings, enhanced oil recovery, and energy storage, etc \cite{}.
They can also serve as a model system for studying biological membranes and intrinsically disordered proteins \cite{Danielsen_2019,Carrick_2026}. The charged groups in BCPs can profoundly impact morphologies and functionalities of their self-assembled nanostructures, rendering them promising candidates for solid-state battery electrolytes \cite{Hallinan_2013,Bouchet_2013,Miller_2017}. Therefore, elucidating how electrostatics governs their self-assembly is crucial to rationally designing BCPs with desired structures and optimal properties.
\end{comment}

%%To the best of our knowledge, P phase has been obtained by ccomplex dendrimer BCP or linear BCP blending systems.

%Self-consistent field theory (SCFT)

\begin{figure*}[t]
\centering
\includegraphics[width=0.8\textwidth]{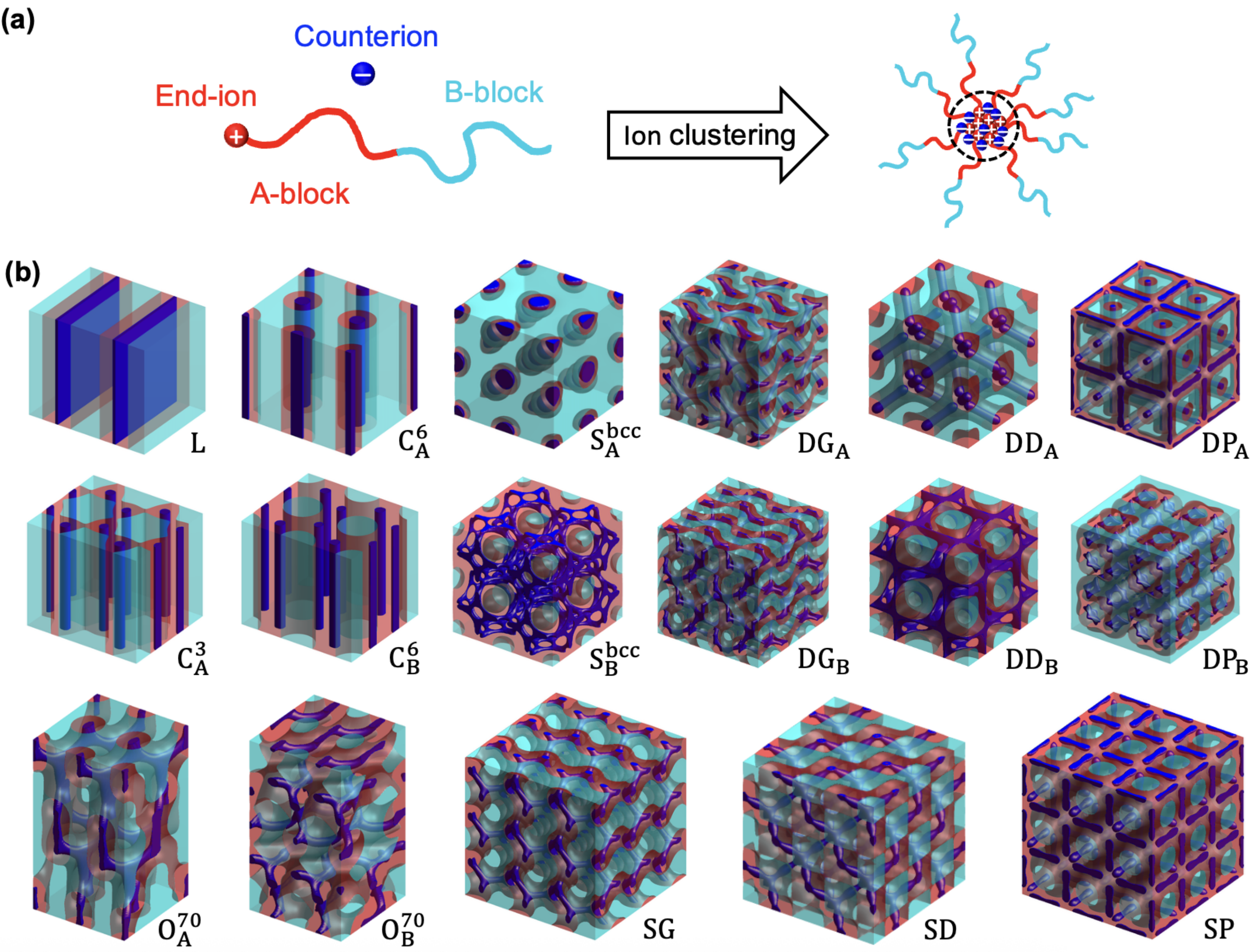}
\caption{(a) Schematic of ion clusters and spontaneous curvature formed in the end-functionalized $AB$ diblock copolymers. (b) Isosurface plots of corresponding polymer density and ion concentration in the candidate ordered morphologies considered in this work,
including lamellae (L), hexagonal cylinders (C$^6_{\rm A}$ and C$^6_{\rm B}$/C$^3_{\rm A}$), body-centered cubic spheres (S$^{\rm bcc}_{\rm A}$ and S$^{\rm bcc}_{\rm B}$), double-gyroid networks (DG$_{\rm A}$ and DG$_{\rm B}$), double-diamond networks (DD$_{\rm A}$ and DD$_{\rm B}$), double-primitive-cubic networks (DP$_{\rm A}$ and DP$_{\rm B}$), orthorhombic $Fddd$ networks (O$^{70}_{\rm A}$ and O$^{70}_{\rm B}$), as well as single-networks of gyroid, diamond and primitive-cubic (SG, SD, and SP).
}

\label{fig1}
\end{figure*}

\section*{Results and Discussion}

%\section*{Results}

We consider an AB diblock copolymer melt with a single monovalent charge placed at the terminus of each A-block, together with its dissociated counterion, and examine the phase behavior as a function of the segregation strength characterized by $\chi N$ and the A-block volume fraction $f_{\rm A}$ at fixed chain length $N = 100$. Statistical segment lengths and volumes are set equal for both blocks ($b = 1.0$ nm, $v = 0.2$ nm$^3$), and the Born radii of the end-ion and counterion are taken equal ($a_+=a_-=a$). To isolate the role of ion correlations from other electrostatic effects,  specifically Born solvation energy and interfacial image charges arising from dielectric contrast, here we assume a uniform dielectric constant.  The Bjerrum length is taken to be a constant $l_B=7.4$ nm independent of other parameters; this value corresponds to $\epsilon_r=7.5$ at room temperature, representative of a low-dielectric polymer melt. 
The coupling parameter characterizing the Coulomb strength can be estimated as $\Gamma= l_B/a \approx 25$ for  ions with radius $a=0.3$nm, indicating the systems are in the strong ion-correlation regime \cite{Levin_2002}. To highlight how strong ion correlations alter the phase behavior of end-ion-functionalized block copolymers relative to the corresponding neutral diblock copolymers, we fix $\Gamma$ and vary $\chi$ independently, so that ion-correlation effects remain consistent throughout the $\chi$-range explored. In real systems, of course, these two parameters both vary with temperature.
Given the dilute ion concentration in this single-charge-per-chain system, excluded volume interactions for the counterions are negligible; however, the ion size still enters the theory through the Born radii for calculating the electrostatic correlation energy.

\subsection*{Candidate Morphologies}

Strong correlations between the end-functionalized charges and their counterions drive the chain ends of the A-block to associate, effectively linking multiple chains at their termini. Since the ions are localized at the ends of the A-blocks, ion concentration is naturally highest in the centroids of the A-domains, generating a curvature preference toward the A-block (Fig. \ref{fig1}a). This curvature preference shapes the accessible morphology space, shown in Fig. \ref{fig1}b. The candidate structures include the classical lamellae (L), body-centered cubic spheres (S$^{\rm bcc}$), hexagonally packed cylinders (C$^6$), orthorhombic $Fddd$ networks (O$^{70}$), and double-gyroid networks (DG), as well as double-diamond networks (DD), double-primitive-cubic networks (DP) and their corresponding single-network counterparts (SG, SD, and SP).
For each candidate structure, the initial polymer density fields are prescribed according to the morphology and symmetry; other fields, including the electrostatic potential and ion concentrations, are then determined self-consistently with the polymer density.

In double-network morphologies (DG, DD, and DP), one block forms two interpenetrating networks embedded in a matrix of the other block; in their single-network counterparts, each block forms its own network, with the two intertwined \cite{Chen_2022,Park_2023,Xie_2022,Cheng_2024,Tian_2025}.
%The curvature preference is reflected in the chain packing: A-chains stretch uniformly when the interface curves toward the A-domain, while B-chains on the concave side experience non-uniform stretching in the DG, DD, DP, O$^{70}$, and S$^{\rm bcc}$ phases.
The curvature preference is reflected in the chain packing: within the ``medial packing'' picture developed by Grason and co-workers \cite{Reddy_2018,Reddy_2021,Reddy_2022,Grason:2023aa,Dimitriyev_2023}, termini of A- and B-blocks are distributed throughout their respective brush-like domains divided by the A-B interface, while the terminal boundaries spread over medial sets within both domains. Spatial variations in medial thickness and chain trajectories therefore govern the chain-stretching costs that contribute to the preference for curvature in the DG, DD, DP, O$^{70}$, and S$^{\rm bcc}$ phases.
At lower A-block fractions, this curvature preference drives a continuous crossover from B-block cylinders C$^6_{\rm B}$ toward the three-coordinated A-block cylinder morphology C$^3_{\rm A}$.

%End-functionalized ions are introduced to manipulate the self-assembled morphology through ion-correlation effects.

\subsection*{Phase Behavior}

\begin{figure}[h]
\centering
\includegraphics[width=0.45\textwidth]{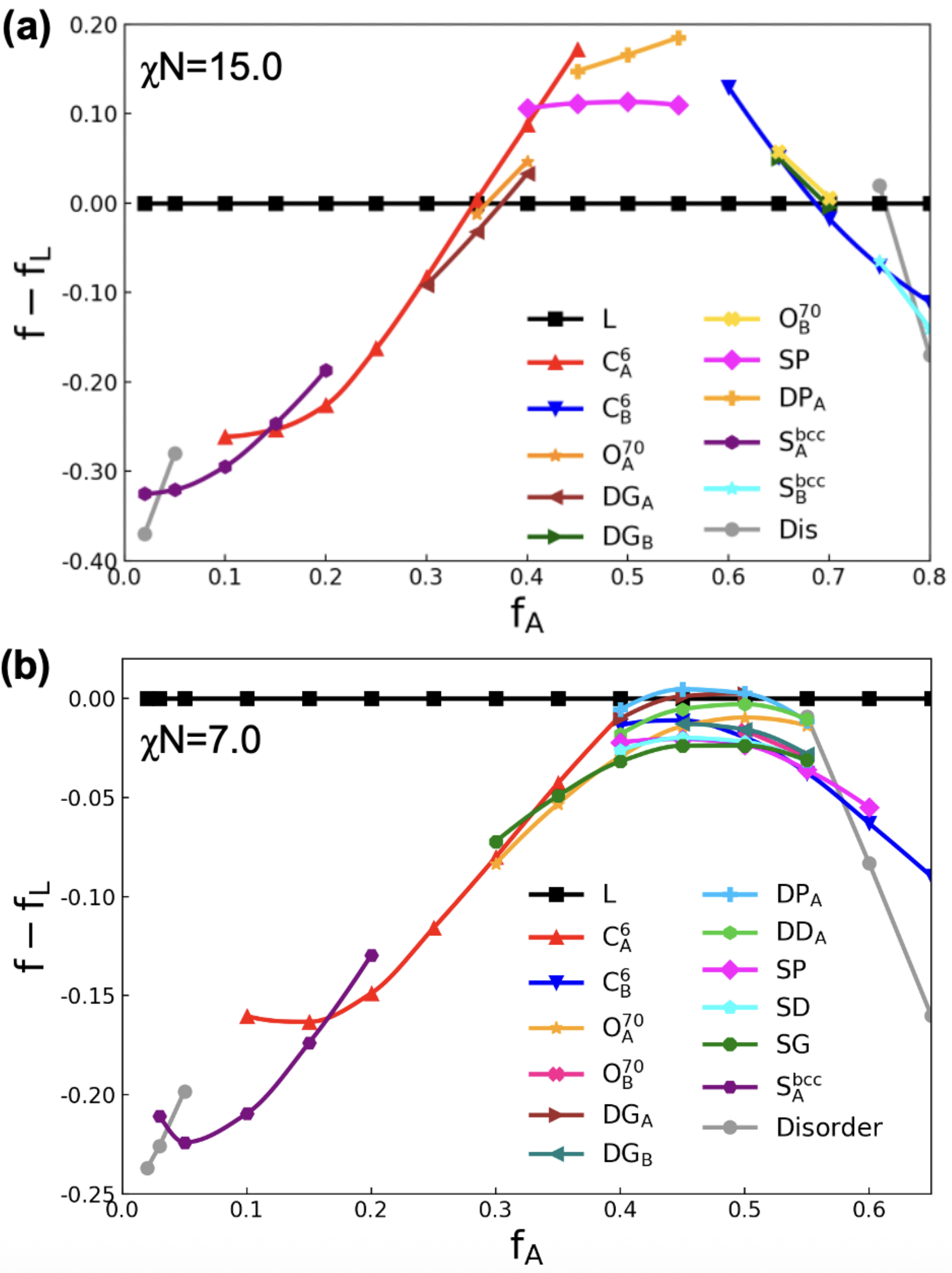}
\caption{Free energy per chain $f$ as a function of the volume fraction of A-block $f_{\rm A}$, relative to that of the L phase: (a) $\chi N=15.0$, and (b) $\chi N=7.0$. $a=0.3$nm. %$a_+=a_-=0.3$nm.
%(c) Schematic illustration comparing the ordered and disordered states.
}
\label{fig2}
\end{figure}

The relative stability of the candidate ordered phases is determined by 
comparing their free energies, each measured relative to the lamellar 
phase. Two representative curves are shown in Fig.~\ref{fig2}. At high 
segregation strength ({$\chi N = 15.0$), only the classical phases 
S$^{\rm bcc}$, C$^6$, DG, and L are stable, while both single- and 
double-network $Im\bar{3}m$ structures are metastable (Fig.~\ref{fig2}a). 
Strikingly, when $\chi N$ decreases to $7.0$---significantly below the 
order-disorder transition of neutral diblock copolymers, 
$(\chi N)_{\rm nc}=10.5$---ordered phases can still form
(Fig.~\ref{fig2}b), because end-association due to ion correlations has similar effect to increasing $N$, thus strenthening the effective segregation between blocks. At this lower 
segregation strength, two single-network structures, SG and SP, emerge 
as the stable phases.
Another single-network, SD ($Pd\bar{3}m$), is found as a metastable state with its free energy very close to those of the SP and SG, particularly near the phase boundary between the latter two.
For these three types of networks, the single-network structures exhibit lower free energies than their corresponding double-network counterparts.

To construct the full phase diagram (Fig.~\ref{fig3}), we compare the free energies of all candidate ordered morphologies with each other and with that of the disordered phase. Because of strong ion clustering, the disordered phase in this system is not a homogeneous melt as in neutral BCPs, but rather a liquid-like collection of ionic clusters in the form of star-like micelles (see inset of Fig.~\ref{fig3}). The free energy of this disordered phase is computed by accounting for the translational entropy of the clusters and their internal free energy, evaluated using the optimal cluster size at each composition---an approach analogous to the treatment of disordered micelles in the neutral BCP melts (see Materials and Methods). The ODT boundary is then determined by the condition at which the free energy of the most stable ordered phase equals that of the disordered phase, reflecting a balance between micellar translational entropy and unfavorable interblock mixing (see Materials and Methods). This analogy with star copolymer melts also explains the overall resemblance between the present phase diagram and that of neutral $n$-arm AB star copolymers with $N$ segments in each diblock arm~\cite{Matsen_2012}: at stronger segregation ($\chi N > 15$) the phase boundaries of the two systems are quite similar, and the ODT in both systems occurs below the well-known mean-field critical value $(\chi N)_{\rm nc}=10.5$ for simple neutral diblock copolymers. However, the 
appearance of stable SP and SG networks at $\chi N < 10$ marks the distinct contribution of end-localized electrostatic interactions.

\begin{figure}[h]
\centering
\includegraphics[width=0.45\textwidth]{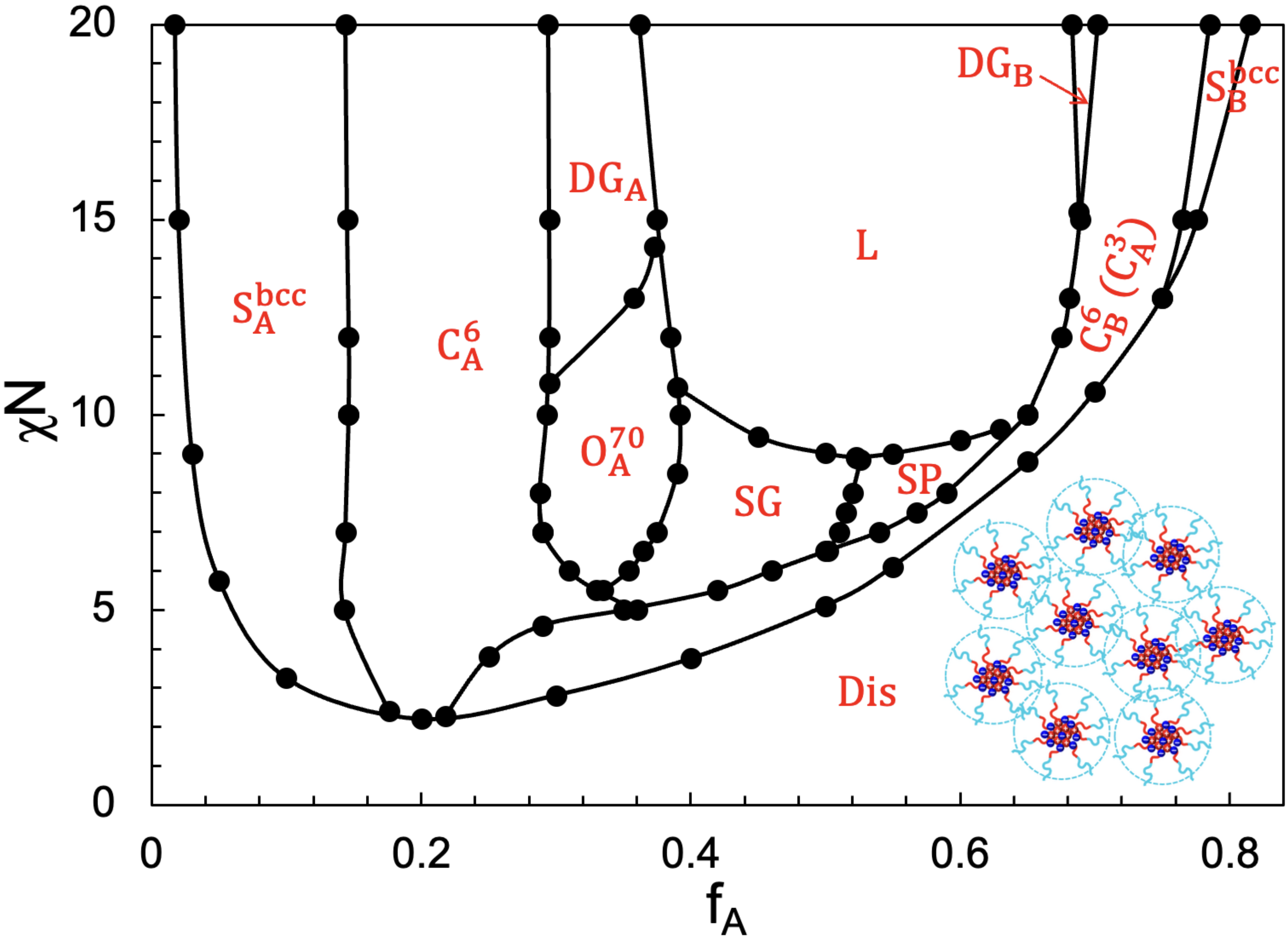}
\caption{Phase diagram in the $f_{\rm A}$-$\chi N$ plane. $a=0.3$nm. %$a_+=a_-=0.3$nm. %$N=100$, 
Symbols denote calculated results, and lines are guides to the eye.
Inset shows a schematic illustration of the disordered state, which is a liquid-like collection of ionic clusters in the form of star-like micelles.
}
\label{fig3}
\end{figure}

The full phase diagram in the $f_{\rm A}$--$\chi N$ plane is shown in 
Fig.~\ref{fig3}. A stable SP region is identified in the window 
$\chi N = 6.0 \sim 9.0$ and $f_{\rm A} = 0.5 \sim 0.6$, centered around nearly 
symmetric compositions and lying entirely below $(\chi N)_{\rm nc} = 10.5$. 
As $\chi N$ increases beyond this window, the SP phase gives way to 
lamellae. Next to the SP phase, and spanning a broader range of stability at lower A-block volume fractions, the SG network emerges.
A comparatively large stability region of the O$^{70}_{\rm A}$ network is located on the low-$f_{\rm A}$ side of the SG phase and at lower segregation strengths than the DG$_{\rm A}$ phase.
Like the SG, the O$^{70}_{\rm A}$ is also a 3-connected network and can be viewed as an anisotropically distorted derivative of the cubic SG structure. Its stabilization therefore appears to arise from the same general preference of end-localized ion clustering for curved network domains.
The SD, DD, and DP networks remain metastable across the entire explored phase diagram.
Beneath the single-network regions, progressing toward the lower-left portion of the diagram, there is a smooth crossover from C$^6_{\rm B}$ to the three-fold coordinated cylinder morphology C$^3_{\rm A}$ as the A-domains withdraw toward the ionic clusters. We note that this crossover is not a true phase transition, as no symmetry breaking occurs going from C$_{\rm B}^6$ to C$_{\rm A}^3$. The phase regions of 
A-block spheres S$^{\rm bcc}_{\rm A}$, cylinders C$^6_{\rm A}$, networks  DG$_{\rm A}$ and O$^{70}_{\rm A}$, are substantially larger than their B-block counterparts, consistent with the preferred spontaneous curvature toward the A-domain. The lowest point of ODT is located at approximately $(f_{\rm A}, \chi N) = (0.2, 2.2)$. It should be noted that this is not a critical point as the transition is of first-order. For $\chi N > 15$ the classical phases characteristic of neutral diblock copolymers are recovered.

The characteristics of the SP phase---stable near symmetric composition, 
below the neutral ODT, replaced by lamellae at stronger segregation---are 
in good qualitative agreement with the experimental observations of the 
Park group~\cite{Lee_2025}.
The notable discrepancy is that the primitive-cubic networks observed experimentally are double-networks, whereas our theory predicts the single-network as the thermodynamically preferred state.
This may reflect several simplifications in the present model: the 
PS-b-PEO/Li$^+$ system is not a true end-functionalized BCP, as Li$^+$ can coordinate with multiple EO groups; the uniform dielectric constant assumption neglects the dielectric contrast between PS and PEO blocks; and
conformational asymmetry between the blocks is not accounted for.
It is also possible that the experimentally observed double-network may be kinetically 
favored over the single-network. Resolving this discrepancy is an important direction for future work.

\subsection*{Stability Mechanism of the Primitive Cubic Network}

In neutral AB diblock copolymers, the stability of self-assembled phases 
is governed by a competition between interfacial energy and chain stretching 
energy~\cite{Matsen_2002,Matsen_Book2005,Matsen_1997,Martinez_Veracoechea_2009,Grason:2023aa,Dimitriyev_2023,Hou_2024,Chang_2024}. The primitive cubic (P) network is 
highly frustrated in this competition: the interfacial curvature varies 
more sharply between struts and nodes in the 6-connected P network than in 
the 4-connected diamond and 3-connected gyroid networks, forcing chains that fill the nodes to stretch 
significantly beyond their equilibrium 
dimensions~\cite{Matsen_1997,Martinez_Veracoechea_2009,Grason_2003,Dimitriyev_2023,Hou_2024,Chang_2024}. This frustration raises the free energy of the P phase distinctly above those of neighboring classical 
phases \cite{Dimitriyev_2023}, making it metastable across virtually the entire neutral phase 
diagram.

In the present system, ion correlations provide a substantial third contribution to the 
free energy that alters this balance. To make this explicit, we decompose the free energy per chain $\Delta f$ into three contributions: the mixing enthalpy contribution responsible for the interfacial 
energy $\Delta u$, chain stretching energy $-T\Delta s$, and electrostatic 
energy $\Delta f_e = \Delta f_{e,\rm MF} + \Delta f_{e,\rm corr}$, each 
measured relative to the completely homogeneous state. The results for 
representative morphologies at $\chi N =7.0$ and $f_{\rm A} = 0.53$ are 
summarized in Table~\ref{tab1}.  
Both the interfacial and stretching energies of the SP phase are higher 
than those of the neighboring SG and L phases as well as the closely related SD phase. Nevertheless, the SP phase 
wins the competition, because its electrostatic energy is the most negative 
of all morphologies considered.

\begin{table}[h]
\centering
%\begin{adjustbox}{max width=\linewidth}
\begin{tabular}{|c|c|c|c|c|c|}
\hline
 Morphology & $\Delta f$ & $\Delta u$ & $-T\Delta s$ & $\Delta f_{\rm e}$ & $\Delta f_{\rm e,corr}$ \\

\hline
SP &-1.842	&-0.378	&6.169 &-7.6323 &-7.6320
\\

\hline
SD & -1.839	&-0.382	&6.075 &-7.5324 &-7.5321
\\

\hline
SG &-1.840	&-0.395	&6.111 &-7.5552 &-7.5549
%\textcolor{red}{SG} &\textcolor{red}{-1.840}	&\textcolor{red}{-0.395}	&\textcolor{red}{6.111} &\textcolor{red}{-7.5552} &\textcolor{red}{-7.5549}
\\

\hline
L &-1.809 &-0.503 &5.993 &-7.2990 &-7.2987
\\

\hline
C$^6_{\rm B}$ (C$^3_{\rm A}$) &-1.838 &-0.317	&6.036 &-7.5578 &-7.5575
\\

\hline
\end{tabular}
%\end{adjustbox}
\caption{Decomposition of the free energy per chain for various morphologies, with each contribution calculated in the excess of completely random-mixing state. $\chi N=7.0$, $f_{\rm A}=0.53$, and $a=0.3 {\rm nm}$.
}
\label{tab1}
\end{table}

Two aspects of Table~\ref{tab1} warrant special attention. First, the mean-field electrostatic term $\Delta f_{e,\rm MF}$ is negligible relative to the correlation term $\Delta f_{e,\rm corr}$ across all morphologies ($| \Delta f_{e,\rm MF}/\Delta f_{e,\rm corr}| < 10^{-4}$); therefore, the electrostatic energy arises almost entirely from correlations. This behavior follows directly from the formation of tight ion clusters under strong electrostatic interactions in these low-dielectric polymers: when ions and counterions aggregate into compact clusters, the charges lie in close proximity within each cluster, greatly reducing net charge separation and leaving negligible net mean electrostatic potential; this effect highlights the necessity of proper treatment of strong correlations.
Second, the ion-correlation energy becomes increasingly negative in the sequence L $\rightarrow$ SG $\rightarrow$ SP, which is the same sequence of increasing interfacial curvature for these morphologies. This pattern directly reflects the geometric preference of ion clusters for compact, curved environments: more highly curved domains correspond to more compact clustering, and thus to a larger gain in correlation energy.
It is noteworthy that ion-correlation energy of the SD network is higher (i.e., less favorable) than that of either SG or SP network, deviating from the trend of increasing interfacial curvature. This unexpected higher ion-correlation energy may arise from additional frustration in ionic arrangement within the SD structure, accounting for its metastability throughout the phase diagram shown in Fig. \ref{fig3}.

\begin{figure}[h]
\centering
\includegraphics[width=0.45\textwidth]{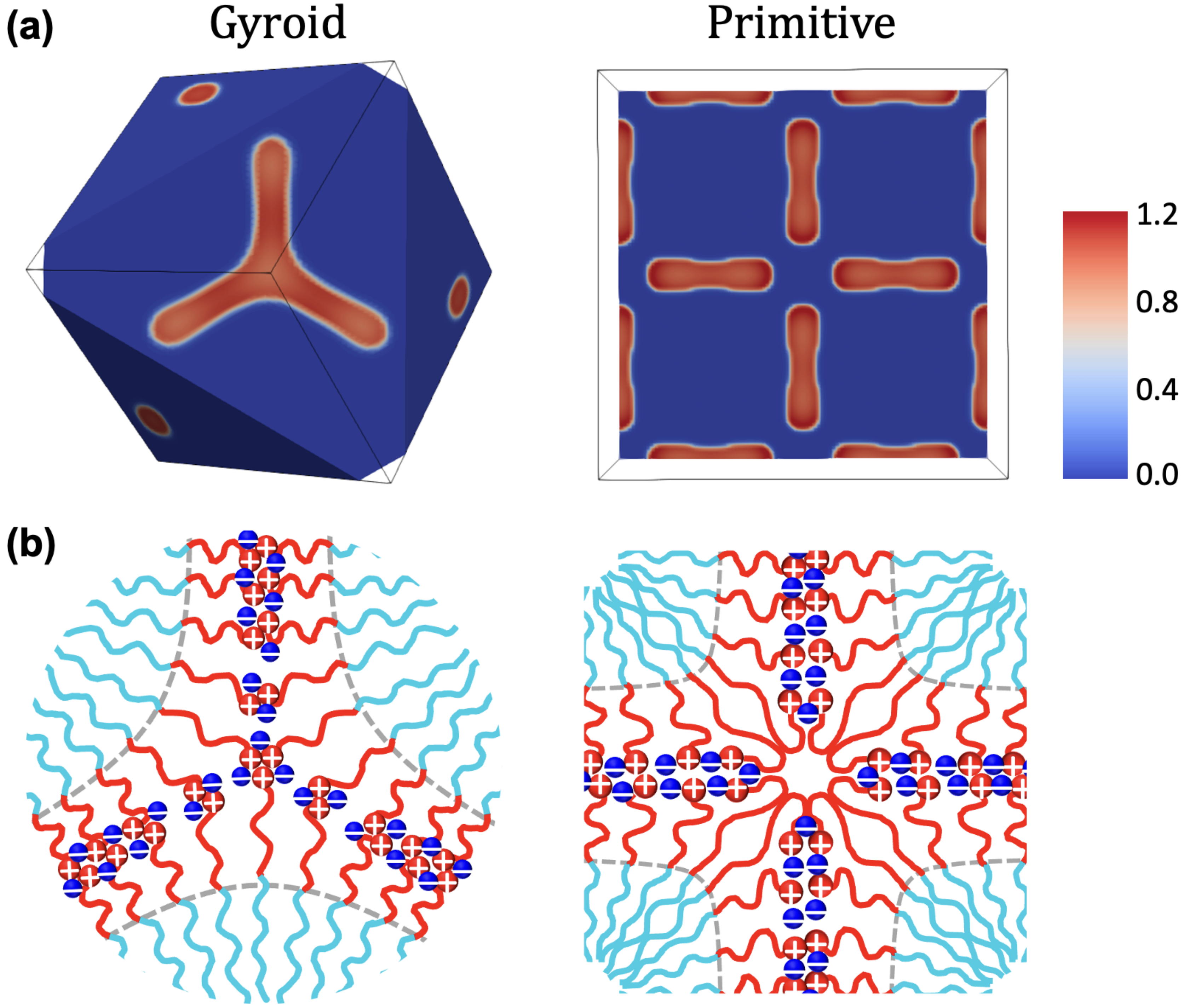}
\caption{(a) 2D ion density profiles and (b) chain arrangements at the cross-sections through the node of networks for gyroid and primitive morphologies.
%$\chi N=7.0$, $f_{\rm A}=0.53$, and $a=$0.3nm.
%$a_+=a_-=$0.3nm.
}
\label{fig4}
\end{figure}

The physical origin of this curvature selectivity is further revealed by comparing the ion distributions in the SG and SP networks (Fig.~\ref{fig4}a). In the gyroid, ions are distributed relatively continuously from struts through nodes, with only a modest accumulation in the struts. In the P network, by contrast, ions are almost entirely confined to the struts---their density in the nodes can be lower by approximately three orders of magnitude. This pronounced local segregation has two coupled consequences, illustrated schematically in Fig.~\ref{fig4}b. On one hand, the absence of ions in the nodes means there is no ionic driving force to keep A-chain ends there; to satisfy incompressibility, A-chains must loop back from the nodes into the struts, amplifying the already substantial stretching penalty of the P network relative to the gyroid. This is reflected in the higher $-T\Delta s$ of SP compared to SG in Table~\ref{tab1}. On the other hand, the concentration of ions into the struts generates highly curved, compact ion clusters, the geometry that maximizes correlation energy gain. The net result is that the reduction in ion-correlation energy through local segregation more than compensates the enhanced stretching penalty, stabilizing the SP phase.
Thus, the geometric frustration that renders the P network metastable in neutral systems becomes now the source of its stability here, as the sharp 
variation in domain geometry actively promotes compact ion clustering.

The relative stability of SG versus SP at shorter A-block fractions 
(Fig.~\ref{fig3}) can be understood from the same physical picture. As 
$f_{\rm A}$ decreases, A-blocks become shorter and have less contour length 
available to simultaneously reach the struts and loop back through the 
nodes of the SP network. The stretching penalty per chain in SP therefore 
grows as $f_{\rm A}$ decreases. In the SG network, the more 
gradual variation in domain thickness between struts and nodes allows chains 
to fill the network more uniformly without requiring many loops-back, so 
the stretching penalty scales more favorably with chain length. Below a 
crossover composition, the stretching cost of SP exceeds the additional 
correlation energy gain from its more pronounced ion segregation, and SG 
becomes the preferred phase.

\subsection*{Effect of Ion Size}

\begin{figure*}[t]
\centering
\includegraphics[width=0.75\textwidth]{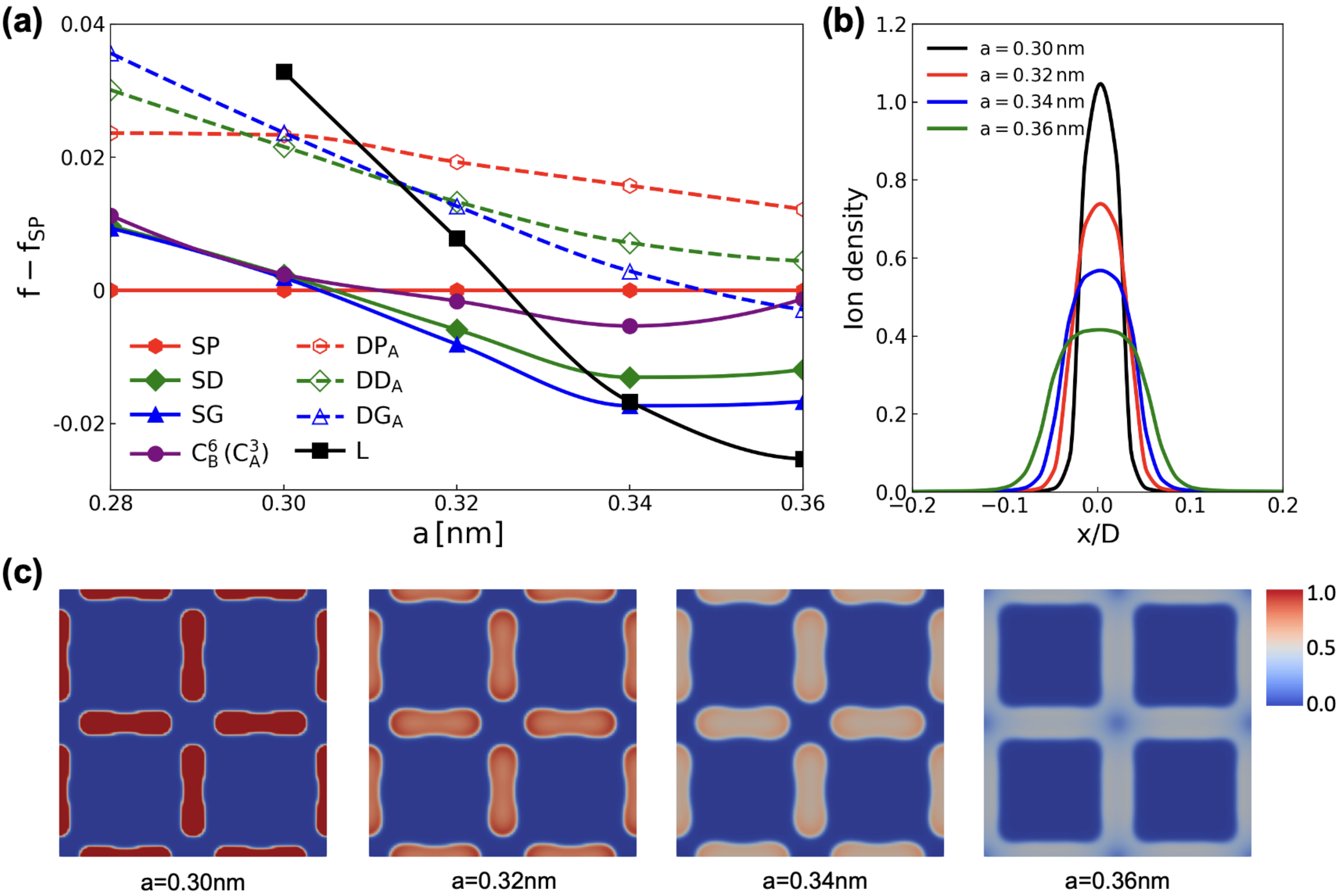}
\caption{Effects of ion size on the relative stability of structures and ion density profiles.
(a) Free energy per chain $f$ as a function of the ion radius $a$, relative to that of the SP phase $f_{\rm SP}$.
(b) Ion density distribution along the normal direction of the L phase with the period length $D$.
(c) 2D ion density profiles of the SP phase for various ion radii $a$. $\chi N=7.0$, $f_{\rm A}=0.53$. 
%, and $a_+=0.3$nm.
%Free energy per chain $f$ as a function of counterion radius $a_-$. $a_+=0.3$nm.
}
\label{fig5}
\end{figure*}

The strength of ion correlations depends sensitively on ion size: smaller ions, characterized by a smaller Born radius, interact more strongly with neighboring charges and form more compact, tightly bound clusters. Varying the ion size therefore 
provides a direct and experimentally accessible means of tuning the 
correlation-driven phase behavior. Fig.~\ref{fig5}a shows the free energies 
of competing morphologies as a function of ion radius $a$ at $\chi N = 7.0$ 
and $f_{\rm A} = 0.53$.
When the ions are sufficiently small (for example, $a<0.28$ nm), the relative stability of both single- and double-network morphologies follows the sequence P $>$ D $>$ G, indicating a preference for network topologies with a larger number of struts connecting the nodes. As $a$ increases beyond $0.30$~nm and $0.34$~nm, the SP and SG networks, respectively, lose their stability, and the lamellar phase becomess the most stable structure. 
This transition reflects the weakening of ion correlations with increasing 
ion size: larger ions form more diffuse, loosely bound clusters, as shown 
directly by the ion density profiles in the lamellar phase 
(Fig.~\ref{fig5}b), where broader distributions are evident for larger $a$. 
As clustering weakens, the interfacial and chain-stretching energies 
regains their dominance, favoring the lamellar phase which carries the 
lowest contributions from these two terms.
For the SP phase, its loss of stability with increasing ion size is accompanied by gradual disappearance of local ion-segregation (Fig. \ref{fig5}c), highlighting the key mechanism of localized ionic interactions in stabilizing this highly frustrated morphology.
This ion-size dependence 
provides a direct point of contact with experiment: Park and coworkes observed that reducing anion size or modifying PEO chain ends with $-\rm PO_3H_2$ groups, both of which promote stronger, more localized 
ionic interactions, significantly expands the stability window of the primitive-cubic network phase \cite{Lee_2025}, in qualitative agreement with our theoretical results.
%Although the DP phase is metastable within the parameter space explored in this work, it exhibits closely competitive free energy with the SP phase for smaller ions ($a<0.25$ nm), where it is the second most stable structure (Fig.~\ref{fig5}a). This near-degeneracy suggests that factors not captured in the present model, such as dielectric contrast between the blocks, conformational asymmetry, or the ability of Li$^+$ to coordinate with multiple EO groups, could tip the balance toward the DP phase, thus helping to reconcile its experimental observation~\cite{Lee_2025} with the metastability predicted here.

Although the DP phase is metastable within the parameter range examined in this study, the trend shown in Fig.~\ref{fig5}a appears to suggest that it could become competitively stable with the SP phase at smaller ion sizes ($a < 0.28$ nm, currently out of reach for our grid size). This trend suggests that features not included in the current model---such as dielectric contrast between blocks, conformational asymmetry, or the capacity of Li$^+$ to coordinate with multiple EO groups---could tip the balance 
toward the DP phase, thereby helping to reconcile its experimental observation~\cite{Lee_2025} with the metastability predicted here. These possibilities remain to be further explored. 

The dielectric constant is an important parameter that warrants further investigation. In this study, we have fixed $\varepsilon_r = 7.5$ and varied the ion radius within the range $a = 0.28$–$0.36$ nm. Reducing $\varepsilon_r$ is expected to produce effects qualitatively similar to those of decreasing the ion size: for $\varepsilon_r = 5$, for example, ion–ion correlations can remain strong even for comparatively larger ions. More generally, any factor that enhances ion correlations---such as smaller ion size, a lower dielectric constant, or higher ion valency---is likely to stabilize highly frustrated network phases. A more systematic investigation of this expanded parameter space will be pursued in future work.

\section*{Conclusions}

We have applied an ion-correlation–augmented self-consistent field theory to AB diblock copolymer melts with ions at the A-block termini. Ion correlations introduce a new control mechanism for self-assembly: strong ion correlations drive chain-end association and favor curvature toward the A domain, stabilizing the highly frustrated single-primitive cubic network ($Pm\bar{3}m$) and single-gyroid network ($I$4$_1$32) as the thermodynamically stable states in a region of parameter space below the neutral order–disorder transition.
Free energy decomposition shows that the electrostatic energy, arising almost entirely from non-mean-field correlation effects, becomes increasingly negative with increasing 
interfacial curvature. In the P network, strong local ion segregation into the struts forms compact curved clusters that more than offset packing frustration, so the geometric frustration that makes the P network metastable in neutral systems becomes the source of its stability here.

While recent theoretical work has shown that frustrated network phases including the DP can in principle be stabilized through elaborate chain  architectures, such as the specially designed $AB$ dendron-like block copolymers and $A'(A''B)_5$ miktoarm star copolymers studied by Li and coworkers ~\cite{Qiang_2020,LiQY_2022,Hou_2024,HChen_2025}, this requires the simultaneous operation of multiple finely tuned mechanisms and remains experimentally unconfirmed. The end-group interaction route identified here offers a 
conceptually simpler and experimentally more accessible alternative \cite{Min_2021b,Lee_2024b}. The key physical requirement is that the interactions be end-localized and  curvature-selective, and strong enough to provide the additional free energy contribution needed to overcome packing  frustration. This curvature-selective end-group mechanism appears to be general beyond ionic interactions: the Park group has demonstrated that hydrogen bonding and dipole-dipole end-group interactions similarly stabilize the $Im\bar{3}m$ network~\cite{Lee_2024a}, suggesting that curvature-selective end-group association is the key physical ingredient, regardless of the specific chemical origin of the end-end attraction~\cite{LEE2025102003}. The 
present theory provides the first mechanistic demonstration of this principle.

A further distinctive feature of the present system is that its
behavior is not determined solely by $\chi N$ and $f_{\rm A}$, in contrast to 
the mean-field phase diagram of conformationally symmetric neutral diblock copolymers. The total chain 
length $N$ enters through several effects: 
increasing $N$ at fixed charge per chain dilutes the ion concentration in 
the A-domain, weakening clustering and destabilizing the single-network 
phases; but increasing $N$ also provides longer A-blocks with more contour 
length to accommodate the loop-back through the nodes of the P network, 
relieving packing frustration and selectively stabilizing SP over SG. The ODT boundary is also affected by the average cluster size, which depends on $N$.
The combined effects will lead to shifts in the relative stability of the SP, SG, and ${\rm C^3_A}$ phases, as well as the ODT boundary, as the chain length is varied. If the polymer chain is sufficiently long such that polymer domain is much larger than the characteristic ion-cluster size, the charges may segregate into discrete clusters in an otherwise nearly uniform surrounding A-domain.} Exploring these chain-length dependencies through systematic calculations represents a natural and experimentally relevant extension of the present work.

The present results also point to other fruitful directions for future 
investigation. Within the current framework, examining the effects of charge 
placement along the A-backbone and the effects of multiple charges per chain \cite{Katkar_2014,Danielsen_2019,Lytle_2019,Li_2021b}, 
would establish the full scope of correlation-driven phase control. As the 
number of charges increases from one toward a fully charged block, there 
will be a crossover between the clustering-dominated regime identified here 
and the uniform-charge regime, and the location and character of that 
crossover are open questions of both theoretical and practical interest. 
Beyond the current model, incorporating dielectric contrast between the A and B blocks, conformational asymmetry, and the capacity of ions such as Li$^+$ to coordinate with multiple repeat units of neutral chains \cite{Ren_2015} could bring the theory into closer contact with specific experimental systems. These additional effects could not only resolve the single- versus double-network discrepancy identified here but also stabilize other nearly degenerate structures, including the single-diamond network.
%% thermal fluctuation in polymer by Matsen_2023
More broadly, the curvature-selective end-group mechanism established in this work 
suggests a general design strategy for designing frustrated network phases with desirable 
transport properties \cite{Min_2021a,Lee_2025}.

Finally, in our model, the charge carried by each ion is assumed to have a Gaussian spatial distribution, which regularizes the short-range Coulomb interaction and provides an effective coarse-grained measure of ion radius. Alternative finite-size ion models are expected to qualitatively preserve the correlation-induced ion-clustering mechanism, although the predicted correlation strength and phase boundaries may differ quantitatively. Accordingly, our present framework should be regarded as a mesoscopic description of ion correlations rather than an exact microscopic representation of local ionic structure. Nevertheless, our theory provides a self-consistent and computationally efficient platform for capturing strong ion-correlation effects in inhomogeneous polymer systems and for systematically exploring the hierarchical phase behavior.

\subsection*{MATERIALS AND METHODS}

\subsection*{Ion-Correlation Augmented SCFT}
We consider an incompressible melt of AB-type diblock polymers consisting 
of $n$ identical polymer chains in a volume $V$. Each A-block carries one 
positive elementary charge $+e$ at its terminus, accompanied by a  
counterion with charge $ -e$; B-blocks are neutral. To focus 
on ion-correlation effects, we assume equal Kuhn lengths $b$ and segment 
volumes $v_0$ for both blocks. The total number of monomers per chain is 
$N = N_A + N_B$, where $N_A$ and $N_B$ are the number of monomers of the A- 
and B-blocks, respectively. The Flory-Huggins parameter $\chi$ describes 
the immiscibility between blocks, and identical dielectric constants 
$\varepsilon_r$ are assumed for both blocks. To avoid divergences of the ion self-energy and overestimation of 
ion correlations from the point-charge model, the charges are described 
by a finite-sized ion model with distribution function $h_K(\mathbf{r}, 
\mathbf{r}')$ and Born radius $a_K$ ($K = \pm $)~\cite{Wang:2010wk}, taken to be Gaussian. The polymer chains are modeled using the 
discrete Gaussian chain model, in which each charged monomer is treated as 
an explicit charged particle with a finite spread of segment charge.

The key idea of the ion-correlation augmented SCFT is to treat the polymer density field and incompressibility at the self-consistent field level~\cite{Fredrickson_Book2006}, while the electrostatic correlations are accounted for by a nonperturbative renormalized Gaussian fluctuation theory~\cite{Wang:2010wk,Agrawal:2022ux,Duan2025PEBrush}.
The effectiveness of this method has been established in several earlier studies, most notably by Wang and coworkers, which reported quantitative or semiquantitative consistency with both experimental data and simulations for systems involving small ions as well as polyelectrolytes~\cite{Agrawal:2022ux,Agrawal_2024,Duan2025PEBrush,Duan_2024a}.
The details of the derivation are given in our earlier papers~\cite{Duan2025PEBrush}. Here we summarize the key self-consistent 
equations for the volume fraction $\phi_P(\mathbf{r})$ and 
conjugate field $\omega_P(\mathbf{r})$ of $P$-blocks ($P = A, B$), the mean 
electrostatic potential $\psi(\mathbf{r})$, the ion concentration $c_K(\mathbf{r})$ 
and self-energy $u_K(\mathbf{r})$ of ion $K$ ($K = \pm$), and 
the Green function (effective interaction between charges) $G(\mathbf{r}, \mathbf{r}')$:

\begin{subequations}\label{SCFTEqs}
\begin{align}
\phi_P({\bf r}) &= \frac{1}{NQ} \sum_{i=1}^{N_P} q_P^{\dagger}({\bf r};i) e^{V_P({\bf r};i)} q_P({\bf r};i) %~~(P=A,B)
\end{align}
\begin{align}
\omega_A({\bf r})  & = \chi \phi_B({\bf r}) + \xi({\bf r})
\end{align}
\begin{align}
\omega_B({\bf r})  & = \chi \phi_A({\bf r}) + \xi({\bf r})
\end{align}
\begin{align}\label{PB}
-\epsilon \nabla^2\psi({\bf r}) &= c_+({\bf r}) - c_-({\bf r})
%\sum_{K=\pm} z_K c_K({\bf r})
\end{align}
\begin{align}
c_+({\bf r}) &= \frac{1}{Nv_0Q}  q_A^{\dagger}({\bf r};1) e^{V_A({\bf r};1)} q_A({\bf r};1)
\end{align}
\begin{align}
c_-({\bf r}) &= \frac{\overline{c}_-}{Q_-} e^ { \psi({\bf r}) - u_-({\bf r}) }
%c_{ctr}({\bf r}) &= \frac{\overline{c}_{ctr}}{Q_{ctr} } \exp \left[-z_{ctr}\psi({\bf r}) - u_{ctr}({\bf r}) - \frac{v_{ctr}}{v_0}\xi({\bf r}) \right]
\end{align}

\begin{align}\label{uK}
u_K({\bf r}) &= \frac{1}{2} \int d{\bf r^{\prime}} d{\bf r^{\prime\prime}} h_K({\bf r},{\bf r}^{\prime}) G({\bf r}^{\prime},{\bf r}^{\prime\prime}) h_K({\bf r^{\prime\prime}},{\bf r})
\end{align}
\begin{align}\label{GreenFunc}
-\epsilon \nabla^2_{\bf r} G({\bf r},{\bf r}^{\prime}) + 2I({\bf r}) G({\bf r},{\bf r}^{\prime}) &= \delta({\bf r}-{\bf r}^{\prime})
\end{align}
\end{subequations}
where $\overline{c}_-$ is the average concentration of counterions. $\epsilon=k_BT \varepsilon_0 \varepsilon_r /e^2$ is the scaled permittivity with $\varepsilon_0$ the vacuum permittivity. %and $\epsilon_r({\bf r})$ the local dielectric constant that can be evaluated based on the local composition \cite{Sing_2014a,Zhuang_2021}.
$I({\bf r})=(1/2)\sum_{K=\pm} c_K({\bf r})$ is the local ionic strength. 
$Q=V^{-1} \int d{\bf r}~ q^{\dagger}_A({\bf r};1)$ is the single-chain partition function of copolymer, and $Q_{-}= V^{-1} \int d{\bf r}~ \exp[\psi({\bf r}) - u_{-}({\bf r})]$ is the partition function of counterions. $\xi({\bf r})$ is the pressure field enforcing the local incompressibility condition.
The forward and backward chain propagators $q_P({\bf r};i)$ and $q_P^{\dagger}({\bf r};i)$ for the $P$-block ($P=A, B$) are determined~\cite{Fredrickson_Book2006}  by 
\begin{subequations}\label{q_P}
\begin{align} %\label{q_forward}
q_P({\bf r};i) &= e^{-V_P({\bf r};i)} \int d{\bf r^{\prime}} \Phi ({\bf r} - {\bf r^{\prime}}) q_P({\bf r^{\prime}};i-1) \\
q_P^{\dagger}({\bf r};i) &= e^{-V_P({\bf r};i)} \int d{\bf r^{\prime}} \Phi ({\bf r} - {\bf r^{\prime}}) q_P^{\dagger}({\bf r^{\prime}};i+1)
\end{align}
\end{subequations}
with the initial conditions $q_P({\bf r};1) = e^{-V_P({\bf r};1)}$ and $q_P^{\dagger}({\bf r};N_P)=e^{-V_P({\bf r};N_P)}$.
$V_P({\bf r};i)$ is the total field experienced by the $i$-th segment of the $P$-block, with
$V_A({\bf r};i)=\omega_A({\bf r})+\psi({\bf r})+u_+({\bf r})$ for $i=1$ and $V_A({\bf r};i)=\omega_A({\bf r})$ otherwise; $V_B({\bf r};i)=\omega_B({\bf r})$.
$\Phi ({\bf r} - {\bf r^{\prime}})$ in Eq. \ref{q_P} is the Gaussian bond transition probability \cite{Fredrickson_Book2006}.

The resulting free energy per chain of the system $f=v_0 N F/(Vk_BT)$ is given by
\begin{align}\label{FreeE}
%&f = \frac{\overline{\phi}}{N} \left[ \ln \left(\frac{\overline{\phi}}{NQ} \right) - 1 \right]
%+ \overline{c}_{ctr}v_0  \left[ \ln \left(\frac{\overline{c}_{ctr}v_{ctr}}{Q_{ctr}}\right) -1 \right] \nonumber\\
&f = - \ln Q
- N \overline{c}_{-}v_0  \ln Q_{-} \nonumber\\
&+ \frac{N}{V} \int d{\bf r} \left [ \chi \phi_A\phi_B -\omega_A \phi_A -\omega_B \phi_B + \xi (\phi_A+\phi_B-1) \right] \nonumber\\
&+ \frac{v_0 N}{V} \int d{\bf r} \left( \frac{\epsilon}{2}\psi\nabla^2\psi \right)
\nonumber\\
&+ \frac{v_0 N}{V} \int d{\bf r} \sum_{K=\pm} c_K({\bf r}) \left[ \int^1_0 d{\tau} u_K({\bf r};\tau)-u_K({\bf r}) \right]
\end{align}
The last term in Eq. \ref{FreeE} is the electrostatic correlation energy derived from the charging method~\cite{Wang_2015,Agrawal:2022ux}, where $\tau$ ($0 \leq \tau \leq 1$) is 
the charging parameter. The intermediate self-energy $u_K(\mathbf{r};\tau)$ 
is calculated from the Green function $G(\mathbf{r},\mathbf{r}';\tau)$, 
obtained by solving Eq.~\ref{GreenFunc} with $I(\mathbf{r})$ replaced by 
$\tau I(\mathbf{r})$.

%%\textcolor{red}{\subsection*{Pair Correlation Function} [to be added]}

\subsection*{Determination of the Order–Disorder Transition (ODT)}

Under the strong ion-correlation conditions considered here, ionic clusters 
are present throughout the system. The disordered (Dis) phase consists of mobile star-like polymer micelles each centered on an ionic cluster. To rigorously describe these liquid-like micelles, one must account for their internal structure, their translational entropy, and their mutual interactions. Considerations of these effects even in neutral systems require accounting for concentration fluctuations beyond the mean-field description of block copolymers~\cite{Wang_2005,Dormidontova_2001}. %Abetz1996 
%Since our treatment of the polymer density and compressibility field is at the level of the self-consistent-field (i.e., mean-field), we make the simplifying approximation of treating these micelles as $n^*$-arm AB star polymers, with $n^*$ being the optimal size of the micelles.
In the present work, we assume that the ODT is primarily governed by an interplay between the translational entropy of micelles and the enthalpic penalty from Flory--Huggins interactions, with the latter modeled at the level of random mixing, consistent with the mean-field treatment of the nonelectrostatic aspects of the problem.
We further assume that the Dis phase is a
monodisperse system of spherical micelles, and the aggregation number of each micelle $n^*$ remains unchanged during the transition from the Dis state to the ordered ${\rm S_{bcc}}$ phase. The latter is consistent with existing works on neutral diblock copolymers, where changes in the aggregation number between the liquid-like micelles and the spheres in the BCC phase were reported to be small \cite{Dormidontova_2001,Schwab1997,Dorfman_2023}. With this assumption the same aggregation number in the Dis and S$_{\rm bcc}$, it is reasonable to neglect changes in other energetic contributions, such as the chain-stretching and electrostatic energies.

Under these approximations, the difference in free energy per chain between the Dis phase and the S$_{\rm bcc}$ phase can be estimated as
\begin{align}\label{dF_Dis&S}
&\Delta f_{\rm Dis/S_{bcc}} = (u_{\rm Dis}-u_{\rm S_{bcc}}) - \frac{1}{n^*}
\end{align}
where $u_{\rm Dis}$ and $u_{\rm S_{bcc}}$ are their respective Flory--Huggins interaction energies. $u_{\rm Dis}=\chi N f_{\rm A}(1-f_{\rm A})$, and $u_{\rm S_{bcc}}$ can be directly extracted from our SCFT calculations.
The last term in Eq. \ref{dF_Dis&S} accounts for the translational entropy of an incompressible system of micelles in the Dis phase with aggregation number $n^*$~\cite{Dormidontova_2001}.
$n^*$ is found to lie in the range $n^* = 15 \sim 25$ in the vicinity of ODT, depending on the block composition. Once the phase boundary between the Dis and S$_{\rm bcc}$ phases is determined by setting the free energy difference in Eq. \ref{dF_Dis&S} to $0$, the phase boundaries between the Dis phase and other ordered phases are then determined indirectly by equating their respective free-energy with the ${\rm S_A^{bcc}}$ phase.

\subsection*{Details of Numerical Calculations}
The equilibrium structure and free energy are obtained by solving Eqs. \ref{SCFTEqs}-\ref{FreeE} iteratively to convergence. For each candidate ordered phase, calculations are carried out under periodic boundary conditions with the space group symmetry imposed as an initial condition, and the free energy is minimized with respect to the unit cell size. Phase boundaries are determined by comparing the converged free energies of all candidate phases at each state point.

Both chain propagators (Eq. \ref{q_P}) and the Poisson equation (Eq. \ref{PB}) are solved by pseudo-spectral method using the FFTW package \cite{Tzeremes:2002aa,Vigil_2022,FFTW}. The Green function $G({\bf r},{\bf r}^{\prime})$ (Eq. \ref{GreenFunc}) is solved using the alternating-direction implicit method (ADI) \cite{Hoffman_2018}. A grid spacing of 0.25b is used, which we have verified gives phase boundaries insensitive to further grid refinement.

To further accelerate the calculations for three-dimensional periodic morphologies, we implemented a crystallographic fast Fourier transform scheme following the approach of Qiang and Li~\cite{Qiang_2020b}. In this method, the pseudo-spectral propagation exploits the space-group symmetries of the target morphology, such that FFT operations are carried out only on symmetry-irreducible grid points. A similar symmetry-adapted procedure is also applied to solving the Poisson equation.
Then full fields can be reconstructed through symmetry operations.
This approach substantially reduces both computational time and memory usage for candidate phases including G, D, P, and O$^{70}$ networks by one order of magnitude, while preserving the numerical accuracy of the equilibrium free energies and phase boundaries.

To improve the computational efficiency in solving the Green function, $G$ is decomposed into a short-range part and a long-range part, $G=G_s+G_l$ \cite{Agrawal:2022ux}. The short-range part $G_s$, which captures the local electrostatic environment, is analytically tractable. The long-range part $G_l$, which accounts for spatially varying ionic strength, is solved numerically. This decomposition also removes the singularity arising from the $\delta$-function on the right hand side of Eq.~\ref{GreenFunc} when consistent numerical treatment of $G$ and $G_s$ is used to 
obtain $G_l = G - G_s$~\cite{Wang_2015}. For three-dimensional morphologies,
% such  as SP, DG, and S$^{\rm bcc}$,
evaluating $G_l$ is computationally demanding due to its six-dimensional character. Since $G_l$ captures electrostatic 
effects acting at length scales much larger than the ion size~\cite{Duan2025PEBrush}, 
its relative change during iteration is much smaller than that of $G_s$, 
enabling $G_l$ to be updated only every $m_G = 400$ iterations without 
affecting the final results. Taken together, these strategies reduce the end-to-end computational cost of solving the Green function to a level comparable to 
that of solving the chain propagators.

Fields conjugate to the polymer densities are updated by simple mixing: 
$\omega_P^{\rm new} \leftarrow \lambda\omega_P^{\rm new} + 
(1-\lambda)\omega_P^{\rm old}$ ($P = A, B$). The same rule is applied to 
the electrostatic potential $\psi$, self-energies $u_K$ ($K = +, -$), and Green function $G$. The incompressibility field is updated as 
$\xi^{\rm new} \leftarrow \xi^{\rm old} + \kappa(\phi_A + \phi_B - 1)$. 
The iteration coefficients are $\lambda = 10^{-2}$ and $\kappa = 10$ for 
$\omega$ and $\xi$; $\lambda = 4\times 10^{-3}$ and $10^{-3}$ for $\psi$ 
and $u_K$, respectively. Convergence is considered to be achieved when the relative errors 
in the free energy, electric potential, and incompressibility condition 
fall below $10^{-8}$, $10^{-7}$, and $10^{-5}$, respectively.

%To further stabilize the convergence, we use the following strategy to update the fields. Fields  conjugate to  the polymer densities are updated by a simple mixing rule, i. e., $\omega^{new}_{P} \leftarrow \lambda \omega^{new}_{P} +(1-\lambda) \omega^{old}_{P}$ $(P=A,B)$. The same rule is adopted for updating the electrostatic potential $\psi$, self-energy $u_K$ ($K=A,ctr$), and Green function $G$. The field conjugated to the incompressibility condition is updated by $\xi^{new} \leftarrow \xi^{old}+\kappa (\phi_A + \phi_B -1)$, where the second term on the r.h.s is adopted to reinforce the incompressibility. The iteration coefficients $\lambda=2\times 10^{-3}$ and $\kappa=10$ are used in the present work. 

%%% It should be noted that we neglect the contribution of intra-chain correlation to self-energy due to its less importance than the individual contribution of a charged monomer $u_K(\bf r)$ under the condition of high ionic strengths (either high salt concentration or high PE density). \cite{Shen_2017,Duan:2024aa} %%%

\section*{Data, Materials, and Software Availability}
Data and code supporting the findings of this study are available at https://doi.org/10.5281/zenodo.20416412. Model parameters and computational algorithms used in this study are included in the article.

\section*{Acknowledgments}

The authors thank Professor Mahesh K. Mahanthappa for helpful discussions.
This research is supported 
by funding from the Hong Kong Quantum AI Lab and AIR@InnoHK of the Hong Kong Government.

\bibliography{Tex_Refs}

\begin{thebibliography}{135}%
\makeatletter
\providecommand \@ifxundefined [1]{%
 \@ifx{#1\undefined}
}%
\providecommand \@ifnum [1]{%
 \ifnum #1\expandafter \@firstoftwo
 \else \expandafter \@secondoftwo
 \fi
}%
\providecommand \@ifx [1]{%
 \ifx #1\expandafter \@firstoftwo
 \else \expandafter \@secondoftwo
 \fi
}%
\providecommand \natexlab [1]{#1}%
\providecommand \enquote  [1]{``#1''}%
\providecommand \bibnamefont  [1]{#1}%
\providecommand \bibfnamefont [1]{#1}%
\providecommand \citenamefont [1]{#1}%
\providecommand \href@noop [0]{\@secondoftwo}%
\providecommand \href [0]{\begingroup \@sanitize@url \@href}%
\providecommand \@href[1]{\@@startlink{#1}\@@href}%
\providecommand \@@href[1]{\endgroup#1\@@endlink}%
\providecommand \@sanitize@url [0]{\catcode `\\12\catcode `\$12\catcode `\&12\catcode `\#12\catcode `\^12\catcode `\_12\catcode `\%12\relax}%
\providecommand \@@startlink[1]{}%
\providecommand \@@endlink[0]{}%
\providecommand \url  [0]{\begingroup\@sanitize@url \@url }%
\providecommand \@url [1]{\endgroup\@href {#1}{\urlprefix }}%
\providecommand \urlprefix  [0]{URL }%
\providecommand \Eprint [0]{\href }%
\providecommand \doibase [0]{https://doi.org/}%
\providecommand \selectlanguage [0]{\@gobble}%
\providecommand \bibinfo  [0]{\@secondoftwo}%
\providecommand \bibfield  [0]{\@secondoftwo}%
\providecommand \translation [1]{[#1]}%
\providecommand \BibitemOpen [0]{}%
\providecommand \bibitemStop [0]{}%
\providecommand \bibitemNoStop [0]{.\EOS\space}%
\providecommand \EOS [0]{\spacefactor3000\relax}%
\providecommand \BibitemShut  [1]{\csname bibitem#1\endcsname}%
\let\auto@bib@innerbib\@empty
%</preamble>
\bibitem [{\citenamefont {Hamley}(1998)}]{Hamley_Book1998}%
  \BibitemOpen
  \bibfield  {author} {\bibinfo {author} {\bibfnamefont {I.~W.}\ \bibnamefont {Hamley}},\ }\href {https://doi.org/10.1093/oso/9780198502180.001.0001} {\emph {\bibinfo {title} {The Physics of Block Copolymers}}}\ (\bibinfo  {publisher} {Oxford University Press},\ \bibinfo {year} {1998})\BibitemShut {NoStop}%
\bibitem [{\citenamefont {Bates}\ and\ \citenamefont {Fredrickson}(1999)}]{Bates_1999}%
  \BibitemOpen
  \bibfield  {author} {\bibinfo {author} {\bibfnamefont {F.~S.}\ \bibnamefont {Bates}}\ and\ \bibinfo {author} {\bibfnamefont {G.~H.}\ \bibnamefont {Fredrickson}},\ }\bibfield  {title} {\bibinfo {title} {Block copolymers---designer soft materials},\ }\href {https://doi.org/10.1063/1.882522} {\bibfield  {journal} {\bibinfo  {journal} {Phys. Today}\ }\textbf {\bibinfo {volume} {52}},\ \bibinfo {pages} {32} (\bibinfo {year} {1999})}\BibitemShut {NoStop}%
\bibitem [{\citenamefont {Bates}\ \emph {et~al.}(2012)\citenamefont {Bates}, \citenamefont {Hillmyer}, \citenamefont {Lodge}, \citenamefont {Bates}, \citenamefont {Delaney},\ and\ \citenamefont {Fredrickson}}]{Bates_2012}%
  \BibitemOpen
  \bibfield  {author} {\bibinfo {author} {\bibfnamefont {F.~S.}\ \bibnamefont {Bates}}, \bibinfo {author} {\bibfnamefont {M.~A.}\ \bibnamefont {Hillmyer}}, \bibinfo {author} {\bibfnamefont {T.~P.}\ \bibnamefont {Lodge}}, \bibinfo {author} {\bibfnamefont {C.~M.}\ \bibnamefont {Bates}}, \bibinfo {author} {\bibfnamefont {K.~T.}\ \bibnamefont {Delaney}},\ and\ \bibinfo {author} {\bibfnamefont {G.~H.}\ \bibnamefont {Fredrickson}},\ }\bibfield  {title} {\bibinfo {title} {Multiblock polymers: Panacea or {Pandora}'s box?},\ }\href {https://doi.org/10.1126/science.1215368} {\bibfield  {journal} {\bibinfo  {journal} {Science}\ }\textbf {\bibinfo {volume} {336}},\ \bibinfo {pages} {434} (\bibinfo {year} {2012})}\BibitemShut {NoStop}%
\bibitem [{\citenamefont {Bates}\ and\ \citenamefont {Bates}(2016)}]{Bates_2016}%
  \BibitemOpen
  \bibfield  {author} {\bibinfo {author} {\bibfnamefont {C.~M.}\ \bibnamefont {Bates}}\ and\ \bibinfo {author} {\bibfnamefont {F.~S.}\ \bibnamefont {Bates}},\ }\bibfield  {title} {\bibinfo {title} {50th anniversary perspective: Block polymers---pure potential},\ }\href {https://doi.org/10.1021/acs.macromol.6b02355} {\bibfield  {journal} {\bibinfo  {journal} {Macromolecules}\ }\textbf {\bibinfo {volume} {50}},\ \bibinfo {pages} {3} (\bibinfo {year} {2016})}\BibitemShut {NoStop}%
\bibitem [{\citenamefont {Matsen}(2002)}]{Matsen_2002}%
  \BibitemOpen
  \bibfield  {author} {\bibinfo {author} {\bibfnamefont {M.~W.}\ \bibnamefont {Matsen}},\ }\bibfield  {title} {\bibinfo {title} {The standard {Gaussian} model for block copolymer melts},\ }\href {https://doi.org/10.1088/0953-8984/14/2/201} {\bibfield  {journal} {\bibinfo  {journal} {J. Phys. Condens. Matter}\ }\textbf {\bibinfo {volume} {14}},\ \bibinfo {pages} {R21} (\bibinfo {year} {2002})}\BibitemShut {NoStop}%
\bibitem [{\citenamefont {Matsen}(2012)}]{Matsen_2012}%
  \BibitemOpen
  \bibfield  {author} {\bibinfo {author} {\bibfnamefont {M.~W.}\ \bibnamefont {Matsen}},\ }\bibfield  {title} {\bibinfo {title} {Effect of architecture on the phase behavior of {AB}-type block copolymer melts},\ }\href {https://doi.org/10.1021/ma202782s} {\bibfield  {journal} {\bibinfo  {journal} {Macromolecules}\ }\textbf {\bibinfo {volume} {45}},\ \bibinfo {pages} {2161} (\bibinfo {year} {2012})}\BibitemShut {NoStop}%
\bibitem [{\citenamefont {Leibler}(1980)}]{Leibler_1980}%
  \BibitemOpen
  \bibfield  {author} {\bibinfo {author} {\bibfnamefont {L.}~\bibnamefont {Leibler}},\ }\bibfield  {title} {\bibinfo {title} {Theory of microphase separation in block copolymers},\ }\href {https://doi.org/10.1021/ma60078a047} {\bibfield  {journal} {\bibinfo  {journal} {Macromolecules}\ }\textbf {\bibinfo {volume} {13}},\ \bibinfo {pages} {1602} (\bibinfo {year} {1980})}\BibitemShut {NoStop}%
\bibitem [{\citenamefont {Bates}\ and\ \citenamefont {Fredrickson}(1990)}]{Bates_1990}%
  \BibitemOpen
  \bibfield  {author} {\bibinfo {author} {\bibfnamefont {F.~S.}\ \bibnamefont {Bates}}\ and\ \bibinfo {author} {\bibfnamefont {G.~H.}\ \bibnamefont {Fredrickson}},\ }\bibfield  {title} {\bibinfo {title} {Block copolymer thermodynamics: Theory and experiment},\ }\href {https://doi.org/10.1146/annurev.pc.41.100190.002521} {\bibfield  {journal} {\bibinfo  {journal} {Ann. Rev. Phys. Chem.}\ }\textbf {\bibinfo {volume} {41}},\ \bibinfo {pages} {525} (\bibinfo {year} {1990})}\BibitemShut {NoStop}%
\bibitem [{\citenamefont {Foerster}\ \emph {et~al.}(1994)\citenamefont {Foerster}, \citenamefont {Khandpur}, \citenamefont {Zhao}, \citenamefont {Bates}, \citenamefont {Hamley}, \citenamefont {Ryan},\ and\ \citenamefont {Bras}}]{Foerster_1994}%
  \BibitemOpen
  \bibfield  {author} {\bibinfo {author} {\bibfnamefont {S.}~\bibnamefont {Foerster}}, \bibinfo {author} {\bibfnamefont {A.~K.}\ \bibnamefont {Khandpur}}, \bibinfo {author} {\bibfnamefont {J.}~\bibnamefont {Zhao}}, \bibinfo {author} {\bibfnamefont {F.~S.}\ \bibnamefont {Bates}}, \bibinfo {author} {\bibfnamefont {I.~W.}\ \bibnamefont {Hamley}}, \bibinfo {author} {\bibfnamefont {A.~J.}\ \bibnamefont {Ryan}},\ and\ \bibinfo {author} {\bibfnamefont {W.}~\bibnamefont {Bras}},\ }\bibfield  {title} {\bibinfo {title} {Complex phase behavior of polyisoprene-polystyrene diblock copolymers near the order-disorder transition},\ }\href {https://doi.org/10.1021/ma00101a033} {\bibfield  {journal} {\bibinfo  {journal} {Macromolecules}\ }\textbf {\bibinfo {volume} {27}},\ \bibinfo {pages} {6922} (\bibinfo {year} {1994})}\BibitemShut {NoStop}%
\bibitem [{\citenamefont {Matsen}\ and\ \citenamefont {Bates}(1996)}]{Matsen_1996}%
  \BibitemOpen
  \bibfield  {author} {\bibinfo {author} {\bibfnamefont {M.~W.}\ \bibnamefont {Matsen}}\ and\ \bibinfo {author} {\bibfnamefont {F.~S.}\ \bibnamefont {Bates}},\ }\bibfield  {title} {\bibinfo {title} {Unifying weak- and strong-segregation block copolymer theories},\ }\href {https://doi.org/10.1021/ma951138i} {\bibfield  {journal} {\bibinfo  {journal} {Macromolecules}\ }\textbf {\bibinfo {volume} {29}},\ \bibinfo {pages} {1091} (\bibinfo {year} {1996})}\BibitemShut {NoStop}%
\bibitem [{\citenamefont {Matsen}\ \emph {et~al.}(2023)\citenamefont {Matsen}, \citenamefont {Beardsley},\ and\ \citenamefont {Willis}}]{Matsen_2023}%
  \BibitemOpen
  \bibfield  {author} {\bibinfo {author} {\bibfnamefont {M.~W.}\ \bibnamefont {Matsen}}, \bibinfo {author} {\bibfnamefont {T.~M.}\ \bibnamefont {Beardsley}},\ and\ \bibinfo {author} {\bibfnamefont {J.~D.}\ \bibnamefont {Willis}},\ }\bibfield  {title} {\bibinfo {title} {Fluctuation-corrected phase diagrams for diblock copolymer melts},\ }\href {https://doi.org/10.1103/PhysRevLett.130.248101} {\bibfield  {journal} {\bibinfo  {journal} {Phys. Rev. Lett.}\ }\textbf {\bibinfo {volume} {130}},\ \bibinfo {pages} {248101} (\bibinfo {year} {2023})}\BibitemShut {NoStop}%
\bibitem [{\citenamefont {Gavrilov}\ \emph {et~al.}(2013)\citenamefont {Gavrilov}, \citenamefont {Kudryavtsev},\ and\ \citenamefont {Chertovich}}]{Gavrilov_2013}%
  \BibitemOpen
  \bibfield  {author} {\bibinfo {author} {\bibfnamefont {A.~A.}\ \bibnamefont {Gavrilov}}, \bibinfo {author} {\bibfnamefont {Y.~V.}\ \bibnamefont {Kudryavtsev}},\ and\ \bibinfo {author} {\bibfnamefont {A.~V.}\ \bibnamefont {Chertovich}},\ }\bibfield  {title} {\bibinfo {title} {Phase diagrams of block copolymer melts by dissipative particle dynamics simulations},\ }\href {https://doi.org/10.1063/1.4837215} {\bibfield  {journal} {\bibinfo  {journal} {J. Chem. Phys.}\ }\textbf {\bibinfo {volume} {139}},\ \bibinfo {pages} {224901} (\bibinfo {year} {2013})}\BibitemShut {NoStop}%
\bibitem [{\citenamefont {Tyler}\ and\ \citenamefont {Morse}(2005)}]{Tyler_2005}%
  \BibitemOpen
  \bibfield  {author} {\bibinfo {author} {\bibfnamefont {C.~A.}\ \bibnamefont {Tyler}}\ and\ \bibinfo {author} {\bibfnamefont {D.~C.}\ \bibnamefont {Morse}},\ }\bibfield  {title} {\bibinfo {title} {Orthorhombic ${Fddd}$ network in triblock and diblock copolymer melts},\ }\href {https://doi.org/10.1103/PhysRevLett.94.208302} {\bibfield  {journal} {\bibinfo  {journal} {Phys. Rev. Lett.}\ }\textbf {\bibinfo {volume} {94}},\ \bibinfo {pages} {208302} (\bibinfo {year} {2005})}\BibitemShut {NoStop}%
\bibitem [{\citenamefont {Takenaka}\ \emph {et~al.}(2007)\citenamefont {Takenaka}, \citenamefont {Wakada}, \citenamefont {Akasaka}, \citenamefont {Nishitsuji}, \citenamefont {Saijo}, \citenamefont {Shimizu}, \citenamefont {Kim},\ and\ \citenamefont {Hasegawa}}]{Takenaka_2007}%
  \BibitemOpen
  \bibfield  {author} {\bibinfo {author} {\bibfnamefont {M.}~\bibnamefont {Takenaka}}, \bibinfo {author} {\bibfnamefont {T.}~\bibnamefont {Wakada}}, \bibinfo {author} {\bibfnamefont {S.}~\bibnamefont {Akasaka}}, \bibinfo {author} {\bibfnamefont {S.}~\bibnamefont {Nishitsuji}}, \bibinfo {author} {\bibfnamefont {K.}~\bibnamefont {Saijo}}, \bibinfo {author} {\bibfnamefont {H.}~\bibnamefont {Shimizu}}, \bibinfo {author} {\bibfnamefont {M.~I.}\ \bibnamefont {Kim}},\ and\ \bibinfo {author} {\bibfnamefont {H.}~\bibnamefont {Hasegawa}},\ }\bibfield  {title} {\bibinfo {title} {Orthorhombic ${Fddd}$ network in diblock copolymer melts},\ }\href {https://doi.org/10.1021/ma070739u} {\bibfield  {journal} {\bibinfo  {journal} {Macromolecules}\ }\textbf {\bibinfo {volume} {40}},\ \bibinfo {pages} {4399} (\bibinfo {year} {2007})}\BibitemShut {NoStop}%
\bibitem [{\citenamefont {Stoykovich}\ \emph {et~al.}(2005)\citenamefont {Stoykovich}, \citenamefont {M{\"u}ller}, \citenamefont {Kim}, \citenamefont {Solak}, \citenamefont {Edwards}, \citenamefont {de~Pablo},\ and\ \citenamefont {Nealey}}]{Stoykovich_2005}%
  \BibitemOpen
  \bibfield  {author} {\bibinfo {author} {\bibfnamefont {M.~P.}\ \bibnamefont {Stoykovich}}, \bibinfo {author} {\bibfnamefont {M.}~\bibnamefont {M{\"u}ller}}, \bibinfo {author} {\bibfnamefont {S.~O.}\ \bibnamefont {Kim}}, \bibinfo {author} {\bibfnamefont {H.~H.}\ \bibnamefont {Solak}}, \bibinfo {author} {\bibfnamefont {E.~W.}\ \bibnamefont {Edwards}}, \bibinfo {author} {\bibfnamefont {J.~J.}\ \bibnamefont {de~Pablo}},\ and\ \bibinfo {author} {\bibfnamefont {P.~F.}\ \bibnamefont {Nealey}},\ }\bibfield  {title} {\bibinfo {title} {Directed assembly of block copolymer blends into nonregular device-oriented structures},\ }\href {https://doi.org/10.1126/science.1111041} {\bibfield  {journal} {\bibinfo  {journal} {Science}\ }\textbf {\bibinfo {volume} {308}},\ \bibinfo {pages} {1442} (\bibinfo {year} {2005})}\BibitemShut {NoStop}%
\bibitem [{\citenamefont {Ruiz}\ \emph {et~al.}(2008)\citenamefont {Ruiz}, \citenamefont {Kang}, \citenamefont {Detcheverry}, \citenamefont {Dobisz}, \citenamefont {Kercher}, \citenamefont {Albrecht}, \citenamefont {de~Pablo},\ and\ \citenamefont {Nealey}}]{Ruiz_2008}%
  \BibitemOpen
  \bibfield  {author} {\bibinfo {author} {\bibfnamefont {R.}~\bibnamefont {Ruiz}}, \bibinfo {author} {\bibfnamefont {H.}~\bibnamefont {Kang}}, \bibinfo {author} {\bibfnamefont {F.~A.}\ \bibnamefont {Detcheverry}}, \bibinfo {author} {\bibfnamefont {E.}~\bibnamefont {Dobisz}}, \bibinfo {author} {\bibfnamefont {D.~S.}\ \bibnamefont {Kercher}}, \bibinfo {author} {\bibfnamefont {T.~R.}\ \bibnamefont {Albrecht}}, \bibinfo {author} {\bibfnamefont {J.~J.}\ \bibnamefont {de~Pablo}},\ and\ \bibinfo {author} {\bibfnamefont {P.~F.}\ \bibnamefont {Nealey}},\ }\bibfield  {title} {\bibinfo {title} {Density multiplication and improved lithography by directed block copolymer assembly},\ }\href {https://doi.org/10.1126/science.1157626} {\bibfield  {journal} {\bibinfo  {journal} {Science}\ }\textbf {\bibinfo {volume} {321}},\ \bibinfo {pages} {936} (\bibinfo {year} {2008})}\BibitemShut {NoStop}%
\bibitem [{\citenamefont {Tang}\ \emph {et~al.}(2008)\citenamefont {Tang}, \citenamefont {Lennon}, \citenamefont {Fredrickson}, \citenamefont {Kramer},\ and\ \citenamefont {Hawker}}]{Tang_2008}%
  \BibitemOpen
  \bibfield  {author} {\bibinfo {author} {\bibfnamefont {C.}~\bibnamefont {Tang}}, \bibinfo {author} {\bibfnamefont {E.~M.}\ \bibnamefont {Lennon}}, \bibinfo {author} {\bibfnamefont {G.~H.}\ \bibnamefont {Fredrickson}}, \bibinfo {author} {\bibfnamefont {E.~J.}\ \bibnamefont {Kramer}},\ and\ \bibinfo {author} {\bibfnamefont {C.~J.}\ \bibnamefont {Hawker}},\ }\bibfield  {title} {\bibinfo {title} {Evolution of block copolymer lithography to highly ordered square arrays},\ }\href {https://doi.org/10.1126/science.1162950} {\bibfield  {journal} {\bibinfo  {journal} {Science}\ }\textbf {\bibinfo {volume} {322}},\ \bibinfo {pages} {429} (\bibinfo {year} {2008})}\BibitemShut {NoStop}%
\bibitem [{\citenamefont {Kim}\ \emph {et~al.}(2009)\citenamefont {Kim}, \citenamefont {Park},\ and\ \citenamefont {Hinsberg}}]{Kim_2009}%
  \BibitemOpen
  \bibfield  {author} {\bibinfo {author} {\bibfnamefont {H.-C.}\ \bibnamefont {Kim}}, \bibinfo {author} {\bibfnamefont {S.-M.}\ \bibnamefont {Park}},\ and\ \bibinfo {author} {\bibfnamefont {W.~D.}\ \bibnamefont {Hinsberg}},\ }\bibfield  {title} {\bibinfo {title} {Block copolymer based nanostructures: Materials, processes, and applications to electronics},\ }\href {https://doi.org/10.1021/cr900159v} {\bibfield  {journal} {\bibinfo  {journal} {Chem. Rev.}\ }\textbf {\bibinfo {volume} {110}},\ \bibinfo {pages} {146} (\bibinfo {year} {2009})}\BibitemShut {NoStop}%
\bibitem [{\citenamefont {Sveinbj{\"o}rnsson}\ \emph {et~al.}(2012)\citenamefont {Sveinbj{\"o}rnsson}, \citenamefont {Weitekamp}, \citenamefont {Miyake}, \citenamefont {Xia}, \citenamefont {Atwater},\ and\ \citenamefont {Grubbs}}]{Sveinbj_rnsson_2012}%
  \BibitemOpen
  \bibfield  {author} {\bibinfo {author} {\bibfnamefont {B.~R.}\ \bibnamefont {Sveinbj{\"o}rnsson}}, \bibinfo {author} {\bibfnamefont {R.~A.}\ \bibnamefont {Weitekamp}}, \bibinfo {author} {\bibfnamefont {G.~M.}\ \bibnamefont {Miyake}}, \bibinfo {author} {\bibfnamefont {Y.}~\bibnamefont {Xia}}, \bibinfo {author} {\bibfnamefont {H.~A.}\ \bibnamefont {Atwater}},\ and\ \bibinfo {author} {\bibfnamefont {R.~H.}\ \bibnamefont {Grubbs}},\ }\bibfield  {title} {\bibinfo {title} {Rapid self-assembly of brush block copolymers to photonic crystals},\ }\href {https://doi.org/10.1073/pnas.1213055109} {\bibfield  {journal} {\bibinfo  {journal} {Proc. Natl. Acad. Sci. U.S.A.}\ }\textbf {\bibinfo {volume} {109}},\ \bibinfo {pages} {14332} (\bibinfo {year} {2012})}\BibitemShut {NoStop}%
\bibitem [{\citenamefont {Stefik}\ \emph {et~al.}(2015)\citenamefont {Stefik}, \citenamefont {Guldin}, \citenamefont {Vignolini}, \citenamefont {Wiesner},\ and\ \citenamefont {Steiner}}]{Stefik_2015}%
  \BibitemOpen
  \bibfield  {author} {\bibinfo {author} {\bibfnamefont {M.}~\bibnamefont {Stefik}}, \bibinfo {author} {\bibfnamefont {S.}~\bibnamefont {Guldin}}, \bibinfo {author} {\bibfnamefont {S.}~\bibnamefont {Vignolini}}, \bibinfo {author} {\bibfnamefont {U.}~\bibnamefont {Wiesner}},\ and\ \bibinfo {author} {\bibfnamefont {U.}~\bibnamefont {Steiner}},\ }\bibfield  {title} {\bibinfo {title} {Block copolymer self-assembly for nanophotonics},\ }\href {https://doi.org/10.1039/c4cs00517a} {\bibfield  {journal} {\bibinfo  {journal} {Chem. Soc. Rev.}\ }\textbf {\bibinfo {volume} {44}},\ \bibinfo {pages} {5076} (\bibinfo {year} {2015})}\BibitemShut {NoStop}%
\bibitem [{\citenamefont {Yoo}\ \emph {et~al.}(2015)\citenamefont {Yoo}, \citenamefont {Kim}, \citenamefont {Shin}, \citenamefont {Park}, \citenamefont {Kim}, \citenamefont {Lee},\ and\ \citenamefont {Park}}]{Yoo_2015}%
  \BibitemOpen
  \bibfield  {author} {\bibinfo {author} {\bibfnamefont {S.}~\bibnamefont {Yoo}}, \bibinfo {author} {\bibfnamefont {J.-H.}\ \bibnamefont {Kim}}, \bibinfo {author} {\bibfnamefont {M.}~\bibnamefont {Shin}}, \bibinfo {author} {\bibfnamefont {H.}~\bibnamefont {Park}}, \bibinfo {author} {\bibfnamefont {J.-H.}\ \bibnamefont {Kim}}, \bibinfo {author} {\bibfnamefont {S.-Y.}\ \bibnamefont {Lee}},\ and\ \bibinfo {author} {\bibfnamefont {S.}~\bibnamefont {Park}},\ }\bibfield  {title} {\bibinfo {title} {Hierarchical multiscale hyperporous block copolymer membranes via tunable dual-phase separation},\ }\href {https://doi.org/10.1126/sciadv.1500101} {\bibfield  {journal} {\bibinfo  {journal} {Sci. Adv.}\ }\textbf {\bibinfo {volume} {1}},\ \bibinfo {pages} {e1500101} (\bibinfo {year} {2015})}\BibitemShut {NoStop}%
\bibitem [{\citenamefont {Rangou}\ \emph {et~al.}(2022)\citenamefont {Rangou}, \citenamefont {Appold}, \citenamefont {Lademann}, \citenamefont {Buhr},\ and\ \citenamefont {Filiz}}]{Rangou_2022}%
  \BibitemOpen
  \bibfield  {author} {\bibinfo {author} {\bibfnamefont {S.}~\bibnamefont {Rangou}}, \bibinfo {author} {\bibfnamefont {M.}~\bibnamefont {Appold}}, \bibinfo {author} {\bibfnamefont {B.}~\bibnamefont {Lademann}}, \bibinfo {author} {\bibfnamefont {K.}~\bibnamefont {Buhr}},\ and\ \bibinfo {author} {\bibfnamefont {V.}~\bibnamefont {Filiz}},\ }\bibfield  {title} {\bibinfo {title} {Thermally and chemically stable isoporous block copolymer membranes},\ }\href {https://doi.org/10.1021/acsmacrolett.2c00352} {\bibfield  {journal} {\bibinfo  {journal} {ACS Macro Lett.}\ }\textbf {\bibinfo {volume} {11}},\ \bibinfo {pages} {1142} (\bibinfo {year} {2022})}\BibitemShut {NoStop}%
\bibitem [{\citenamefont {Robbins}\ \emph {et~al.}(2016)\citenamefont {Robbins}, \citenamefont {Beaucage}, \citenamefont {Sai}, \citenamefont {Tan}, \citenamefont {Werner}, \citenamefont {Sethna}, \citenamefont {DiSalvo}, \citenamefont {Gruner}, \citenamefont {Van~Dover},\ and\ \citenamefont {Wiesner}}]{Robbins_2016}%
  \BibitemOpen
  \bibfield  {author} {\bibinfo {author} {\bibfnamefont {S.~W.}\ \bibnamefont {Robbins}}, \bibinfo {author} {\bibfnamefont {P.~A.}\ \bibnamefont {Beaucage}}, \bibinfo {author} {\bibfnamefont {H.}~\bibnamefont {Sai}}, \bibinfo {author} {\bibfnamefont {K.~W.}\ \bibnamefont {Tan}}, \bibinfo {author} {\bibfnamefont {J.~G.}\ \bibnamefont {Werner}}, \bibinfo {author} {\bibfnamefont {J.~P.}\ \bibnamefont {Sethna}}, \bibinfo {author} {\bibfnamefont {F.~J.}\ \bibnamefont {DiSalvo}}, \bibinfo {author} {\bibfnamefont {S.~M.}\ \bibnamefont {Gruner}}, \bibinfo {author} {\bibfnamefont {R.~B.}\ \bibnamefont {Van~Dover}},\ and\ \bibinfo {author} {\bibfnamefont {U.}~\bibnamefont {Wiesner}},\ }\bibfield  {title} {\bibinfo {title} {Block copolymer self-assembly--directed synthesis of mesoporous gyroidal superconductors},\ }\href {https://doi.org/10.1126/sciadv.1501119} {\bibfield  {journal} {\bibinfo  {journal} {Sci. Adv.}\ }\textbf {\bibinfo {volume} {2}},\ \bibinfo {pages} {e1501119} (\bibinfo {year} {2016})}\BibitemShut
  {NoStop}%
\bibitem [{\citenamefont {Yang}\ \emph {et~al.}(2022)\citenamefont {Yang}, \citenamefont {Choi}, \citenamefont {Han}, \citenamefont {Kim}, \citenamefont {Lee}, \citenamefont {Jung}, \citenamefont {Jin},\ and\ \citenamefont {Kim}}]{Yang_2022}%
  \BibitemOpen
  \bibfield  {author} {\bibinfo {author} {\bibfnamefont {G.~G.}\ \bibnamefont {Yang}}, \bibinfo {author} {\bibfnamefont {H.~J.}\ \bibnamefont {Choi}}, \bibinfo {author} {\bibfnamefont {K.~H.}\ \bibnamefont {Han}}, \bibinfo {author} {\bibfnamefont {J.~H.}\ \bibnamefont {Kim}}, \bibinfo {author} {\bibfnamefont {C.~W.}\ \bibnamefont {Lee}}, \bibinfo {author} {\bibfnamefont {E.~I.}\ \bibnamefont {Jung}}, \bibinfo {author} {\bibfnamefont {H.~M.}\ \bibnamefont {Jin}},\ and\ \bibinfo {author} {\bibfnamefont {S.~O.}\ \bibnamefont {Kim}},\ }\bibfield  {title} {\bibinfo {title} {Block copolymer nanopatterning for nonsemiconductor device applications},\ }\href {https://doi.org/10.1021/acsami.1c22836} {\bibfield  {journal} {\bibinfo  {journal} {ACS Appl. Mater. Interfaces}\ }\textbf {\bibinfo {volume} {14}},\ \bibinfo {pages} {12011} (\bibinfo {year} {2022})}\BibitemShut {NoStop}%
\bibitem [{\citenamefont {Matsen}\ and\ \citenamefont {Schick}(1994)}]{Matsen_1994}%
  \BibitemOpen
  \bibfield  {author} {\bibinfo {author} {\bibfnamefont {M.~W.}\ \bibnamefont {Matsen}}\ and\ \bibinfo {author} {\bibfnamefont {M.}~\bibnamefont {Schick}},\ }\bibfield  {title} {\bibinfo {title} {Stable and unstable phases of a diblock copolymer melt},\ }\href {https://doi.org/10.1103/PhysRevLett.72.2660} {\bibfield  {journal} {\bibinfo  {journal} {Phys. Rev. Lett.}\ }\textbf {\bibinfo {volume} {72}},\ \bibinfo {pages} {2660} (\bibinfo {year} {1994})}\BibitemShut {NoStop}%
\bibitem [{\citenamefont {Drolet}\ and\ \citenamefont {Fredrickson}(1999)}]{Drolet_1999}%
  \BibitemOpen
  \bibfield  {author} {\bibinfo {author} {\bibfnamefont {F.}~\bibnamefont {Drolet}}\ and\ \bibinfo {author} {\bibfnamefont {G.~H.}\ \bibnamefont {Fredrickson}},\ }\bibfield  {title} {\bibinfo {title} {Combinatorial screening of complex block copolymer assembly with self-consistent field theory},\ }\href {https://doi.org/10.1103/PhysRevLett.83.4317} {\bibfield  {journal} {\bibinfo  {journal} {Phys. Rev. Lett.}\ }\textbf {\bibinfo {volume} {83}},\ \bibinfo {pages} {4317} (\bibinfo {year} {1999})}\BibitemShut {NoStop}%
\bibitem [{\citenamefont {Bohbot-Raviv}\ and\ \citenamefont {Wang}(2000)}]{Bohbot-Raviv:2000aa}%
  \BibitemOpen
  \bibfield  {author} {\bibinfo {author} {\bibfnamefont {Y.}~\bibnamefont {Bohbot-Raviv}}\ and\ \bibinfo {author} {\bibfnamefont {Z.-G.}\ \bibnamefont {Wang}},\ }\bibfield  {title} {\bibinfo {title} {Discovering new ordered phases of block copolymers},\ }\href {https://doi.org/10.1103/PhysRevLett.85.3428} {\bibfield  {journal} {\bibinfo  {journal} {Phys. Rev. Lett.}\ }\textbf {\bibinfo {volume} {85}},\ \bibinfo {pages} {3428} (\bibinfo {year} {2000})}\BibitemShut {NoStop}%
\bibitem [{\citenamefont {Fredrickson}(2006)}]{Fredrickson_Book2006}%
  \BibitemOpen
  \bibfield  {author} {\bibinfo {author} {\bibfnamefont {G.~H.}\ \bibnamefont {Fredrickson}},\ }\href {https://books.google.com/books?id=ZiwTDAAAQBAJ} {\emph {\bibinfo {title} {The Equilibrium Theory of Inhomogeneous Polymers}}},\ International Series of Monographs on Physics\ (\bibinfo  {publisher} {OUP Oxford},\ \bibinfo {year} {2006})\BibitemShut {NoStop}%
\bibitem [{\citenamefont {Xu}\ \emph {et~al.}(2024)\citenamefont {Xu}, \citenamefont {Dong},\ and\ \citenamefont {Li}}]{Xu_2024}%
  \BibitemOpen
  \bibfield  {author} {\bibinfo {author} {\bibfnamefont {Z.}~\bibnamefont {Xu}}, \bibinfo {author} {\bibfnamefont {Q.}~\bibnamefont {Dong}},\ and\ \bibinfo {author} {\bibfnamefont {W.}~\bibnamefont {Li}},\ }\bibfield  {title} {\bibinfo {title} {Architectural design of block copolymers},\ }\href {https://doi.org/10.1021/acs.macromol.3c01730} {\bibfield  {journal} {\bibinfo  {journal} {Macromolecules}\ }\textbf {\bibinfo {volume} {57}},\ \bibinfo {pages} {1869} (\bibinfo {year} {2024})}\BibitemShut {NoStop}%
\bibitem [{\citenamefont {Grason}\ \emph {et~al.}(2003)\citenamefont {Grason}, \citenamefont {DiDonna},\ and\ \citenamefont {Kamien}}]{Grason_2003}%
  \BibitemOpen
  \bibfield  {author} {\bibinfo {author} {\bibfnamefont {G.~M.}\ \bibnamefont {Grason}}, \bibinfo {author} {\bibfnamefont {B.~A.}\ \bibnamefont {DiDonna}},\ and\ \bibinfo {author} {\bibfnamefont {R.~D.}\ \bibnamefont {Kamien}},\ }\bibfield  {title} {\bibinfo {title} {Geometric theory of diblock copolymer phases},\ }\href {https://doi.org/10.1103/PhysRevLett.91.058304} {\bibfield  {journal} {\bibinfo  {journal} {Phys. Rev. Lett.}\ }\textbf {\bibinfo {volume} {91}},\ \bibinfo {pages} {058304} (\bibinfo {year} {2003})}\BibitemShut {NoStop}%
\bibitem [{\citenamefont {Lee}\ \emph {et~al.}(2010)\citenamefont {Lee}, \citenamefont {Bluemle},\ and\ \citenamefont {Bates}}]{Lee_2010}%
  \BibitemOpen
  \bibfield  {author} {\bibinfo {author} {\bibfnamefont {S.}~\bibnamefont {Lee}}, \bibinfo {author} {\bibfnamefont {M.~J.}\ \bibnamefont {Bluemle}},\ and\ \bibinfo {author} {\bibfnamefont {F.~S.}\ \bibnamefont {Bates}},\ }\bibfield  {title} {\bibinfo {title} {Discovery of a {Frank--Kasper} $\sigma$ phase in sphere-forming block copolymer melts},\ }\href {https://doi.org/10.1126/science.1195552} {\bibfield  {journal} {\bibinfo  {journal} {Science}\ }\textbf {\bibinfo {volume} {330}},\ \bibinfo {pages} {349} (\bibinfo {year} {2010})}\BibitemShut {NoStop}%
\bibitem [{\citenamefont {Xie}\ \emph {et~al.}(2014)\citenamefont {Xie}, \citenamefont {Li}, \citenamefont {Qiu},\ and\ \citenamefont {Shi}}]{Xie_2014}%
  \BibitemOpen
  \bibfield  {author} {\bibinfo {author} {\bibfnamefont {N.}~\bibnamefont {Xie}}, \bibinfo {author} {\bibfnamefont {W.}~\bibnamefont {Li}}, \bibinfo {author} {\bibfnamefont {F.}~\bibnamefont {Qiu}},\ and\ \bibinfo {author} {\bibfnamefont {A.-C.}\ \bibnamefont {Shi}},\ }\bibfield  {title} {\bibinfo {title} {$\sigma$ phase formed in conformationally asymmetric {AB}-type block copolymers},\ }\href {https://doi.org/10.1021/mz500445v} {\bibfield  {journal} {\bibinfo  {journal} {ACS Macro Lett.}\ }\textbf {\bibinfo {volume} {3}},\ \bibinfo {pages} {906} (\bibinfo {year} {2014})}\BibitemShut {NoStop}%
\bibitem [{\citenamefont {Liu}\ \emph {et~al.}(2016)\citenamefont {Liu}, \citenamefont {Qiang}, \citenamefont {Li}, \citenamefont {Qiu},\ and\ \citenamefont {Shi}}]{Liu_2016}%
  \BibitemOpen
  \bibfield  {author} {\bibinfo {author} {\bibfnamefont {M.}~\bibnamefont {Liu}}, \bibinfo {author} {\bibfnamefont {Y.}~\bibnamefont {Qiang}}, \bibinfo {author} {\bibfnamefont {W.}~\bibnamefont {Li}}, \bibinfo {author} {\bibfnamefont {F.}~\bibnamefont {Qiu}},\ and\ \bibinfo {author} {\bibfnamefont {A.-C.}\ \bibnamefont {Shi}},\ }\bibfield  {title} {\bibinfo {title} {Stabilizing the {Frank--Kasper} phases via binary blends of {AB} diblock copolymers},\ }\href {https://doi.org/10.1021/acsmacrolett.6b00685} {\bibfield  {journal} {\bibinfo  {journal} {ACS Macro Lett.}\ }\textbf {\bibinfo {volume} {5}},\ \bibinfo {pages} {1167} (\bibinfo {year} {2016})}\BibitemShut {NoStop}%
\bibitem [{\citenamefont {Zhao}\ and\ \citenamefont {Li}(2019)}]{Zhao_2019}%
  \BibitemOpen
  \bibfield  {author} {\bibinfo {author} {\bibfnamefont {M.}~\bibnamefont {Zhao}}\ and\ \bibinfo {author} {\bibfnamefont {W.}~\bibnamefont {Li}},\ }\bibfield  {title} {\bibinfo {title} {Laves phases formed in the binary blend of {AB}$_4$ miktoarm star copolymer and a-homopolymer},\ }\href {https://doi.org/10.1021/acs.macromol.8b02407} {\bibfield  {journal} {\bibinfo  {journal} {Macromolecules}\ }\textbf {\bibinfo {volume} {52}},\ \bibinfo {pages} {1832} (\bibinfo {year} {2019})}\BibitemShut {NoStop}%
\bibitem [{\citenamefont {Kim}\ \emph {et~al.}(2017{\natexlab{a}})\citenamefont {Kim}, \citenamefont {Schulze}, \citenamefont {Arora}, \citenamefont {Lewis}, \citenamefont {Hillmyer}, \citenamefont {Dorfman},\ and\ \citenamefont {Bates}}]{Kim_2017}%
  \BibitemOpen
  \bibfield  {author} {\bibinfo {author} {\bibfnamefont {K.}~\bibnamefont {Kim}}, \bibinfo {author} {\bibfnamefont {M.~W.}\ \bibnamefont {Schulze}}, \bibinfo {author} {\bibfnamefont {A.}~\bibnamefont {Arora}}, \bibinfo {author} {\bibfnamefont {R.~M.}\ \bibnamefont {Lewis}}, \bibinfo {author} {\bibfnamefont {M.~A.}\ \bibnamefont {Hillmyer}}, \bibinfo {author} {\bibfnamefont {K.~D.}\ \bibnamefont {Dorfman}},\ and\ \bibinfo {author} {\bibfnamefont {F.~S.}\ \bibnamefont {Bates}},\ }\bibfield  {title} {\bibinfo {title} {Thermal processing of diblock copolymer melts mimics metallurgy},\ }\href {https://doi.org/10.1126/science.aam7212} {\bibfield  {journal} {\bibinfo  {journal} {Science}\ }\textbf {\bibinfo {volume} {356}},\ \bibinfo {pages} {520} (\bibinfo {year} {2017}{\natexlab{a}})}\BibitemShut {NoStop}%
\bibitem [{\citenamefont {Kim}\ \emph {et~al.}(2018)\citenamefont {Kim}, \citenamefont {Arora}, \citenamefont {Lewis}, \citenamefont {Liu}, \citenamefont {Li}, \citenamefont {Shi}, \citenamefont {Dorfman},\ and\ \citenamefont {Bates}}]{Kim_2018}%
  \BibitemOpen
  \bibfield  {author} {\bibinfo {author} {\bibfnamefont {K.}~\bibnamefont {Kim}}, \bibinfo {author} {\bibfnamefont {A.}~\bibnamefont {Arora}}, \bibinfo {author} {\bibfnamefont {R.~M.}\ \bibnamefont {Lewis}}, \bibinfo {author} {\bibfnamefont {M.}~\bibnamefont {Liu}}, \bibinfo {author} {\bibfnamefont {W.}~\bibnamefont {Li}}, \bibinfo {author} {\bibfnamefont {A.-C.}\ \bibnamefont {Shi}}, \bibinfo {author} {\bibfnamefont {K.~D.}\ \bibnamefont {Dorfman}},\ and\ \bibinfo {author} {\bibfnamefont {F.~S.}\ \bibnamefont {Bates}},\ }\bibfield  {title} {\bibinfo {title} {Origins of low-symmetry phases in asymmetric diblock copolymer melts},\ }\href {https://doi.org/10.1073/pnas.1717850115} {\bibfield  {journal} {\bibinfo  {journal} {Proc. Natl. Acad. Sci. U.S.A.}\ }\textbf {\bibinfo {volume} {115}},\ \bibinfo {pages} {847} (\bibinfo {year} {2018})}\BibitemShut {NoStop}%
\bibitem [{\citenamefont {Bates}\ \emph {et~al.}(2019)\citenamefont {Bates}, \citenamefont {Lequieu}, \citenamefont {Barbon}, \citenamefont {Lewis}, \citenamefont {Delaney}, \citenamefont {Anastasaki}, \citenamefont {Hawker}, \citenamefont {Fredrickson},\ and\ \citenamefont {Bates}}]{Bates_2019}%
  \BibitemOpen
  \bibfield  {author} {\bibinfo {author} {\bibfnamefont {M.~W.}\ \bibnamefont {Bates}}, \bibinfo {author} {\bibfnamefont {J.}~\bibnamefont {Lequieu}}, \bibinfo {author} {\bibfnamefont {S.~M.}\ \bibnamefont {Barbon}}, \bibinfo {author} {\bibfnamefont {R.~M.}\ \bibnamefont {Lewis}}, \bibinfo {author} {\bibfnamefont {K.~T.}\ \bibnamefont {Delaney}}, \bibinfo {author} {\bibfnamefont {A.}~\bibnamefont {Anastasaki}}, \bibinfo {author} {\bibfnamefont {C.~J.}\ \bibnamefont {Hawker}}, \bibinfo {author} {\bibfnamefont {G.~H.}\ \bibnamefont {Fredrickson}},\ and\ \bibinfo {author} {\bibfnamefont {C.~M.}\ \bibnamefont {Bates}},\ }\bibfield  {title} {\bibinfo {title} {Stability of the {A15} phase in diblock copolymer melts},\ }\href {https://doi.org/10.1073/pnas.1900121116} {\bibfield  {journal} {\bibinfo  {journal} {Proc. Natl. Acad. Sci. U.S.A.}\ }\textbf {\bibinfo {volume} {116}},\ \bibinfo {pages} {13194} (\bibinfo {year} {2019})}\BibitemShut {NoStop}%
\bibitem [{\citenamefont {Cheong}\ \emph {et~al.}(2020)\citenamefont {Cheong}, \citenamefont {Bates},\ and\ \citenamefont {Dorfman}}]{Cheong_2020}%
  \BibitemOpen
  \bibfield  {author} {\bibinfo {author} {\bibfnamefont {G.~K.}\ \bibnamefont {Cheong}}, \bibinfo {author} {\bibfnamefont {F.~S.}\ \bibnamefont {Bates}},\ and\ \bibinfo {author} {\bibfnamefont {K.~D.}\ \bibnamefont {Dorfman}},\ }\bibfield  {title} {\bibinfo {title} {Symmetry breaking in particle-forming diblock polymer/homopolymer blends},\ }\href {https://doi.org/10.1073/pnas.2006079117} {\bibfield  {journal} {\bibinfo  {journal} {Proc. Natl. Acad. Sci. U.S.A.}\ }\textbf {\bibinfo {volume} {117}},\ \bibinfo {pages} {16764} (\bibinfo {year} {2020})}\BibitemShut {NoStop}%
\bibitem [{\citenamefont {Dorfman}(2021)}]{Dorfman_2021}%
  \BibitemOpen
  \bibfield  {author} {\bibinfo {author} {\bibfnamefont {K.~D.}\ \bibnamefont {Dorfman}},\ }\bibfield  {title} {\bibinfo {title} {{Frank--Kasper} phases in block polymers},\ }\href {https://doi.org/10.1021/acs.macromol.1c01650} {\bibfield  {journal} {\bibinfo  {journal} {Macromolecules}\ }\textbf {\bibinfo {volume} {54}},\ \bibinfo {pages} {10251} (\bibinfo {year} {2021})}\BibitemShut {NoStop}%
\bibitem [{\citenamefont {Zhuang}\ \emph {et~al.}(2024)\citenamefont {Zhuang}, \citenamefont {He}, \citenamefont {Tang},\ and\ \citenamefont {Wang}}]{Zhuang_2024}%
  \BibitemOpen
  \bibfield  {author} {\bibinfo {author} {\bibfnamefont {Z.}~\bibnamefont {Zhuang}}, \bibinfo {author} {\bibfnamefont {J.}~\bibnamefont {He}}, \bibinfo {author} {\bibfnamefont {J.}~\bibnamefont {Tang}},\ and\ \bibinfo {author} {\bibfnamefont {Q.}~\bibnamefont {Wang}},\ }\bibfield  {title} {\bibinfo {title} {Toward the relative stability of {Frank--Kasper} phases formed by neat diblock copolymer melts and binary blends},\ }\href {https://doi.org/10.1021/acs.macromol.4c02119} {\bibfield  {journal} {\bibinfo  {journal} {Macromolecules}\ }\textbf {\bibinfo {volume} {58}},\ \bibinfo {pages} {415} (\bibinfo {year} {2024})}\BibitemShut {NoStop}%
\bibitem [{\citenamefont {Zhao}\ \emph {et~al.}(2025)\citenamefont {Zhao}, \citenamefont {Wang}, \citenamefont {Li},\ and\ \citenamefont {Qiang}}]{Zhao_2025}%
  \BibitemOpen
  \bibfield  {author} {\bibinfo {author} {\bibfnamefont {B.}~\bibnamefont {Zhao}}, \bibinfo {author} {\bibfnamefont {C.}~\bibnamefont {Wang}}, \bibinfo {author} {\bibfnamefont {W.}~\bibnamefont {Li}},\ and\ \bibinfo {author} {\bibfnamefont {Y.}~\bibnamefont {Qiang}},\ }\bibfield  {title} {\bibinfo {title} {Laves phases emerge in neat {AB}-type block copolymer as hybrid spherical phases},\ }\href {https://doi.org/10.1021/acsmacrolett.5c00084} {\bibfield  {journal} {\bibinfo  {journal} {ACS Macro Lett.}\ }\textbf {\bibinfo {volume} {14}},\ \bibinfo {pages} {721} (\bibinfo {year} {2025})}\BibitemShut {NoStop}%
\bibitem [{\citenamefont {Geng}\ \emph {et~al.}(2025)\citenamefont {Geng}, \citenamefont {Li}, \citenamefont {Yang}, \citenamefont {An}, \citenamefont {M\"uller},\ and\ \citenamefont {Sun}}]{Geng_2025}%
  \BibitemOpen
  \bibfield  {author} {\bibinfo {author} {\bibfnamefont {X.-J.}\ \bibnamefont {Geng}}, \bibinfo {author} {\bibfnamefont {H.}~\bibnamefont {Li}}, \bibinfo {author} {\bibfnamefont {X.}~\bibnamefont {Yang}}, \bibinfo {author} {\bibfnamefont {L.-J.}\ \bibnamefont {An}}, \bibinfo {author} {\bibfnamefont {M.}~\bibnamefont {M\"uller}},\ and\ \bibinfo {author} {\bibfnamefont {D.-W.}\ \bibnamefont {Sun}},\ }\bibfield  {title} {\bibinfo {title} {Process-directed self-assembly of the {Frank--Kasper A15} structure in linear, conformationally symmetric block copolymers},\ }\href {https://doi.org/10.1103/PhysRevLett.134.118102} {\bibfield  {journal} {\bibinfo  {journal} {Phys. Rev. Lett.}\ }\textbf {\bibinfo {volume} {134}},\ \bibinfo {pages} {118102} (\bibinfo {year} {2025})}\BibitemShut {NoStop}%
\bibitem [{\citenamefont {Gillard}\ \emph {et~al.}(2016)\citenamefont {Gillard}, \citenamefont {Lee},\ and\ \citenamefont {Bates}}]{Gillard_2016}%
  \BibitemOpen
  \bibfield  {author} {\bibinfo {author} {\bibfnamefont {T.~M.}\ \bibnamefont {Gillard}}, \bibinfo {author} {\bibfnamefont {S.}~\bibnamefont {Lee}},\ and\ \bibinfo {author} {\bibfnamefont {F.~S.}\ \bibnamefont {Bates}},\ }\bibfield  {title} {\bibinfo {title} {Dodecagonal quasicrystalline order in a diblock copolymer melt},\ }\href {https://doi.org/10.1073/pnas.1601692113} {\bibfield  {journal} {\bibinfo  {journal} {Proc. Natl. Acad. Sci. U.S.A.}\ }\textbf {\bibinfo {volume} {113}},\ \bibinfo {pages} {5167} (\bibinfo {year} {2016})}\BibitemShut {NoStop}%
\bibitem [{\citenamefont {Schulze}\ \emph {et~al.}(2017)\citenamefont {Schulze}, \citenamefont {Lewis}, \citenamefont {Lettow}, \citenamefont {Hickey}, \citenamefont {Gillard}, \citenamefont {Hillmyer},\ and\ \citenamefont {Bates}}]{Schulze_2017}%
  \BibitemOpen
  \bibfield  {author} {\bibinfo {author} {\bibfnamefont {M.~W.}\ \bibnamefont {Schulze}}, \bibinfo {author} {\bibfnamefont {R.~M.}\ \bibnamefont {Lewis}}, \bibinfo {author} {\bibfnamefont {J.~H.}\ \bibnamefont {Lettow}}, \bibinfo {author} {\bibfnamefont {R.~J.}\ \bibnamefont {Hickey}}, \bibinfo {author} {\bibfnamefont {T.~M.}\ \bibnamefont {Gillard}}, \bibinfo {author} {\bibfnamefont {M.~A.}\ \bibnamefont {Hillmyer}},\ and\ \bibinfo {author} {\bibfnamefont {F.~S.}\ \bibnamefont {Bates}},\ }\bibfield  {title} {\bibinfo {title} {Conformational asymmetry and quasicrystal approximants in linear diblock copolymers},\ }\href {https://doi.org/10.1103/PhysRevLett.118.207801} {\bibfield  {journal} {\bibinfo  {journal} {Phys. Rev. Lett.}\ }\textbf {\bibinfo {volume} {118}},\ \bibinfo {pages} {207801} (\bibinfo {year} {2017})}\BibitemShut {NoStop}%
\bibitem [{\citenamefont {Lindsay}\ \emph {et~al.}(2020)\citenamefont {Lindsay}, \citenamefont {Lewis}, \citenamefont {Lee}, \citenamefont {Peterson}, \citenamefont {Lodge},\ and\ \citenamefont {Bates}}]{Lindsay_2020}%
  \BibitemOpen
  \bibfield  {author} {\bibinfo {author} {\bibfnamefont {A.~P.}\ \bibnamefont {Lindsay}}, \bibinfo {author} {\bibfnamefont {R.~M.}\ \bibnamefont {Lewis}}, \bibinfo {author} {\bibfnamefont {B.}~\bibnamefont {Lee}}, \bibinfo {author} {\bibfnamefont {A.~J.}\ \bibnamefont {Peterson}}, \bibinfo {author} {\bibfnamefont {T.~P.}\ \bibnamefont {Lodge}},\ and\ \bibinfo {author} {\bibfnamefont {F.~S.}\ \bibnamefont {Bates}},\ }\bibfield  {title} {\bibinfo {title} {A15, $\sigma$, and a quasicrystal: Access to complex particle packings via bidisperse diblock copolymer blends},\ }\href {https://doi.org/10.1021/acsmacrolett.9b01026} {\bibfield  {journal} {\bibinfo  {journal} {ACS Macro Lett.}\ }\textbf {\bibinfo {volume} {9}},\ \bibinfo {pages} {197} (\bibinfo {year} {2020})}\BibitemShut {NoStop}%
\bibitem [{\citenamefont {Mueller}\ \emph {et~al.}(2021)\citenamefont {Mueller}, \citenamefont {Lindsay}, \citenamefont {Jayaraman}, \citenamefont {Lodge}, \citenamefont {Mahanthappa},\ and\ \citenamefont {Bates}}]{Mueller_2021}%
  \BibitemOpen
  \bibfield  {author} {\bibinfo {author} {\bibfnamefont {A.~J.}\ \bibnamefont {Mueller}}, \bibinfo {author} {\bibfnamefont {A.~P.}\ \bibnamefont {Lindsay}}, \bibinfo {author} {\bibfnamefont {A.}~\bibnamefont {Jayaraman}}, \bibinfo {author} {\bibfnamefont {T.~P.}\ \bibnamefont {Lodge}}, \bibinfo {author} {\bibfnamefont {M.~K.}\ \bibnamefont {Mahanthappa}},\ and\ \bibinfo {author} {\bibfnamefont {F.~S.}\ \bibnamefont {Bates}},\ }\bibfield  {title} {\bibinfo {title} {Quasicrystals and their approximants in a crystalline--amorphous diblock copolymer},\ }\href {https://doi.org/10.1021/acs.macromol.0c02871} {\bibfield  {journal} {\bibinfo  {journal} {Macromolecules}\ }\textbf {\bibinfo {volume} {54}},\ \bibinfo {pages} {2647} (\bibinfo {year} {2021})}\BibitemShut {NoStop}%
\bibitem [{\citenamefont {Mueller}\ \emph {et~al.}(2024)\citenamefont {Mueller}, \citenamefont {Lindsay}, \citenamefont {Lewis}, \citenamefont {Zhang}, \citenamefont {Narayanan}, \citenamefont {Lodge}, \citenamefont {Mahanthappa},\ and\ \citenamefont {Bates}}]{Mueller_2024}%
  \BibitemOpen
  \bibfield  {author} {\bibinfo {author} {\bibfnamefont {A.~J.}\ \bibnamefont {Mueller}}, \bibinfo {author} {\bibfnamefont {A.~P.}\ \bibnamefont {Lindsay}}, \bibinfo {author} {\bibfnamefont {R.~M.}\ \bibnamefont {Lewis}}, \bibinfo {author} {\bibfnamefont {Q.}~\bibnamefont {Zhang}}, \bibinfo {author} {\bibfnamefont {S.}~\bibnamefont {Narayanan}}, \bibinfo {author} {\bibfnamefont {T.~P.}\ \bibnamefont {Lodge}}, \bibinfo {author} {\bibfnamefont {M.~K.}\ \bibnamefont {Mahanthappa}},\ and\ \bibinfo {author} {\bibfnamefont {F.~S.}\ \bibnamefont {Bates}},\ }\bibfield  {title} {\bibinfo {title} {Particle dynamics in a diblock-copolymer-based dodecagonal quasicrystal and its periodic approximant by {X}-ray photon correlation spectroscopy},\ }\href {https://doi.org/10.1103/PhysRevLett.132.158101} {\bibfield  {journal} {\bibinfo  {journal} {Phys. Rev. Lett.}\ }\textbf {\bibinfo {volume} {132}},\ \bibinfo {pages} {158101} (\bibinfo {year} {2024})}\BibitemShut {NoStop}%
\bibitem [{\citenamefont {Li}\ \emph {et~al.}(2025)\citenamefont {Li}, \citenamefont {Xie}, \citenamefont {Gan}, \citenamefont {Ma}, \citenamefont {Shi},\ and\ \citenamefont {Dong}}]{Li_2025}%
  \BibitemOpen
  \bibfield  {author} {\bibinfo {author} {\bibfnamefont {J.}~\bibnamefont {Li}}, \bibinfo {author} {\bibfnamefont {J.}~\bibnamefont {Xie}}, \bibinfo {author} {\bibfnamefont {Z.}~\bibnamefont {Gan}}, \bibinfo {author} {\bibfnamefont {Z.}~\bibnamefont {Ma}}, \bibinfo {author} {\bibfnamefont {A.-C.}\ \bibnamefont {Shi}},\ and\ \bibinfo {author} {\bibfnamefont {X.-H.}\ \bibnamefont {Dong}},\ }\bibfield  {title} {\bibinfo {title} {Synergistic impact of intra- and interchain dispersity on block copolymer self-assembly},\ }\href {https://doi.org/10.1021/acs.macromol.5c00433} {\bibfield  {journal} {\bibinfo  {journal} {Macromolecules}\ }\textbf {\bibinfo {volume} {58}},\ \bibinfo {pages} {3860} (\bibinfo {year} {2025})}\BibitemShut {NoStop}%
\bibitem [{\citenamefont {Lin}\ \emph {et~al.}(2018)\citenamefont {Lin}, \citenamefont {Higuchi}, \citenamefont {Chen}, \citenamefont {Tsai}, \citenamefont {Jinnai},\ and\ \citenamefont {Hashimoto}}]{Lin_2018}%
  \BibitemOpen
  \bibfield  {author} {\bibinfo {author} {\bibfnamefont {C.-H.}\ \bibnamefont {Lin}}, \bibinfo {author} {\bibfnamefont {T.}~\bibnamefont {Higuchi}}, \bibinfo {author} {\bibfnamefont {H.-L.}\ \bibnamefont {Chen}}, \bibinfo {author} {\bibfnamefont {J.-C.}\ \bibnamefont {Tsai}}, \bibinfo {author} {\bibfnamefont {H.}~\bibnamefont {Jinnai}},\ and\ \bibinfo {author} {\bibfnamefont {T.}~\bibnamefont {Hashimoto}},\ }\bibfield  {title} {\bibinfo {title} {Stabilizing the ordered bicontinuous double diamond structure of diblock copolymer by configurational regularity},\ }\href {https://doi.org/10.1021/acs.macromol.7b02404} {\bibfield  {journal} {\bibinfo  {journal} {Macromolecules}\ }\textbf {\bibinfo {volume} {51}},\ \bibinfo {pages} {4049} (\bibinfo {year} {2018})}\BibitemShut {NoStop}%
\bibitem [{\citenamefont {Takagi}\ \emph {et~al.}(2019)\citenamefont {Takagi}, \citenamefont {Suzuki}, \citenamefont {Aoyama}, \citenamefont {Mihira}, \citenamefont {Takano},\ and\ \citenamefont {Matsushita}}]{Takagi_2019}%
  \BibitemOpen
  \bibfield  {author} {\bibinfo {author} {\bibfnamefont {W.}~\bibnamefont {Takagi}}, \bibinfo {author} {\bibfnamefont {J.}~\bibnamefont {Suzuki}}, \bibinfo {author} {\bibfnamefont {Y.}~\bibnamefont {Aoyama}}, \bibinfo {author} {\bibfnamefont {T.}~\bibnamefont {Mihira}}, \bibinfo {author} {\bibfnamefont {A.}~\bibnamefont {Takano}},\ and\ \bibinfo {author} {\bibfnamefont {Y.}~\bibnamefont {Matsushita}},\ }\bibfield  {title} {\bibinfo {title} {Bicontinuous double-diamond structures formed in ternary blends of {AB} diblock copolymers with block chains of different lengths},\ }\href {https://doi.org/10.1021/acs.macromol.9b00724} {\bibfield  {journal} {\bibinfo  {journal} {Macromolecules}\ }\textbf {\bibinfo {volume} {52}},\ \bibinfo {pages} {6633} (\bibinfo {year} {2019})}\BibitemShut {NoStop}%
\bibitem [{\citenamefont {Takagi}\ and\ \citenamefont {Yamamoto}(2021)}]{Takagi_2021}%
  \BibitemOpen
  \bibfield  {author} {\bibinfo {author} {\bibfnamefont {H.}~\bibnamefont {Takagi}}\ and\ \bibinfo {author} {\bibfnamefont {K.}~\bibnamefont {Yamamoto}},\ }\bibfield  {title} {\bibinfo {title} {Effect of block copolymer composition and homopolymer molecular weight on ordered bicontinuous double-diamond structures in binary blends of polystyrene--polyisoprene block copolymer and polyisoprene homopolymer},\ }\href {https://doi.org/10.1021/acs.macromol.1c00429} {\bibfield  {journal} {\bibinfo  {journal} {Macromolecules}\ }\textbf {\bibinfo {volume} {54}},\ \bibinfo {pages} {5136} (\bibinfo {year} {2021})}\BibitemShut {NoStop}%
\bibitem [{\citenamefont {Lai}\ and\ \citenamefont {Shi}(2021)}]{Lai_2021}%
  \BibitemOpen
  \bibfield  {author} {\bibinfo {author} {\bibfnamefont {C.~T.}\ \bibnamefont {Lai}}\ and\ \bibinfo {author} {\bibfnamefont {A.-C.}\ \bibnamefont {Shi}},\ }\bibfield  {title} {\bibinfo {title} {Binary blends of diblock copolymers: An effective route to novel bicontinuous phases},\ }\href {https://doi.org/https://doi.org/10.1002/mats.202100019} {\bibfield  {journal} {\bibinfo  {journal} {Macromol. Theory Simul.}\ }\textbf {\bibinfo {volume} {30}},\ \bibinfo {pages} {2100019} (\bibinfo {year} {2021})}\BibitemShut {NoStop}%
\bibitem [{\citenamefont {Wang}\ \emph {et~al.}(2025)\citenamefont {Wang}, \citenamefont {Feng}, \citenamefont {Lu}, \citenamefont {Feng}, \citenamefont {Wang},\ and\ \citenamefont {Li}}]{Wang_2025}%
  \BibitemOpen
  \bibfield  {author} {\bibinfo {author} {\bibfnamefont {X.}~\bibnamefont {Wang}}, \bibinfo {author} {\bibfnamefont {Y.}~\bibnamefont {Feng}}, \bibinfo {author} {\bibfnamefont {S.}~\bibnamefont {Lu}}, \bibinfo {author} {\bibfnamefont {X.}~\bibnamefont {Feng}}, \bibinfo {author} {\bibfnamefont {G.}~\bibnamefont {Wang}},\ and\ \bibinfo {author} {\bibfnamefont {W.}~\bibnamefont {Li}},\ }\bibfield  {title} {\bibinfo {title} {{SCFT}-guided experimental fabrication of double-diamond structures in {A$_1$B/A$_2$B} block copolymer binary blends},\ }\href {https://doi.org/10.1021/acs.macromol.5c01972} {\bibfield  {journal} {\bibinfo  {journal} {Macromolecules}\ }\textbf {\bibinfo {volume} {58}},\ \bibinfo {pages} {9776} (\bibinfo {year} {2025})}\BibitemShut {NoStop}%
\bibitem [{\citenamefont {Mart{\'\i}nez-Veracoechea}\ and\ \citenamefont {Escobedo}(2007)}]{Mart_nez_Veracoechea_2007}%
  \BibitemOpen
  \bibfield  {author} {\bibinfo {author} {\bibfnamefont {F.~J.}\ \bibnamefont {Mart{\'\i}nez-Veracoechea}}\ and\ \bibinfo {author} {\bibfnamefont {F.~A.}\ \bibnamefont {Escobedo}},\ }\bibfield  {title} {\bibinfo {title} {{Monte Carlo} study of the stabilization of complex bicontinuous phases in diblock copolymer systems},\ }\href {https://doi.org/10.1021/ma071449g} {\bibfield  {journal} {\bibinfo  {journal} {Macromolecules}\ }\textbf {\bibinfo {volume} {40}},\ \bibinfo {pages} {7354} (\bibinfo {year} {2007})}\BibitemShut {NoStop}%
\bibitem [{\citenamefont {Martinez-Veracoechea}\ and\ \citenamefont {Escobedo}(2009)}]{Martinez_Veracoechea_2009}%
  \BibitemOpen
  \bibfield  {author} {\bibinfo {author} {\bibfnamefont {F.~J.}\ \bibnamefont {Martinez-Veracoechea}}\ and\ \bibinfo {author} {\bibfnamefont {F.~A.}\ \bibnamefont {Escobedo}},\ }\bibfield  {title} {\bibinfo {title} {The plumber's nightmare phase in diblock copolymer/homopolymer blends. {A} self-consistent field theory study.},\ }\href {https://doi.org/10.1021/ma901591r} {\bibfield  {journal} {\bibinfo  {journal} {Macromolecules}\ }\textbf {\bibinfo {volume} {42}},\ \bibinfo {pages} {9058} (\bibinfo {year} {2009})}\BibitemShut {NoStop}%
\bibitem [{\citenamefont {Qiang}\ \emph {et~al.}(2020)\citenamefont {Qiang}, \citenamefont {Li},\ and\ \citenamefont {Shi}}]{Qiang_2020}%
  \BibitemOpen
  \bibfield  {author} {\bibinfo {author} {\bibfnamefont {Y.}~\bibnamefont {Qiang}}, \bibinfo {author} {\bibfnamefont {W.}~\bibnamefont {Li}},\ and\ \bibinfo {author} {\bibfnamefont {A.-C.}\ \bibnamefont {Shi}},\ }\bibfield  {title} {\bibinfo {title} {Stabilizing phases of block copolymers with gigantic spheres via designed chain architectures},\ }\href {https://doi.org/10.1021/acsmacrolett.0c00193} {\bibfield  {journal} {\bibinfo  {journal} {ACS Macro Lett.}\ }\textbf {\bibinfo {volume} {9}},\ \bibinfo {pages} {668} (\bibinfo {year} {2020})}\BibitemShut {NoStop}%
\bibitem [{\citenamefont {Chang}\ \emph {et~al.}(2021)\citenamefont {Chang}, \citenamefont {Manesi}, \citenamefont {Yang}, \citenamefont {Hung}, \citenamefont {Yang}, \citenamefont {Chiu}, \citenamefont {Avgeropoulos},\ and\ \citenamefont {Ho}}]{Chang_2021}%
  \BibitemOpen
  \bibfield  {author} {\bibinfo {author} {\bibfnamefont {C.-Y.}\ \bibnamefont {Chang}}, \bibinfo {author} {\bibfnamefont {G.-M.}\ \bibnamefont {Manesi}}, \bibinfo {author} {\bibfnamefont {C.-Y.}\ \bibnamefont {Yang}}, \bibinfo {author} {\bibfnamefont {Y.-C.}\ \bibnamefont {Hung}}, \bibinfo {author} {\bibfnamefont {K.-C.}\ \bibnamefont {Yang}}, \bibinfo {author} {\bibfnamefont {P.-T.}\ \bibnamefont {Chiu}}, \bibinfo {author} {\bibfnamefont {A.}~\bibnamefont {Avgeropoulos}},\ and\ \bibinfo {author} {\bibfnamefont {R.-M.}\ \bibnamefont {Ho}},\ }\bibfield  {title} {\bibinfo {title} {Mesoscale networks and corresponding transitions from self-assembly of block copolymers},\ }\href {https://doi.org/10.1073/pnas.2022275118} {\bibfield  {journal} {\bibinfo  {journal} {Proc. Natl. Acad. Sci. U.S.A.}\ }\textbf {\bibinfo {volume} {118}},\ \bibinfo {pages} {e2022275118} (\bibinfo {year} {2021})}\BibitemShut {NoStop}%
\bibitem [{\citenamefont {Li}\ \emph {et~al.}(2022)\citenamefont {Li}, \citenamefont {Woo}, \citenamefont {Kim},\ and\ \citenamefont {Li}}]{LiQY_2022}%
  \BibitemOpen
  \bibfield  {author} {\bibinfo {author} {\bibfnamefont {Q.}~\bibnamefont {Li}}, \bibinfo {author} {\bibfnamefont {D.}~\bibnamefont {Woo}}, \bibinfo {author} {\bibfnamefont {J.-K.}\ \bibnamefont {Kim}},\ and\ \bibinfo {author} {\bibfnamefont {W.}~\bibnamefont {Li}},\ }\bibfield  {title} {\bibinfo {title} {Truly ``inverted'' cylinders and spheres formed in the {A(AB)$_3$/AC} blends of {B/C} hydrogen bonding interactions},\ }\href {https://doi.org/10.1021/acs.macromol.2c01084} {\bibfield  {journal} {\bibinfo  {journal} {Macromolecules}\ }\textbf {\bibinfo {volume} {55}},\ \bibinfo {pages} {6525} (\bibinfo {year} {2022})}\BibitemShut {NoStop}%
\bibitem [{\citenamefont {Chen}\ \emph {et~al.}(2025)\citenamefont {Chen}, \citenamefont {Dong}, \citenamefont {Qiang}, \citenamefont {Peng}, \citenamefont {Huang},\ and\ \citenamefont {Li}}]{HChen_2025}%
  \BibitemOpen
  \bibfield  {author} {\bibinfo {author} {\bibfnamefont {H.}~\bibnamefont {Chen}}, \bibinfo {author} {\bibfnamefont {Q.}~\bibnamefont {Dong}}, \bibinfo {author} {\bibfnamefont {Y.}~\bibnamefont {Qiang}}, \bibinfo {author} {\bibfnamefont {L.}~\bibnamefont {Peng}}, \bibinfo {author} {\bibfnamefont {X.}~\bibnamefont {Huang}},\ and\ \bibinfo {author} {\bibfnamefont {W.}~\bibnamefont {Li}},\ }\bibfield  {title} {\bibinfo {title} {Mechanisms of multiple reentrant transitions between {Frank--Kasper} and classical spherical phases in {AB}-type dendron-like copolymer},\ }\href {https://doi.org/10.1021/acs.macromol.5c02113} {\bibfield  {journal} {\bibinfo  {journal} {Macromolecules}\ }\textbf {\bibinfo {volume} {58}},\ \bibinfo {pages} {10192} (\bibinfo {year} {2025})}\BibitemShut {NoStop}%
\bibitem [{\citenamefont {Lee}\ \emph {et~al.}(2024{\natexlab{a}})\citenamefont {Lee}, \citenamefont {Kwon}, \citenamefont {Min}, \citenamefont {Jin}, \citenamefont {Hwang}, \citenamefont {Lee}, \citenamefont {Lee},\ and\ \citenamefont {Park}}]{Lee_2024a}%
  \BibitemOpen
  \bibfield  {author} {\bibinfo {author} {\bibfnamefont {H.}~\bibnamefont {Lee}}, \bibinfo {author} {\bibfnamefont {S.}~\bibnamefont {Kwon}}, \bibinfo {author} {\bibfnamefont {J.}~\bibnamefont {Min}}, \bibinfo {author} {\bibfnamefont {S.-M.}\ \bibnamefont {Jin}}, \bibinfo {author} {\bibfnamefont {J.~H.}\ \bibnamefont {Hwang}}, \bibinfo {author} {\bibfnamefont {E.}~\bibnamefont {Lee}}, \bibinfo {author} {\bibfnamefont {W.~B.}\ \bibnamefont {Lee}},\ and\ \bibinfo {author} {\bibfnamefont {M.~J.}\ \bibnamefont {Park}},\ }\bibfield  {title} {\bibinfo {title} {Thermodynamically stable plumber's nightmare structures in block copolymers},\ }\href {https://doi.org/10.1126/science.adh0483} {\bibfield  {journal} {\bibinfo  {journal} {Science}\ }\textbf {\bibinfo {volume} {383}},\ \bibinfo {pages} {70} (\bibinfo {year} {2024}{\natexlab{a}})}\BibitemShut {NoStop}%
\bibitem [{\citenamefont {Chang}\ \emph {et~al.}(2024)\citenamefont {Chang}, \citenamefont {Chen},\ and\ \citenamefont {Ho}}]{Chang_2024}%
  \BibitemOpen
  \bibfield  {author} {\bibinfo {author} {\bibfnamefont {C.-Y.}\ \bibnamefont {Chang}}, \bibinfo {author} {\bibfnamefont {Y.-H.}\ \bibnamefont {Chen}},\ and\ \bibinfo {author} {\bibfnamefont {R.-M.}\ \bibnamefont {Ho}},\ }\bibfield  {title} {\bibinfo {title} {Metastable network phases from controlled self-assembly of high-{$\chi$} block copolymers},\ }\href {https://doi.org/10.1103/PhysRevMaterials.8.030301} {\bibfield  {journal} {\bibinfo  {journal} {Phys. Rev. Mater.}\ }\textbf {\bibinfo {volume} {8}},\ \bibinfo {pages} {030301} (\bibinfo {year} {2024})}\BibitemShut {NoStop}%
\bibitem [{\citenamefont {Chang}\ \emph {et~al.}(2025)\citenamefont {Chang}, \citenamefont {Manesi}, \citenamefont {Chen}, \citenamefont {Tsai}, \citenamefont {Su}, \citenamefont {Avgeropoulos},\ and\ \citenamefont {Ho}}]{Chang_2025}%
  \BibitemOpen
  \bibfield  {author} {\bibinfo {author} {\bibfnamefont {C.-Y.}\ \bibnamefont {Chang}}, \bibinfo {author} {\bibfnamefont {G.-M.}\ \bibnamefont {Manesi}}, \bibinfo {author} {\bibfnamefont {Y.-H.}\ \bibnamefont {Chen}}, \bibinfo {author} {\bibfnamefont {Y.-J.}\ \bibnamefont {Tsai}}, \bibinfo {author} {\bibfnamefont {H.-Y.}\ \bibnamefont {Su}}, \bibinfo {author} {\bibfnamefont {A.}~\bibnamefont {Avgeropoulos}},\ and\ \bibinfo {author} {\bibfnamefont {R.-M.}\ \bibnamefont {Ho}},\ }\bibfield  {title} {\bibinfo {title} {Architecture effect on network phase formation from controlled self-assembly of high-$\chi$ block copolymers},\ }\href {https://doi.org/10.1021/acs.macromol.5c02152} {\bibfield  {journal} {\bibinfo  {journal} {Macromolecules}\ }\textbf {\bibinfo {volume} {58}},\ \bibinfo {pages} {12574} (\bibinfo {year} {2025})}\BibitemShut {NoStop}%
\bibitem [{\citenamefont {Reddy}\ \emph {et~al.}(2018)\citenamefont {Reddy}, \citenamefont {Buckley}, \citenamefont {Arora}, \citenamefont {Bates}, \citenamefont {Dorfman},\ and\ \citenamefont {Grason}}]{Reddy_2018}%
  \BibitemOpen
  \bibfield  {author} {\bibinfo {author} {\bibfnamefont {A.}~\bibnamefont {Reddy}}, \bibinfo {author} {\bibfnamefont {M.~B.}\ \bibnamefont {Buckley}}, \bibinfo {author} {\bibfnamefont {A.}~\bibnamefont {Arora}}, \bibinfo {author} {\bibfnamefont {F.~S.}\ \bibnamefont {Bates}}, \bibinfo {author} {\bibfnamefont {K.~D.}\ \bibnamefont {Dorfman}},\ and\ \bibinfo {author} {\bibfnamefont {G.~M.}\ \bibnamefont {Grason}},\ }\bibfield  {title} {\bibinfo {title} {Stable {Frank--Kasper} phases of self-assembled, soft matter spheres},\ }\href {https://doi.org/10.1073/pnas.1809655115} {\bibfield  {journal} {\bibinfo  {journal} {Proc. Natl. Acad. Sci. U.S.A.}\ }\textbf {\bibinfo {volume} {115}},\ \bibinfo {pages} {10233} (\bibinfo {year} {2018})}\BibitemShut {NoStop}%
\bibitem [{\citenamefont {Reddy}\ \emph {et~al.}(2021)\citenamefont {Reddy}, \citenamefont {Feng}, \citenamefont {Thomas},\ and\ \citenamefont {Grason}}]{Reddy_2021}%
  \BibitemOpen
  \bibfield  {author} {\bibinfo {author} {\bibfnamefont {A.}~\bibnamefont {Reddy}}, \bibinfo {author} {\bibfnamefont {X.}~\bibnamefont {Feng}}, \bibinfo {author} {\bibfnamefont {E.~L.}\ \bibnamefont {Thomas}},\ and\ \bibinfo {author} {\bibfnamefont {G.~M.}\ \bibnamefont {Grason}},\ }\bibfield  {title} {\bibinfo {title} {Block copolymers beneath the surface: Measuring and modeling complex morphology at the subdomain scale},\ }\href {https://doi.org/10.1021/acs.macromol.1c00958} {\bibfield  {journal} {\bibinfo  {journal} {Macromolecules}\ }\textbf {\bibinfo {volume} {54}},\ \bibinfo {pages} {9223} (\bibinfo {year} {2021})}\BibitemShut {NoStop}%
\bibitem [{\citenamefont {Reddy}\ \emph {et~al.}(2022)\citenamefont {Reddy}, \citenamefont {Dimitriyev},\ and\ \citenamefont {Grason}}]{Reddy_2022}%
  \BibitemOpen
  \bibfield  {author} {\bibinfo {author} {\bibfnamefont {A.}~\bibnamefont {Reddy}}, \bibinfo {author} {\bibfnamefont {M.~S.}\ \bibnamefont {Dimitriyev}},\ and\ \bibinfo {author} {\bibfnamefont {G.~M.}\ \bibnamefont {Grason}},\ }\bibfield  {title} {\bibinfo {title} {Medial packing and elastic asymmetry stabilize the double-gyroid in block copolymers},\ }\href {https://doi.org/10.1038/s41467-022-30343-2} {\bibfield  {journal} {\bibinfo  {journal} {Nat. Commun.}\ }\textbf {\bibinfo {volume} {13}},\ \bibinfo {pages} {2629} (\bibinfo {year} {2022})}\BibitemShut {NoStop}%
\bibitem [{\citenamefont {Grason}\ and\ \citenamefont {Thomas}(2023)}]{Grason:2023aa}%
  \BibitemOpen
  \bibfield  {author} {\bibinfo {author} {\bibfnamefont {G.~M.}\ \bibnamefont {Grason}}\ and\ \bibinfo {author} {\bibfnamefont {E.~L.}\ \bibnamefont {Thomas}},\ }\bibfield  {title} {\bibinfo {title} {How does your gyroid grow? {A} mesoatomic perspective on supramolecular, soft matter network crystals},\ }\href {https://doi.org/10.1103/PhysRevMaterials.7.045603} {\bibfield  {journal} {\bibinfo  {journal} {Phys. Rev. Mater.}\ }\textbf {\bibinfo {volume} {7}},\ \bibinfo {pages} {045603} (\bibinfo {year} {2023})}\BibitemShut {NoStop}%
\bibitem [{\citenamefont {Dimitriyev}\ \emph {et~al.}(2023)\citenamefont {Dimitriyev}, \citenamefont {Reddy},\ and\ \citenamefont {Grason}}]{Dimitriyev_2023}%
  \BibitemOpen
  \bibfield  {author} {\bibinfo {author} {\bibfnamefont {M.~S.}\ \bibnamefont {Dimitriyev}}, \bibinfo {author} {\bibfnamefont {A.}~\bibnamefont {Reddy}},\ and\ \bibinfo {author} {\bibfnamefont {G.~M.}\ \bibnamefont {Grason}},\ }\bibfield  {title} {\bibinfo {title} {Medial packing, frustration, and competing network phases in strongly segregated block copolymers},\ }\href {https://doi.org/10.1021/acs.macromol.3c01098} {\bibfield  {journal} {\bibinfo  {journal} {Macromolecules}\ }\textbf {\bibinfo {volume} {56}},\ \bibinfo {pages} {7184} (\bibinfo {year} {2023})}\BibitemShut {NoStop}%
\bibitem [{\citenamefont {Matsen}\ and\ \citenamefont {Bates}(1997)}]{Matsen_1997}%
  \BibitemOpen
  \bibfield  {author} {\bibinfo {author} {\bibfnamefont {M.~W.}\ \bibnamefont {Matsen}}\ and\ \bibinfo {author} {\bibfnamefont {F.~S.}\ \bibnamefont {Bates}},\ }\bibfield  {title} {\bibinfo {title} {Block copolymer microstructures in the intermediate-segregation regime},\ }\href {https://doi.org/10.1063/1.473153} {\bibfield  {journal} {\bibinfo  {journal} {J. Chem. Phys.}\ }\textbf {\bibinfo {volume} {106}},\ \bibinfo {pages} {2436} (\bibinfo {year} {1997})}\BibitemShut {NoStop}%
\bibitem [{\citenamefont {Hou}\ \emph {et~al.}(2024)\citenamefont {Hou}, \citenamefont {Xu}, \citenamefont {Dong}, \citenamefont {Shi},\ and\ \citenamefont {Li}}]{Hou_2024}%
  \BibitemOpen
  \bibfield  {author} {\bibinfo {author} {\bibfnamefont {L.}~\bibnamefont {Hou}}, \bibinfo {author} {\bibfnamefont {Z.}~\bibnamefont {Xu}}, \bibinfo {author} {\bibfnamefont {Q.}~\bibnamefont {Dong}}, \bibinfo {author} {\bibfnamefont {A.-C.}\ \bibnamefont {Shi}},\ and\ \bibinfo {author} {\bibfnamefont {W.}~\bibnamefont {Li}},\ }\bibfield  {title} {\bibinfo {title} {Stabilizing {DG}, {DD}, and {DP} bicontinuous network phases in pure {AB}-type block copolymers beyond relieving packing frustration},\ }\href {https://doi.org/10.1021/acs.macromol.3c02428} {\bibfield  {journal} {\bibinfo  {journal} {Macromolecules}\ }\textbf {\bibinfo {volume} {57}},\ \bibinfo {pages} {2165} (\bibinfo {year} {2024})}\BibitemShut {NoStop}%
\bibitem [{\citenamefont {Weber}\ \emph {et~al.}(2011)\citenamefont {Weber}, \citenamefont {Ye}, \citenamefont {Schmitt}, \citenamefont {Banik}, \citenamefont {Elabd},\ and\ \citenamefont {Mahanthappa}}]{Weber_2011}%
  \BibitemOpen
  \bibfield  {author} {\bibinfo {author} {\bibfnamefont {R.~L.}\ \bibnamefont {Weber}}, \bibinfo {author} {\bibfnamefont {Y.}~\bibnamefont {Ye}}, \bibinfo {author} {\bibfnamefont {A.~L.}\ \bibnamefont {Schmitt}}, \bibinfo {author} {\bibfnamefont {S.~M.}\ \bibnamefont {Banik}}, \bibinfo {author} {\bibfnamefont {Y.~A.}\ \bibnamefont {Elabd}},\ and\ \bibinfo {author} {\bibfnamefont {M.~K.}\ \bibnamefont {Mahanthappa}},\ }\bibfield  {title} {\bibinfo {title} {Effect of nanoscale morphology on the conductivity of polymerized ionic liquid block copolymers},\ }\href {https://doi.org/10.1021/ma201067h} {\bibfield  {journal} {\bibinfo  {journal} {Macromolecules}\ }\textbf {\bibinfo {volume} {44}},\ \bibinfo {pages} {5727} (\bibinfo {year} {2011})}\BibitemShut {NoStop}%
\bibitem [{\citenamefont {Nakamura}\ \emph {et~al.}(2011)\citenamefont {Nakamura}, \citenamefont {Balsara},\ and\ \citenamefont {Wang}}]{Nakamura_2011}%
  \BibitemOpen
  \bibfield  {author} {\bibinfo {author} {\bibfnamefont {I.}~\bibnamefont {Nakamura}}, \bibinfo {author} {\bibfnamefont {N.~P.}\ \bibnamefont {Balsara}},\ and\ \bibinfo {author} {\bibfnamefont {Z.-G.}\ \bibnamefont {Wang}},\ }\bibfield  {title} {\bibinfo {title} {Thermodynamics of ion-containing polymer blends and block copolymers},\ }\href {https://doi.org/10.1103/PhysRevLett.107.198301} {\bibfield  {journal} {\bibinfo  {journal} {Phys. Rev. Lett.}\ }\textbf {\bibinfo {volume} {107}},\ \bibinfo {pages} {198301} (\bibinfo {year} {2011})}\BibitemShut {NoStop}%
\bibitem [{\citenamefont {Scalfani}\ \emph {et~al.}(2012)\citenamefont {Scalfani}, \citenamefont {Wiesenauer}, \citenamefont {Ekblad}, \citenamefont {Edwards}, \citenamefont {Gin},\ and\ \citenamefont {Bailey}}]{Scalfani_2012}%
  \BibitemOpen
  \bibfield  {author} {\bibinfo {author} {\bibfnamefont {V.~F.}\ \bibnamefont {Scalfani}}, \bibinfo {author} {\bibfnamefont {E.~F.}\ \bibnamefont {Wiesenauer}}, \bibinfo {author} {\bibfnamefont {J.~R.}\ \bibnamefont {Ekblad}}, \bibinfo {author} {\bibfnamefont {J.~P.}\ \bibnamefont {Edwards}}, \bibinfo {author} {\bibfnamefont {D.~L.}\ \bibnamefont {Gin}},\ and\ \bibinfo {author} {\bibfnamefont {T.~S.}\ \bibnamefont {Bailey}},\ }\bibfield  {title} {\bibinfo {title} {Morphological phase behavior of poly({RTIL})-containing diblock copolymer melts},\ }\href {https://doi.org/10.1021/ma300328u} {\bibfield  {journal} {\bibinfo  {journal} {Macromolecules}\ }\textbf {\bibinfo {volume} {45}},\ \bibinfo {pages} {4262} (\bibinfo {year} {2012})}\BibitemShut {NoStop}%
\bibitem [{\citenamefont {Qin}\ and\ \citenamefont {de~Pablo}(2016)}]{Qin_2016}%
  \BibitemOpen
  \bibfield  {author} {\bibinfo {author} {\bibfnamefont {J.}~\bibnamefont {Qin}}\ and\ \bibinfo {author} {\bibfnamefont {J.~J.}\ \bibnamefont {de~Pablo}},\ }\bibfield  {title} {\bibinfo {title} {Ordering transition in salt-doped diblock copolymers},\ }\href {https://doi.org/10.1021/acs.macromol.5b02643} {\bibfield  {journal} {\bibinfo  {journal} {Macromolecules}\ }\textbf {\bibinfo {volume} {49}},\ \bibinfo {pages} {3630} (\bibinfo {year} {2016})}\BibitemShut {NoStop}%
\bibitem [{\citenamefont {Hou}\ and\ \citenamefont {Qin}(2018)}]{Hou_2018}%
  \BibitemOpen
  \bibfield  {author} {\bibinfo {author} {\bibfnamefont {K.~J.}\ \bibnamefont {Hou}}\ and\ \bibinfo {author} {\bibfnamefont {J.}~\bibnamefont {Qin}},\ }\bibfield  {title} {\bibinfo {title} {Solvation and entropic regimes in ion-containing block copolymers},\ }\href {https://doi.org/10.1021/acs.macromol.8b01616} {\bibfield  {journal} {\bibinfo  {journal} {Macromolecules}\ }\textbf {\bibinfo {volume} {51}},\ \bibinfo {pages} {7463} (\bibinfo {year} {2018})}\BibitemShut {NoStop}%
\bibitem [{\citenamefont {Kong}\ \emph {et~al.}(2021)\citenamefont {Kong}, \citenamefont {Hou},\ and\ \citenamefont {Qin}}]{Kong_2021}%
  \BibitemOpen
  \bibfield  {author} {\bibinfo {author} {\bibfnamefont {X.}~\bibnamefont {Kong}}, \bibinfo {author} {\bibfnamefont {K.~J.-Y.}\ \bibnamefont {Hou}},\ and\ \bibinfo {author} {\bibfnamefont {J.}~\bibnamefont {Qin}},\ }\bibfield  {title} {\bibinfo {title} {Weakening of solvation-induced ordering by composition fluctuation in salt-doped block polymers},\ }\href {https://doi.org/10.1021/acsmacrolett.1c00107} {\bibfield  {journal} {\bibinfo  {journal} {ACS Macro Lett.}\ }\textbf {\bibinfo {volume} {10}},\ \bibinfo {pages} {545} (\bibinfo {year} {2021})}\BibitemShut {NoStop}%
\bibitem [{\citenamefont {Kim}\ \emph {et~al.}(2017{\natexlab{b}})\citenamefont {Kim}, \citenamefont {Jeong}, \citenamefont {Yethiraj},\ and\ \citenamefont {Mahanthappa}}]{Kim_2017PNAS}%
  \BibitemOpen
  \bibfield  {author} {\bibinfo {author} {\bibfnamefont {S.~A.}\ \bibnamefont {Kim}}, \bibinfo {author} {\bibfnamefont {K.-J.}\ \bibnamefont {Jeong}}, \bibinfo {author} {\bibfnamefont {A.}~\bibnamefont {Yethiraj}},\ and\ \bibinfo {author} {\bibfnamefont {M.~K.}\ \bibnamefont {Mahanthappa}},\ }\bibfield  {title} {\bibinfo {title} {Low-symmetry sphere packings of simple surfactant micelles induced by ionic sphericity},\ }\href {https://doi.org/10.1073/pnas.1701608114} {\bibfield  {journal} {\bibinfo  {journal} {Proc. Natl. Acad. Sci. U.S.A.}\ }\textbf {\bibinfo {volume} {114}},\ \bibinfo {pages} {4072} (\bibinfo {year} {2017}{\natexlab{b}})}\BibitemShut {NoStop}%
\bibitem [{\citenamefont {Brown}\ \emph {et~al.}(2018)\citenamefont {Brown}, \citenamefont {Seo},\ and\ \citenamefont {Hall}}]{Brown_2018}%
  \BibitemOpen
  \bibfield  {author} {\bibinfo {author} {\bibfnamefont {J.~R.}\ \bibnamefont {Brown}}, \bibinfo {author} {\bibfnamefont {Y.}~\bibnamefont {Seo}},\ and\ \bibinfo {author} {\bibfnamefont {L.~M.}\ \bibnamefont {Hall}},\ }\bibfield  {title} {\bibinfo {title} {Ion correlation effects in salt-doped block copolymers},\ }\href {https://doi.org/10.1103/PhysRevLett.120.127801} {\bibfield  {journal} {\bibinfo  {journal} {Phys. Rev. Lett.}\ }\textbf {\bibinfo {volume} {120}},\ \bibinfo {pages} {127801} (\bibinfo {year} {2018})}\BibitemShut {NoStop}%
\bibitem [{\citenamefont {Zhang}\ \emph {et~al.}(2021{\natexlab{a}})\citenamefont {Zhang}, \citenamefont {Zheng}, \citenamefont {Sims}, \citenamefont {Bates},\ and\ \citenamefont {Lodge}}]{Zhang_2021}%
  \BibitemOpen
  \bibfield  {author} {\bibinfo {author} {\bibfnamefont {B.}~\bibnamefont {Zhang}}, \bibinfo {author} {\bibfnamefont {C.}~\bibnamefont {Zheng}}, \bibinfo {author} {\bibfnamefont {M.~B.}\ \bibnamefont {Sims}}, \bibinfo {author} {\bibfnamefont {F.~S.}\ \bibnamefont {Bates}},\ and\ \bibinfo {author} {\bibfnamefont {T.~P.}\ \bibnamefont {Lodge}},\ }\bibfield  {title} {\bibinfo {title} {Influence of charge fraction on the phase behavior of symmetric single-ion conducting diblock copolymers},\ }\href {https://doi.org/10.1021/acsmacrolett.1c00393} {\bibfield  {journal} {\bibinfo  {journal} {ACS Macro Lett.}\ }\textbf {\bibinfo {volume} {10}},\ \bibinfo {pages} {1035} (\bibinfo {year} {2021}{\natexlab{a}})}\BibitemShut {NoStop}%
\bibitem [{\citenamefont {Zhang}\ \emph {et~al.}(2021{\natexlab{b}})\citenamefont {Zhang}, \citenamefont {Krajniak},\ and\ \citenamefont {Ganesan}}]{ZhangZ_2021}%
  \BibitemOpen
  \bibfield  {author} {\bibinfo {author} {\bibfnamefont {Z.}~\bibnamefont {Zhang}}, \bibinfo {author} {\bibfnamefont {J.}~\bibnamefont {Krajniak}},\ and\ \bibinfo {author} {\bibfnamefont {V.}~\bibnamefont {Ganesan}},\ }\bibfield  {title} {\bibinfo {title} {A multiscale simulation study of influence of morphology on ion transport in block copolymeric ionic liquids},\ }\href {https://doi.org/10.1021/acs.macromol.1c00025} {\bibfield  {journal} {\bibinfo  {journal} {Macromolecules}\ }\textbf {\bibinfo {volume} {54}},\ \bibinfo {pages} {4997} (\bibinfo {year} {2021}{\natexlab{b}})}\BibitemShut {NoStop}%
\bibitem [{\citenamefont {Min}\ \emph {et~al.}(2021{\natexlab{a}})\citenamefont {Min}, \citenamefont {Barpuzary}, \citenamefont {Ham}, \citenamefont {Kang},\ and\ \citenamefont {Park}}]{Min_2021b}%
  \BibitemOpen
  \bibfield  {author} {\bibinfo {author} {\bibfnamefont {J.}~\bibnamefont {Min}}, \bibinfo {author} {\bibfnamefont {D.}~\bibnamefont {Barpuzary}}, \bibinfo {author} {\bibfnamefont {H.}~\bibnamefont {Ham}}, \bibinfo {author} {\bibfnamefont {G.-C.}\ \bibnamefont {Kang}},\ and\ \bibinfo {author} {\bibfnamefont {M.~J.}\ \bibnamefont {Park}},\ }\bibfield  {title} {\bibinfo {title} {Charged block copolymers: From fundamentals to electromechanical applications},\ }\href {https://doi.org/10.1021/acs.accounts.1c00423} {\bibfield  {journal} {\bibinfo  {journal} {Acc. Chem. Res.}\ }\textbf {\bibinfo {volume} {54}},\ \bibinfo {pages} {4024} (\bibinfo {year} {2021}{\natexlab{a}})}\BibitemShut {NoStop}%
\bibitem [{\citenamefont {Lee}\ \emph {et~al.}(2024{\natexlab{b}})\citenamefont {Lee}, \citenamefont {Kim},\ and\ \citenamefont {Park}}]{Lee_2024b}%
  \BibitemOpen
  \bibfield  {author} {\bibinfo {author} {\bibfnamefont {H.}~\bibnamefont {Lee}}, \bibinfo {author} {\bibfnamefont {J.}~\bibnamefont {Kim}},\ and\ \bibinfo {author} {\bibfnamefont {M.~J.}\ \bibnamefont {Park}},\ }\bibfield  {title} {\bibinfo {title} {Exploration of complex nanostructures in block copolymers},\ }\href {https://doi.org/10.1103/PhysRevMaterials.8.020302} {\bibfield  {journal} {\bibinfo  {journal} {Phys. Rev. Mater.}\ }\textbf {\bibinfo {volume} {8}},\ \bibinfo {pages} {020302} (\bibinfo {year} {2024}{\natexlab{b}})}\BibitemShut {NoStop}%
\bibitem [{\citenamefont {Huo}\ \emph {et~al.}(2023)\citenamefont {Huo}, \citenamefont {Zhao}, \citenamefont {Duan},\ and\ \citenamefont {Sun}}]{Huo_2023}%
  \BibitemOpen
  \bibfield  {author} {\bibinfo {author} {\bibfnamefont {H.}~\bibnamefont {Huo}}, \bibinfo {author} {\bibfnamefont {W.}~\bibnamefont {Zhao}}, \bibinfo {author} {\bibfnamefont {X.}~\bibnamefont {Duan}},\ and\ \bibinfo {author} {\bibfnamefont {Z.-Y.}\ \bibnamefont {Sun}},\ }\bibfield  {title} {\bibinfo {title} {Control of diblock copolyelectrolyte morphology through electric field application},\ }\href {https://doi.org/10.1021/acs.macromol.2c01780} {\bibfield  {journal} {\bibinfo  {journal} {Macromolecules}\ }\textbf {\bibinfo {volume} {56}},\ \bibinfo {pages} {1065} (\bibinfo {year} {2023})}\BibitemShut {NoStop}%
\bibitem [{\citenamefont {Zhang}\ \emph {et~al.}(2023)\citenamefont {Zhang}, \citenamefont {Zheng}, \citenamefont {Bates},\ and\ \citenamefont {Lodge}}]{Zhang_2023}%
  \BibitemOpen
  \bibfield  {author} {\bibinfo {author} {\bibfnamefont {B.}~\bibnamefont {Zhang}}, \bibinfo {author} {\bibfnamefont {C.}~\bibnamefont {Zheng}}, \bibinfo {author} {\bibfnamefont {F.~S.}\ \bibnamefont {Bates}},\ and\ \bibinfo {author} {\bibfnamefont {T.~P.}\ \bibnamefont {Lodge}},\ }\bibfield  {title} {\bibinfo {title} {Self-assembly of charged diblock copolymers with reduced backbone polarity},\ }\href {https://doi.org/10.1021/acsapm.2c02220} {\bibfield  {journal} {\bibinfo  {journal} {ACS Appl. Polym. Mater.}\ }\textbf {\bibinfo {volume} {5}},\ \bibinfo {pages} {2223} (\bibinfo {year} {2023})}\BibitemShut {NoStop}%
\bibitem [{\citenamefont {Tsai}\ \emph {et~al.}(2025)\citenamefont {Tsai}, \citenamefont {Fan}, \citenamefont {Zhou},\ and\ \citenamefont {Xie}}]{Tsai_2025}%
  \BibitemOpen
  \bibfield  {author} {\bibinfo {author} {\bibfnamefont {C.-C.}\ \bibnamefont {Tsai}}, \bibinfo {author} {\bibfnamefont {H.}~\bibnamefont {Fan}}, \bibinfo {author} {\bibfnamefont {Y.}~\bibnamefont {Zhou}},\ and\ \bibinfo {author} {\bibfnamefont {S.}~\bibnamefont {Xie}},\ }\bibfield  {title} {\bibinfo {title} {Rational design of ionomer microstructures for thermally reprocessable materials with creep resistance and recoverability},\ }\href {https://doi.org/10.1021/jacsau.5c01317} {\bibfield  {journal} {\bibinfo  {journal} {JACS Au}\ }\textbf {\bibinfo {volume} {5}},\ \bibinfo {pages} {6324} (\bibinfo {year} {2025})}\BibitemShut {NoStop}%
\bibitem [{\citenamefont {Hou}\ \emph {et~al.}(2026)\citenamefont {Hou}, \citenamefont {Zhao}, \citenamefont {Wei}, \citenamefont {Zhou}, \citenamefont {Shen},\ and\ \citenamefont {Shi}}]{Hou_2026}%
  \BibitemOpen
  \bibfield  {author} {\bibinfo {author} {\bibfnamefont {L.}~\bibnamefont {Hou}}, \bibinfo {author} {\bibfnamefont {X.}~\bibnamefont {Zhao}}, \bibinfo {author} {\bibfnamefont {Z.}~\bibnamefont {Wei}}, \bibinfo {author} {\bibfnamefont {N.}~\bibnamefont {Zhou}}, \bibinfo {author} {\bibfnamefont {L.}~\bibnamefont {Shen}},\ and\ \bibinfo {author} {\bibfnamefont {W.}~\bibnamefont {Shi}},\ }\bibfield  {title} {\bibinfo {title} {Asymmetric self-assembly of functional ionic block copolymers with tailored dense charge modification},\ }\href {https://doi.org/10.1021/jacs.6c04169} {\bibfield  {journal} {\bibinfo  {journal} {J. Am. Chem. Soc.}\ }\textbf {\bibinfo {volume} {148}},\ \bibinfo {pages} {24041} (\bibinfo {year} {2026})}\BibitemShut {NoStop}%
\bibitem [{\citenamefont {Park}\ and\ \citenamefont {Balsara}(2008)}]{Park_2008}%
  \BibitemOpen
  \bibfield  {author} {\bibinfo {author} {\bibfnamefont {M.~J.}\ \bibnamefont {Park}}\ and\ \bibinfo {author} {\bibfnamefont {N.~P.}\ \bibnamefont {Balsara}},\ }\bibfield  {title} {\bibinfo {title} {Phase behavior of symmetric sulfonated block copolymers},\ }\href {https://doi.org/10.1021/ma702733f} {\bibfield  {journal} {\bibinfo  {journal} {Macromolecules}\ }\textbf {\bibinfo {volume} {41}},\ \bibinfo {pages} {3678} (\bibinfo {year} {2008})}\BibitemShut {NoStop}%
\bibitem [{\citenamefont {Shim}\ \emph {et~al.}(2019)\citenamefont {Shim}, \citenamefont {Bates},\ and\ \citenamefont {Lodge}}]{Shim_2019}%
  \BibitemOpen
  \bibfield  {author} {\bibinfo {author} {\bibfnamefont {J.}~\bibnamefont {Shim}}, \bibinfo {author} {\bibfnamefont {F.~S.}\ \bibnamefont {Bates}},\ and\ \bibinfo {author} {\bibfnamefont {T.~P.}\ \bibnamefont {Lodge}},\ }\bibfield  {title} {\bibinfo {title} {Superlattice by charged block copolymer self-assembly},\ }\href {https://doi.org/10.1038/s41467-019-10141-z} {\bibfield  {journal} {\bibinfo  {journal} {Nat. Commun.}\ }\textbf {\bibinfo {volume} {10}},\ \bibinfo {pages} {2108} (\bibinfo {year} {2019})}\BibitemShut {NoStop}%
\bibitem [{\citenamefont {Min}\ \emph {et~al.}(2021{\natexlab{b}})\citenamefont {Min}, \citenamefont {Jung}, \citenamefont {Jeong}, \citenamefont {Lee}, \citenamefont {Son},\ and\ \citenamefont {Park}}]{Min_2021a}%
  \BibitemOpen
  \bibfield  {author} {\bibinfo {author} {\bibfnamefont {J.}~\bibnamefont {Min}}, \bibinfo {author} {\bibfnamefont {H.~Y.}\ \bibnamefont {Jung}}, \bibinfo {author} {\bibfnamefont {S.}~\bibnamefont {Jeong}}, \bibinfo {author} {\bibfnamefont {B.}~\bibnamefont {Lee}}, \bibinfo {author} {\bibfnamefont {C.~Y.}\ \bibnamefont {Son}},\ and\ \bibinfo {author} {\bibfnamefont {M.~J.}\ \bibnamefont {Park}},\ }\bibfield  {title} {\bibinfo {title} {Enhancing ion transport in charged block copolymers by stabilizing low symmetry morphology: Electrostatic control of interfaces},\ }\href {https://doi.org/10.1073/pnas.2107987118} {\bibfield  {journal} {\bibinfo  {journal} {Proc. Natl. Acad. Sci. U.S.A.}\ }\textbf {\bibinfo {volume} {118}},\ \bibinfo {pages} {e2107987118} (\bibinfo {year} {2021}{\natexlab{b}})}\BibitemShut {NoStop}%
\bibitem [{\citenamefont {Yan}\ \emph {et~al.}(2020)\citenamefont {Yan}, \citenamefont {Rank}, \citenamefont {Mecking},\ and\ \citenamefont {Winey}}]{Yan_2020}%
  \BibitemOpen
  \bibfield  {author} {\bibinfo {author} {\bibfnamefont {L.}~\bibnamefont {Yan}}, \bibinfo {author} {\bibfnamefont {C.}~\bibnamefont {Rank}}, \bibinfo {author} {\bibfnamefont {S.}~\bibnamefont {Mecking}},\ and\ \bibinfo {author} {\bibfnamefont {K.~I.}\ \bibnamefont {Winey}},\ }\bibfield  {title} {\bibinfo {title} {Gyroid and other ordered morphologies in single-ion conducting polymers and their impact on ion conductivity},\ }\href {https://doi.org/10.1021/jacs.9b09701} {\bibfield  {journal} {\bibinfo  {journal} {J. Am. Chem. Soc.}\ }\textbf {\bibinfo {volume} {142}},\ \bibinfo {pages} {857} (\bibinfo {year} {2020})}\BibitemShut {NoStop}%
\bibitem [{\citenamefont {Park}\ \emph {et~al.}(2021{\natexlab{a}})\citenamefont {Park}, \citenamefont {Staiger}, \citenamefont {Mecking},\ and\ \citenamefont {Winey}}]{Park_2021AN}%
  \BibitemOpen
  \bibfield  {author} {\bibinfo {author} {\bibfnamefont {J.}~\bibnamefont {Park}}, \bibinfo {author} {\bibfnamefont {A.}~\bibnamefont {Staiger}}, \bibinfo {author} {\bibfnamefont {S.}~\bibnamefont {Mecking}},\ and\ \bibinfo {author} {\bibfnamefont {K.~I.}\ \bibnamefont {Winey}},\ }\bibfield  {title} {\bibinfo {title} {Sub-3-nanometer domain spacings of ultrahigh-{$\chi$} multiblock copolymers with pendant ionic groups},\ }\href {https://doi.org/10.1021/acsnano.1c06734} {\bibfield  {journal} {\bibinfo  {journal} {ACS Nano}\ }\textbf {\bibinfo {volume} {15}},\ \bibinfo {pages} {16738} (\bibinfo {year} {2021}{\natexlab{a}})}\BibitemShut {NoStop}%
\bibitem [{\citenamefont {Park}\ \emph {et~al.}(2021{\natexlab{b}})\citenamefont {Park}, \citenamefont {Staiger}, \citenamefont {Mecking},\ and\ \citenamefont {Winey}}]{Park_2021}%
  \BibitemOpen
  \bibfield  {author} {\bibinfo {author} {\bibfnamefont {J.}~\bibnamefont {Park}}, \bibinfo {author} {\bibfnamefont {A.}~\bibnamefont {Staiger}}, \bibinfo {author} {\bibfnamefont {S.}~\bibnamefont {Mecking}},\ and\ \bibinfo {author} {\bibfnamefont {K.~I.}\ \bibnamefont {Winey}},\ }\bibfield  {title} {\bibinfo {title} {Structure--property relationships in single-ion conducting multiblock copolymers: A phase diagram and ionic conductivities},\ }\href {https://doi.org/10.1021/acs.macromol.1c00493} {\bibfield  {journal} {\bibinfo  {journal} {Macromolecules}\ }\textbf {\bibinfo {volume} {54}},\ \bibinfo {pages} {4269} (\bibinfo {year} {2021}{\natexlab{b}})}\BibitemShut {NoStop}%
\bibitem [{\citenamefont {Brown}\ \emph {et~al.}(2025)\citenamefont {Brown}, \citenamefont {Burlein}, \citenamefont {Ferko}, \citenamefont {Saumer}, \citenamefont {Mecking},\ and\ \citenamefont {Winey}}]{Brown_2025}%
  \BibitemOpen
  \bibfield  {author} {\bibinfo {author} {\bibfnamefont {M.~K.}\ \bibnamefont {Brown}}, \bibinfo {author} {\bibfnamefont {V.~A.}\ \bibnamefont {Burlein}}, \bibinfo {author} {\bibfnamefont {B.~T.}\ \bibnamefont {Ferko}}, \bibinfo {author} {\bibfnamefont {A.}~\bibnamefont {Saumer}}, \bibinfo {author} {\bibfnamefont {S.}~\bibnamefont {Mecking}},\ and\ \bibinfo {author} {\bibfnamefont {K.~I.}\ \bibnamefont {Winey}},\ }\bibfield  {title} {\bibinfo {title} {Pendant groups in the majority domain of aliphatic multiblock copolymers prohibit the double gyroid morphology},\ }\href {https://doi.org/10.1021/acs.macromol.5c01982} {\bibfield  {journal} {\bibinfo  {journal} {Macromolecules}\ }\textbf {\bibinfo {volume} {58}},\ \bibinfo {pages} {12283} (\bibinfo {year} {2025})}\BibitemShut {NoStop}%
\bibitem [{\citenamefont {Williams}\ \emph {et~al.}(1986)\citenamefont {Williams}, \citenamefont {Russell}, \citenamefont {Jerome},\ and\ \citenamefont {Horrion}}]{Williams_1986}%
  \BibitemOpen
  \bibfield  {author} {\bibinfo {author} {\bibfnamefont {C.~E.}\ \bibnamefont {Williams}}, \bibinfo {author} {\bibfnamefont {T.~P.}\ \bibnamefont {Russell}}, \bibinfo {author} {\bibfnamefont {R.}~\bibnamefont {Jerome}},\ and\ \bibinfo {author} {\bibfnamefont {J.}~\bibnamefont {Horrion}},\ }\bibfield  {title} {\bibinfo {title} {Ionic aggregation in model ionomers},\ }\href {https://doi.org/10.1021/ma00165a036} {\bibfield  {journal} {\bibinfo  {journal} {Macromolecules}\ }\textbf {\bibinfo {volume} {19}},\ \bibinfo {pages} {2877} (\bibinfo {year} {1986})}\BibitemShut {NoStop}%
\bibitem [{\citenamefont {Eisenberg}\ \emph {et~al.}(1990)\citenamefont {Eisenberg}, \citenamefont {Hird},\ and\ \citenamefont {Moore}}]{Eisenberg_1990}%
  \BibitemOpen
  \bibfield  {author} {\bibinfo {author} {\bibfnamefont {A.}~\bibnamefont {Eisenberg}}, \bibinfo {author} {\bibfnamefont {B.}~\bibnamefont {Hird}},\ and\ \bibinfo {author} {\bibfnamefont {R.~B.}\ \bibnamefont {Moore}},\ }\bibfield  {title} {\bibinfo {title} {A new multiplet-cluster model for the morphology of random ionomers},\ }\href {https://doi.org/10.1021/ma00220a012} {\bibfield  {journal} {\bibinfo  {journal} {Macromolecules}\ }\textbf {\bibinfo {volume} {23}},\ \bibinfo {pages} {4098} (\bibinfo {year} {1990})}\BibitemShut {NoStop}%
\bibitem [{\citenamefont {Hall}\ \emph {et~al.}(2011)\citenamefont {Hall}, \citenamefont {Seitz}, \citenamefont {Winey}, \citenamefont {Opper}, \citenamefont {Wagener}, \citenamefont {Stevens},\ and\ \citenamefont {Frischknecht}}]{Hall_2011}%
  \BibitemOpen
  \bibfield  {author} {\bibinfo {author} {\bibfnamefont {L.~M.}\ \bibnamefont {Hall}}, \bibinfo {author} {\bibfnamefont {M.~E.}\ \bibnamefont {Seitz}}, \bibinfo {author} {\bibfnamefont {K.~I.}\ \bibnamefont {Winey}}, \bibinfo {author} {\bibfnamefont {K.~L.}\ \bibnamefont {Opper}}, \bibinfo {author} {\bibfnamefont {K.~B.}\ \bibnamefont {Wagener}}, \bibinfo {author} {\bibfnamefont {M.~J.}\ \bibnamefont {Stevens}},\ and\ \bibinfo {author} {\bibfnamefont {A.~L.}\ \bibnamefont {Frischknecht}},\ }\bibfield  {title} {\bibinfo {title} {Ionic aggregate structure in ionomer melts: Effect of molecular architecture on aggregates and the ionomer peak},\ }\href {https://doi.org/10.1021/ja209142b} {\bibfield  {journal} {\bibinfo  {journal} {J. Am. Chem. Soc.}\ }\textbf {\bibinfo {volume} {134}},\ \bibinfo {pages} {574} (\bibinfo {year} {2011})}\BibitemShut {NoStop}%
\bibitem [{\citenamefont {Lee}\ \emph {et~al.}(2025{\natexlab{a}})\citenamefont {Lee}, \citenamefont {Kim},\ and\ \citenamefont {Park}}]{Lee_2025}%
  \BibitemOpen
  \bibfield  {author} {\bibinfo {author} {\bibfnamefont {H.}~\bibnamefont {Lee}}, \bibinfo {author} {\bibfnamefont {J.}~\bibnamefont {Kim}},\ and\ \bibinfo {author} {\bibfnamefont {M.~J.}\ \bibnamefont {Park}},\ }\bibfield  {title} {\bibinfo {title} {Block copolymer electrolytes with double primitive cubic structures: Enhancing solid-state lithium conduction via lithium salt localization},\ }\href {https://doi.org/10.1021/acsnano.4c13442} {\bibfield  {journal} {\bibinfo  {journal} {ACS Nano}\ }\textbf {\bibinfo {volume} {19}},\ \bibinfo {pages} {1251} (\bibinfo {year} {2025}{\natexlab{a}})}\BibitemShut {NoStop}%
\bibitem [{\citenamefont {Shi}\ and\ \citenamefont {Noolandi}(1999)}]{Shi_1999}%
  \BibitemOpen
  \bibfield  {author} {\bibinfo {author} {\bibfnamefont {A.-C.}\ \bibnamefont {Shi}}\ and\ \bibinfo {author} {\bibfnamefont {J.}~\bibnamefont {Noolandi}},\ }\bibfield  {title} {\bibinfo {title} {Theory of inhomogeneous weakly charged polyelectrolytes},\ }\href {https://doi.org/10.1002/(sici)1521-3919(19990501)8:3<214::aid-mats214>3.0.co;2-u} {\bibfield  {journal} {\bibinfo  {journal} {Macromol. Theory Simul.}\ }\textbf {\bibinfo {volume} {8}},\ \bibinfo {pages} {214} (\bibinfo {year} {1999})}\BibitemShut {NoStop}%
\bibitem [{\citenamefont {Wang}\ \emph {et~al.}(2004)\citenamefont {Wang}, \citenamefont {Taniguchi},\ and\ \citenamefont {Fredrickson}}]{Wang_2004}%
  \BibitemOpen
  \bibfield  {author} {\bibinfo {author} {\bibfnamefont {Q.}~\bibnamefont {Wang}}, \bibinfo {author} {\bibfnamefont {T.}~\bibnamefont {Taniguchi}},\ and\ \bibinfo {author} {\bibfnamefont {G.~H.}\ \bibnamefont {Fredrickson}},\ }\bibfield  {title} {\bibinfo {title} {Self-consistent field theory of polyelectrolyte systems},\ }\href {https://doi.org/10.1021/jp037053y} {\bibfield  {journal} {\bibinfo  {journal} {J. Phys. Chem. B}\ }\textbf {\bibinfo {volume} {108}},\ \bibinfo {pages} {6733} (\bibinfo {year} {2004})}\BibitemShut {NoStop}%
\bibitem [{\citenamefont {Kumar}\ and\ \citenamefont {Muthukumar}(2007)}]{Kumar_2007}%
  \BibitemOpen
  \bibfield  {author} {\bibinfo {author} {\bibfnamefont {R.}~\bibnamefont {Kumar}}\ and\ \bibinfo {author} {\bibfnamefont {M.}~\bibnamefont {Muthukumar}},\ }\bibfield  {title} {\bibinfo {title} {Microphase separation in polyelectrolytic diblock copolymer melt: Weak segregation limit},\ }\href {https://doi.org/10.1063/1.2737049} {\bibfield  {journal} {\bibinfo  {journal} {J. Chem. Phys.}\ }\textbf {\bibinfo {volume} {126}},\ \bibinfo {pages} {214902} (\bibinfo {year} {2007})}\BibitemShut {NoStop}%
\bibitem [{\citenamefont {Yang}\ \emph {et~al.}(2011)\citenamefont {Yang}, \citenamefont {Vishnyakov},\ and\ \citenamefont {Neimark}}]{Yang_2011}%
  \BibitemOpen
  \bibfield  {author} {\bibinfo {author} {\bibfnamefont {S.}~\bibnamefont {Yang}}, \bibinfo {author} {\bibfnamefont {A.}~\bibnamefont {Vishnyakov}},\ and\ \bibinfo {author} {\bibfnamefont {A.~V.}\ \bibnamefont {Neimark}},\ }\bibfield  {title} {\bibinfo {title} {Self-assembly in block polyelectrolytes},\ }\href {https://doi.org/10.1063/1.3532831} {\bibfield  {journal} {\bibinfo  {journal} {J. Chem. Phys.}\ }\textbf {\bibinfo {volume} {134}},\ \bibinfo {pages} {054104} (\bibinfo {year} {2011})}\BibitemShut {NoStop}%
\bibitem [{\citenamefont {Liu}\ \emph {et~al.}(2011)\citenamefont {Liu}, \citenamefont {Zhang}, \citenamefont {Tong},\ and\ \citenamefont {Yang}}]{Liu_2011}%
  \BibitemOpen
  \bibfield  {author} {\bibinfo {author} {\bibfnamefont {Y.-X.}\ \bibnamefont {Liu}}, \bibinfo {author} {\bibfnamefont {H.-D.}\ \bibnamefont {Zhang}}, \bibinfo {author} {\bibfnamefont {C.-H.}\ \bibnamefont {Tong}},\ and\ \bibinfo {author} {\bibfnamefont {Y.-L.}\ \bibnamefont {Yang}},\ }\bibfield  {title} {\bibinfo {title} {Microphase separation and phase diagram of concentrated diblock copolyelectrolyte solutions studied by self-consistent field theory calculations in two-dimensional space},\ }\href {https://doi.org/10.1021/ma2010266} {\bibfield  {journal} {\bibinfo  {journal} {Macromolecules}\ }\textbf {\bibinfo {volume} {44}},\ \bibinfo {pages} {8261} (\bibinfo {year} {2011})}\BibitemShut {NoStop}%
\bibitem [{\citenamefont {Young}\ and\ \citenamefont {Epps}(2009)}]{Young_2009}%
  \BibitemOpen
  \bibfield  {author} {\bibinfo {author} {\bibfnamefont {W.-S.}\ \bibnamefont {Young}}\ and\ \bibinfo {author} {\bibfnamefont {T.~H.}\ \bibnamefont {Epps}},\ }\bibfield  {title} {\bibinfo {title} {Salt doping in {PEO}-containing block copolymers: Counterion and concentration effects},\ }\href {https://doi.org/10.1021/ma802799p} {\bibfield  {journal} {\bibinfo  {journal} {Macromolecules}\ }\textbf {\bibinfo {volume} {42}},\ \bibinfo {pages} {2672} (\bibinfo {year} {2009})}\BibitemShut {NoStop}%
\bibitem [{\citenamefont {Loo}\ \emph {et~al.}(2018)\citenamefont {Loo}, \citenamefont {Galluzzo}, \citenamefont {Li}, \citenamefont {Maslyn}, \citenamefont {Oh}, \citenamefont {Mongcopa}, \citenamefont {Zhu}, \citenamefont {Wang}, \citenamefont {Wang}, \citenamefont {Garetz},\ and\ \citenamefont {Balsara}}]{Loo_2018}%
  \BibitemOpen
  \bibfield  {author} {\bibinfo {author} {\bibfnamefont {W.~S.}\ \bibnamefont {Loo}}, \bibinfo {author} {\bibfnamefont {M.~D.}\ \bibnamefont {Galluzzo}}, \bibinfo {author} {\bibfnamefont {X.}~\bibnamefont {Li}}, \bibinfo {author} {\bibfnamefont {J.~A.}\ \bibnamefont {Maslyn}}, \bibinfo {author} {\bibfnamefont {H.~J.}\ \bibnamefont {Oh}}, \bibinfo {author} {\bibfnamefont {K.~I.}\ \bibnamefont {Mongcopa}}, \bibinfo {author} {\bibfnamefont {C.}~\bibnamefont {Zhu}}, \bibinfo {author} {\bibfnamefont {A.~A.}\ \bibnamefont {Wang}}, \bibinfo {author} {\bibfnamefont {X.}~\bibnamefont {Wang}}, \bibinfo {author} {\bibfnamefont {B.~A.}\ \bibnamefont {Garetz}},\ and\ \bibinfo {author} {\bibfnamefont {N.~P.}\ \bibnamefont {Balsara}},\ }\bibfield  {title} {\bibinfo {title} {Phase behavior of mixtures of block copolymers and a lithium salt},\ }\href {https://doi.org/10.1021/acs.jpcb.8b04189} {\bibfield  {journal} {\bibinfo  {journal} {J. Phys. Chem. B}\ }\textbf {\bibinfo {volume} {122}},\ \bibinfo {pages} {8065} (\bibinfo
  {year} {2018})}\BibitemShut {NoStop}%
\bibitem [{\citenamefont {Sing}\ \emph {et~al.}(2014)\citenamefont {Sing}, \citenamefont {Zwanikken},\ and\ \citenamefont {de~la Cruz}}]{Sing_2014a}%
  \BibitemOpen
  \bibfield  {author} {\bibinfo {author} {\bibfnamefont {C.~E.}\ \bibnamefont {Sing}}, \bibinfo {author} {\bibfnamefont {J.~W.}\ \bibnamefont {Zwanikken}},\ and\ \bibinfo {author} {\bibfnamefont {M.~O.}\ \bibnamefont {de~la Cruz}},\ }\bibfield  {title} {\bibinfo {title} {Electrostatic control of block copolymer~morphology},\ }\href {https://doi.org/10.1038/nmat4001} {\bibfield  {journal} {\bibinfo  {journal} {Nat. Mater.}\ }\textbf {\bibinfo {volume} {13}},\ \bibinfo {pages} {694} (\bibinfo {year} {2014})}\BibitemShut {NoStop}%
\bibitem [{\citenamefont {Sing}\ and\ \citenamefont {de~la Cruz}(2014)}]{Sing_2014b}%
  \BibitemOpen
  \bibfield  {author} {\bibinfo {author} {\bibfnamefont {C.~E.}\ \bibnamefont {Sing}}\ and\ \bibinfo {author} {\bibfnamefont {M.~O.}\ \bibnamefont {de~la Cruz}},\ }\bibfield  {title} {\bibinfo {title} {Polyelectrolyte blends and nontrivial behavior in effective {Flory--Huggins} parameters},\ }\href {https://doi.org/10.1021/mz500202n} {\bibfield  {journal} {\bibinfo  {journal} {ACS Macro Lett.}\ }\textbf {\bibinfo {volume} {3}},\ \bibinfo {pages} {698} (\bibinfo {year} {2014})}\BibitemShut {NoStop}%
\bibitem [{\citenamefont {Sing}\ \emph {et~al.}(2015)\citenamefont {Sing}, \citenamefont {Zwanikken},\ and\ \citenamefont {de~la Cruz}}]{Sing_2015}%
  \BibitemOpen
  \bibfield  {author} {\bibinfo {author} {\bibfnamefont {C.~E.}\ \bibnamefont {Sing}}, \bibinfo {author} {\bibfnamefont {J.~W.}\ \bibnamefont {Zwanikken}},\ and\ \bibinfo {author} {\bibfnamefont {M.~O.}\ \bibnamefont {de~la Cruz}},\ }\bibfield  {title} {\bibinfo {title} {Theory of melt polyelectrolyte blends and block copolymers: Phase behavior, surface tension, and microphase periodicity},\ }\href {https://doi.org/10.1063/1.4905830} {\bibfield  {journal} {\bibinfo  {journal} {J. Chem. Phys.}\ }\textbf {\bibinfo {volume} {142}},\ \bibinfo {pages} {034902} (\bibinfo {year} {2015})}\BibitemShut {NoStop}%
\bibitem [{\citenamefont {Audus}\ \emph {et~al.}(2015)\citenamefont {Audus}, \citenamefont {Gopez}, \citenamefont {Krogstad}, \citenamefont {Lynd}, \citenamefont {Kramer}, \citenamefont {Hawker},\ and\ \citenamefont {Fredrickson}}]{Audus_2015}%
  \BibitemOpen
  \bibfield  {author} {\bibinfo {author} {\bibfnamefont {D.~J.}\ \bibnamefont {Audus}}, \bibinfo {author} {\bibfnamefont {J.~D.}\ \bibnamefont {Gopez}}, \bibinfo {author} {\bibfnamefont {D.~V.}\ \bibnamefont {Krogstad}}, \bibinfo {author} {\bibfnamefont {N.~A.}\ \bibnamefont {Lynd}}, \bibinfo {author} {\bibfnamefont {E.~J.}\ \bibnamefont {Kramer}}, \bibinfo {author} {\bibfnamefont {C.~J.}\ \bibnamefont {Hawker}},\ and\ \bibinfo {author} {\bibfnamefont {G.~H.}\ \bibnamefont {Fredrickson}},\ }\bibfield  {title} {\bibinfo {title} {Phase behavior of electrostatically complexed polyelectrolyte gels using an embedded fluctuation model},\ }\href {https://doi.org/10.1039/c4sm02299h} {\bibfield  {journal} {\bibinfo  {journal} {Soft Matter}\ }\textbf {\bibinfo {volume} {11}},\ \bibinfo {pages} {1214} (\bibinfo {year} {2015})}\BibitemShut {NoStop}%
\bibitem [{\citenamefont {Duan}\ \emph {et~al.}(2025)\citenamefont {Duan}, \citenamefont {Agrawal},\ and\ \citenamefont {Wang}}]{Duan2025PEBrush}%
  \BibitemOpen
  \bibfield  {author} {\bibinfo {author} {\bibfnamefont {C.}~\bibnamefont {Duan}}, \bibinfo {author} {\bibfnamefont {N.~R.}\ \bibnamefont {Agrawal}},\ and\ \bibinfo {author} {\bibfnamefont {R.}~\bibnamefont {Wang}},\ }\bibfield  {title} {\bibinfo {title} {Electrostatic correlation augmented self-consistent field theory and its application to polyelectrolyte brushes},\ }\href {https://doi.org/10.1103/PhysRevLett.134.048101} {\bibfield  {journal} {\bibinfo  {journal} {Phys. Rev. Lett.}\ }\textbf {\bibinfo {volume} {134}},\ \bibinfo {pages} {048101} (\bibinfo {year} {2025})}\BibitemShut {NoStop}%
\bibitem [{\citenamefont {Wang}(2010)}]{Wang:2010wk}%
  \BibitemOpen
  \bibfield  {author} {\bibinfo {author} {\bibfnamefont {Z.-G.}\ \bibnamefont {Wang}},\ }\bibfield  {title} {\bibinfo {title} {Fluctuation in electrolyte solutions: The self energy},\ }\href {https://doi.org/10.1103/PhysRevE.81.021501} {\bibfield  {journal} {\bibinfo  {journal} {Phys. Rev. E}\ }\textbf {\bibinfo {volume} {81}},\ \bibinfo {pages} {021501} (\bibinfo {year} {2010})}\BibitemShut {NoStop}%
\bibitem [{\citenamefont {Agrawal}\ and\ \citenamefont {Wang}(2022)}]{Agrawal:2022ux}%
  \BibitemOpen
  \bibfield  {author} {\bibinfo {author} {\bibfnamefont {N.~R.}\ \bibnamefont {Agrawal}}\ and\ \bibinfo {author} {\bibfnamefont {R.}~\bibnamefont {Wang}},\ }\bibfield  {title} {\bibinfo {title} {Self-consistent description of vapor-liquid interface in ionic fluids},\ }\href {https://doi.org/10.1103/PhysRevLett.129.228001} {\bibfield  {journal} {\bibinfo  {journal} {Phys. Rev. Lett.}\ }\textbf {\bibinfo {volume} {129}},\ \bibinfo {pages} {228001} (\bibinfo {year} {2022})}\BibitemShut {NoStop}%
\bibitem [{\citenamefont {Levin}(2002)}]{Levin_2002}%
  \BibitemOpen
  \bibfield  {author} {\bibinfo {author} {\bibfnamefont {Y.}~\bibnamefont {Levin}},\ }\bibfield  {title} {\bibinfo {title} {Electrostatic correlations: from plasma to biology},\ }\href {https://doi.org/10.1088/0034-4885/65/11/201} {\bibfield  {journal} {\bibinfo  {journal} {Rep. Prog. Phys.}\ }\textbf {\bibinfo {volume} {65}},\ \bibinfo {pages} {1577} (\bibinfo {year} {2002})}\BibitemShut {NoStop}%
\bibitem [{\citenamefont {Chen}\ \emph {et~al.}(2022)\citenamefont {Chen}, \citenamefont {Mahanthappa},\ and\ \citenamefont {Dorfman}}]{Chen_2022}%
  \BibitemOpen
  \bibfield  {author} {\bibinfo {author} {\bibfnamefont {P.}~\bibnamefont {Chen}}, \bibinfo {author} {\bibfnamefont {M.~K.}\ \bibnamefont {Mahanthappa}},\ and\ \bibinfo {author} {\bibfnamefont {K.~D.}\ \bibnamefont {Dorfman}},\ }\bibfield  {title} {\bibinfo {title} {Stability of cubic single network phases in diblock copolymer melts},\ }\href {https://doi.org/https://doi.org/10.1002/pol.20220318} {\bibfield  {journal} {\bibinfo  {journal} {J. Polym. Sci.}\ }\textbf {\bibinfo {volume} {60}},\ \bibinfo {pages} {2543} (\bibinfo {year} {2022})}\BibitemShut {NoStop}%
\bibitem [{\citenamefont {Park}\ \emph {et~al.}(2023)\citenamefont {Park}, \citenamefont {Bates},\ and\ \citenamefont {Dorfman}}]{Park_2023}%
  \BibitemOpen
  \bibfield  {author} {\bibinfo {author} {\bibfnamefont {S.~J.}\ \bibnamefont {Park}}, \bibinfo {author} {\bibfnamefont {F.~S.}\ \bibnamefont {Bates}},\ and\ \bibinfo {author} {\bibfnamefont {K.~D.}\ \bibnamefont {Dorfman}},\ }\bibfield  {title} {\bibinfo {title} {Single gyroid in {H}-shaped block copolymers},\ }\href {https://doi.org/10.1103/PhysRevMaterials.7.105601} {\bibfield  {journal} {\bibinfo  {journal} {Phys. Rev. Mater.}\ }\textbf {\bibinfo {volume} {7}},\ \bibinfo {pages} {105601} (\bibinfo {year} {2023})}\BibitemShut {NoStop}%
\bibitem [{\citenamefont {Xie}\ \emph {et~al.}(2022)\citenamefont {Xie}, \citenamefont {Qiang},\ and\ \citenamefont {Li}}]{Xie_2022}%
  \BibitemOpen
  \bibfield  {author} {\bibinfo {author} {\bibfnamefont {Q.}~\bibnamefont {Xie}}, \bibinfo {author} {\bibfnamefont {Y.}~\bibnamefont {Qiang}},\ and\ \bibinfo {author} {\bibfnamefont {W.}~\bibnamefont {Li}},\ }\bibfield  {title} {\bibinfo {title} {Single gyroid self-assembled by linear {BABAB} pentablock copolymer},\ }\href {https://doi.org/10.1021/acsmacrolett.1c00656} {\bibfield  {journal} {\bibinfo  {journal} {ACS Macro Lett.}\ }\textbf {\bibinfo {volume} {11}},\ \bibinfo {pages} {205} (\bibinfo {year} {2022})}\BibitemShut {NoStop}%
\bibitem [{\citenamefont {Cheng}\ \emph {et~al.}(2024)\citenamefont {Cheng}, \citenamefont {Xu},\ and\ \citenamefont {Li}}]{Cheng_2024}%
  \BibitemOpen
  \bibfield  {author} {\bibinfo {author} {\bibfnamefont {Y.}~\bibnamefont {Cheng}}, \bibinfo {author} {\bibfnamefont {Z.}~\bibnamefont {Xu}},\ and\ \bibinfo {author} {\bibfnamefont {W.}~\bibnamefont {Li}},\ }\bibfield  {title} {\bibinfo {title} {Understand the relative stability of single-gyroid to double-gyroid in {AB}-type block copolymer: Chain packing and geometric analysis},\ }\href {https://doi.org/10.1021/acs.macromol.4c00899} {\bibfield  {journal} {\bibinfo  {journal} {Macromolecules}\ }\textbf {\bibinfo {volume} {57}},\ \bibinfo {pages} {9167} (\bibinfo {year} {2024})}\BibitemShut {NoStop}%
\bibitem [{\citenamefont {Tian}\ \emph {et~al.}(2025)\citenamefont {Tian}, \citenamefont {Dong}, \citenamefont {Li},\ and\ \citenamefont {Li}}]{Tian_2025}%
  \BibitemOpen
  \bibfield  {author} {\bibinfo {author} {\bibfnamefont {K.}~\bibnamefont {Tian}}, \bibinfo {author} {\bibfnamefont {Q.}~\bibnamefont {Dong}}, \bibinfo {author} {\bibfnamefont {L.}~\bibnamefont {Li}},\ and\ \bibinfo {author} {\bibfnamefont {W.}~\bibnamefont {Li}},\ }\bibfield  {title} {\bibinfo {title} {Bidispersity-induced novel self-assembly behaviors in homopolymer-tethered {AB} diblock copolymers},\ }\href {https://doi.org/10.1021/acs.macromol.5c02401} {\bibfield  {journal} {\bibinfo  {journal} {Macromolecules}\ }\textbf {\bibinfo {volume} {58}},\ \bibinfo {pages} {11495} (\bibinfo {year} {2025})}\BibitemShut {NoStop}%
\bibitem [{\citenamefont {Matsen}(2005)}]{Matsen_Book2005}%
  \BibitemOpen
  \bibfield  {author} {\bibinfo {author} {\bibfnamefont {M.~W.}\ \bibnamefont {Matsen}},\ }\href {https://onlinelibrary.wiley.com/doi/abs/10.1002/9783527617050.ch2} {\emph {\bibinfo {title} {Self-Consistent Field Theory and Its Applications}}}\ (\bibinfo  {publisher} {John Wiley \& Sons, Ltd},\ \bibinfo {year} {2005})\ Chap.~\bibinfo {chapter} {2}, pp.\ \bibinfo {pages} {87--178}\BibitemShut {NoStop}%
\bibitem [{\citenamefont {Lee}\ \emph {et~al.}(2025{\natexlab{b}})\citenamefont {Lee}, \citenamefont {Lee}, \citenamefont {Kim},\ and\ \citenamefont {Park}}]{LEE2025102003}%
  \BibitemOpen
  \bibfield  {author} {\bibinfo {author} {\bibfnamefont {H.}~\bibnamefont {Lee}}, \bibinfo {author} {\bibfnamefont {Y.}~\bibnamefont {Lee}}, \bibinfo {author} {\bibfnamefont {N.}~\bibnamefont {Kim}},\ and\ \bibinfo {author} {\bibfnamefont {M.~J.}\ \bibnamefont {Park}},\ }\bibfield  {title} {\bibinfo {title} {Polymer chain-end chemistry: Unlocking next-generation functional materials},\ }\href {https://doi.org/https://doi.org/10.1016/j.progpolymsci.2025.102003} {\bibfield  {journal} {\bibinfo  {journal} {Prog. Polym. Sci.}\ }\textbf {\bibinfo {volume} {168}},\ \bibinfo {pages} {102003} (\bibinfo {year} {2025}{\natexlab{b}})}\BibitemShut {NoStop}%
\bibitem [{\citenamefont {Katkar}\ and\ \citenamefont {Muthukumar}(2014)}]{Katkar_2014}%
  \BibitemOpen
  \bibfield  {author} {\bibinfo {author} {\bibfnamefont {H.~H.}\ \bibnamefont {Katkar}}\ and\ \bibinfo {author} {\bibfnamefont {M.}~\bibnamefont {Muthukumar}},\ }\bibfield  {title} {\bibinfo {title} {Effect of charge patterns along a solid-state nanopore on polyelectrolyte translocation},\ }\href {https://doi.org/10.1063/1.4869862} {\bibfield  {journal} {\bibinfo  {journal} {J. Chem. Phys.}\ }\textbf {\bibinfo {volume} {140}},\ \bibinfo {pages} {135102} (\bibinfo {year} {2014})}\BibitemShut {NoStop}%
\bibitem [{\citenamefont {Danielsen}\ \emph {et~al.}(2019)\citenamefont {Danielsen}, \citenamefont {McCarty}, \citenamefont {Shea}, \citenamefont {Delaney},\ and\ \citenamefont {Fredrickson}}]{Danielsen_2019}%
  \BibitemOpen
  \bibfield  {author} {\bibinfo {author} {\bibfnamefont {S.~P.~O.}\ \bibnamefont {Danielsen}}, \bibinfo {author} {\bibfnamefont {J.}~\bibnamefont {McCarty}}, \bibinfo {author} {\bibfnamefont {J.-E.}\ \bibnamefont {Shea}}, \bibinfo {author} {\bibfnamefont {K.~T.}\ \bibnamefont {Delaney}},\ and\ \bibinfo {author} {\bibfnamefont {G.~H.}\ \bibnamefont {Fredrickson}},\ }\bibfield  {title} {\bibinfo {title} {Molecular design of self-coacervation phenomena in block polyampholytes},\ }\href {https://doi.org/10.1073/pnas.1900435116} {\bibfield  {journal} {\bibinfo  {journal} {Proc. Natl. Acad. Sci. U.S.A.}\ }\textbf {\bibinfo {volume} {116}},\ \bibinfo {pages} {8224} (\bibinfo {year} {2019})}\BibitemShut {NoStop}%
\bibitem [{\citenamefont {Lytle}\ \emph {et~al.}(2019)\citenamefont {Lytle}, \citenamefont {Chang}, \citenamefont {Markiewicz}, \citenamefont {Perry},\ and\ \citenamefont {Sing}}]{Lytle_2019}%
  \BibitemOpen
  \bibfield  {author} {\bibinfo {author} {\bibfnamefont {T.~K.}\ \bibnamefont {Lytle}}, \bibinfo {author} {\bibfnamefont {L.-W.}\ \bibnamefont {Chang}}, \bibinfo {author} {\bibfnamefont {N.}~\bibnamefont {Markiewicz}}, \bibinfo {author} {\bibfnamefont {S.~L.}\ \bibnamefont {Perry}},\ and\ \bibinfo {author} {\bibfnamefont {C.~E.}\ \bibnamefont {Sing}},\ }\bibfield  {title} {\bibinfo {title} {Designing electrostatic interactions via polyelectrolyte monomer sequence},\ }\href {https://doi.org/10.1021/acscentsci.9b00087} {\bibfield  {journal} {\bibinfo  {journal} {ACS Cent. Sci.}\ }\textbf {\bibinfo {volume} {5}},\ \bibinfo {pages} {709} (\bibinfo {year} {2019})}\BibitemShut {NoStop}%
\bibitem [{\citenamefont {Li}\ \emph {et~al.}(2021)\citenamefont {Li}, \citenamefont {Zhuang},\ and\ \citenamefont {Yu}}]{Li_2021b}%
  \BibitemOpen
  \bibfield  {author} {\bibinfo {author} {\bibfnamefont {M.}~\bibnamefont {Li}}, \bibinfo {author} {\bibfnamefont {B.}~\bibnamefont {Zhuang}},\ and\ \bibinfo {author} {\bibfnamefont {J.}~\bibnamefont {Yu}},\ }\bibfield  {title} {\bibinfo {title} {Sequence--conformation relationship of zwitterionic peptide brushes: Theories and simulations},\ }\href {https://doi.org/10.1021/acs.macromol.1c01229} {\bibfield  {journal} {\bibinfo  {journal} {Macromolecules}\ }\textbf {\bibinfo {volume} {54}},\ \bibinfo {pages} {9565} (\bibinfo {year} {2021})}\BibitemShut {NoStop}%
\bibitem [{\citenamefont {Ren}\ \emph {et~al.}(2015)\citenamefont {Ren}, \citenamefont {Nakamura},\ and\ \citenamefont {Wang}}]{Ren_2015}%
  \BibitemOpen
  \bibfield  {author} {\bibinfo {author} {\bibfnamefont {C.-L.}\ \bibnamefont {Ren}}, \bibinfo {author} {\bibfnamefont {I.}~\bibnamefont {Nakamura}},\ and\ \bibinfo {author} {\bibfnamefont {Z.-G.}\ \bibnamefont {Wang}},\ }\bibfield  {title} {\bibinfo {title} {Effects of ion-induced cross-linking on the phase behavior in salt-doped polymer blends},\ }\href {https://doi.org/10.1021/acs.macromol.5b02229} {\bibfield  {journal} {\bibinfo  {journal} {Macromolecules}\ }\textbf {\bibinfo {volume} {49}},\ \bibinfo {pages} {425} (\bibinfo {year} {2015})}\BibitemShut {NoStop}%
\bibitem [{\citenamefont {Agrawal}\ \emph {et~al.}(2024)\citenamefont {Agrawal}, \citenamefont {Duan},\ and\ \citenamefont {Wang}}]{Agrawal_2024}%
  \BibitemOpen
  \bibfield  {author} {\bibinfo {author} {\bibfnamefont {N.~R.}\ \bibnamefont {Agrawal}}, \bibinfo {author} {\bibfnamefont {C.}~\bibnamefont {Duan}},\ and\ \bibinfo {author} {\bibfnamefont {R.}~\bibnamefont {Wang}},\ }\bibfield  {title} {\bibinfo {title} {Nature of overcharging and charge inversion in electrical double layers},\ }\href {https://doi.org/10.1021/acs.jpcb.3c04739} {\bibfield  {journal} {\bibinfo  {journal} {J. Phys. Chem. B}\ }\textbf {\bibinfo {volume} {128}},\ \bibinfo {pages} {303} (\bibinfo {year} {2024})}\BibitemShut {NoStop}%
\bibitem [{\citenamefont {Duan}\ and\ \citenamefont {Wang}(2024)}]{Duan_2024a}%
  \BibitemOpen
  \bibfield  {author} {\bibinfo {author} {\bibfnamefont {C.}~\bibnamefont {Duan}}\ and\ \bibinfo {author} {\bibfnamefont {R.}~\bibnamefont {Wang}},\ }\bibfield  {title} {\bibinfo {title} {Ion correlation-driven hysteretic adhesion and repulsion between opposing polyelectrolyte brushes},\ }\href {https://doi.org/10.1021/acsmacrolett.4c00426} {\bibfield  {journal} {\bibinfo  {journal} {ACS Macro Lett.}\ }\textbf {\bibinfo {volume} {13}},\ \bibinfo {pages} {1127} (\bibinfo {year} {2024})}\BibitemShut {NoStop}%
\bibitem [{\citenamefont {Wang}\ and\ \citenamefont {Wang}(2015)}]{Wang_2015}%
  \BibitemOpen
  \bibfield  {author} {\bibinfo {author} {\bibfnamefont {R.}~\bibnamefont {Wang}}\ and\ \bibinfo {author} {\bibfnamefont {Z.-G.}\ \bibnamefont {Wang}},\ }\bibfield  {title} {\bibinfo {title} {On the theoretical description of weakly charged surfaces},\ }\href {https://doi.org/10.1063/1.4914170} {\bibfield  {journal} {\bibinfo  {journal} {J. Chem. Phys.}\ }\textbf {\bibinfo {volume} {142}},\ \bibinfo {pages} {104705} (\bibinfo {year} {2015})}\BibitemShut {NoStop}%
\bibitem [{\citenamefont {Wang}\ \emph {et~al.}(2005)\citenamefont {Wang}, \citenamefont {Wang},\ and\ \citenamefont {Yang}}]{Wang_2005}%
  \BibitemOpen
  \bibfield  {author} {\bibinfo {author} {\bibfnamefont {J.}~\bibnamefont {Wang}}, \bibinfo {author} {\bibfnamefont {Z.-G.}\ \bibnamefont {Wang}},\ and\ \bibinfo {author} {\bibfnamefont {Y.}~\bibnamefont {Yang}},\ }\bibfield  {title} {\bibinfo {title} {Nature of disordered micelles in sphere-forming block copolymer melts},\ }\href {https://doi.org/10.1021/ma047990j} {\bibfield  {journal} {\bibinfo  {journal} {Macromolecules}\ }\textbf {\bibinfo {volume} {38}},\ \bibinfo {pages} {1979} (\bibinfo {year} {2005})}\BibitemShut {NoStop}%
\bibitem [{\citenamefont {Dormidontova}\ and\ \citenamefont {Lodge}(2001)}]{Dormidontova_2001}%
  \BibitemOpen
  \bibfield  {author} {\bibinfo {author} {\bibfnamefont {E.~E.}\ \bibnamefont {Dormidontova}}\ and\ \bibinfo {author} {\bibfnamefont {T.~P.}\ \bibnamefont {Lodge}},\ }\bibfield  {title} {\bibinfo {title} {The order-disorder transition and the disordered micelle regime in sphere-forming block copolymer melts},\ }\href {https://doi.org/10.1021/ma010098h} {\bibfield  {journal} {\bibinfo  {journal} {Macromolecules}\ }\textbf {\bibinfo {volume} {34}},\ \bibinfo {pages} {9143} (\bibinfo {year} {2001})}\BibitemShut {NoStop}%
\bibitem [{\citenamefont {Schwab}\ and\ \citenamefont {St{\"u}hn}(1997)}]{Schwab1997}%
  \BibitemOpen
  \bibfield  {author} {\bibinfo {author} {\bibfnamefont {M.}~\bibnamefont {Schwab}}\ and\ \bibinfo {author} {\bibfnamefont {B.}~\bibnamefont {St{\"u}hn}},\ }\bibfield  {title} {\bibinfo {title} {Asymmetric diblock copolymers-phase behaviour and kinetics of structure formation},\ }\href {https://doi.org/10.1007/s003960050091} {\bibfield  {journal} {\bibinfo  {journal} {Colloid Polym. Sci.}\ }\textbf {\bibinfo {volume} {275}},\ \bibinfo {pages} {341} (\bibinfo {year} {1997})}\BibitemShut {NoStop}%
\bibitem [{\citenamefont {Dorfman}\ and\ \citenamefont {Wang}(2023)}]{Dorfman_2023}%
  \BibitemOpen
  \bibfield  {author} {\bibinfo {author} {\bibfnamefont {K.~D.}\ \bibnamefont {Dorfman}}\ and\ \bibinfo {author} {\bibfnamefont {Z.-G.}\ \bibnamefont {Wang}},\ }\bibfield  {title} {\bibinfo {title} {Liquid-like states in micelle-forming diblock copolymer melts},\ }\href {https://doi.org/10.1021/acsmacrolett.3c00259} {\bibfield  {journal} {\bibinfo  {journal} {ACS Macro Lett.}\ }\textbf {\bibinfo {volume} {12}},\ \bibinfo {pages} {980} (\bibinfo {year} {2023})}\BibitemShut {NoStop}%
\bibitem [{\citenamefont {Tzeremes}\ \emph {et~al.}(2002)\citenamefont {Tzeremes}, \citenamefont {Rasmussen}, \citenamefont {Lookman},\ and\ \citenamefont {Saxena}}]{Tzeremes:2002aa}%
  \BibitemOpen
  \bibfield  {author} {\bibinfo {author} {\bibfnamefont {G.}~\bibnamefont {Tzeremes}}, \bibinfo {author} {\bibfnamefont {K.~O.}\ \bibnamefont {Rasmussen}}, \bibinfo {author} {\bibfnamefont {T.}~\bibnamefont {Lookman}},\ and\ \bibinfo {author} {\bibfnamefont {A.}~\bibnamefont {Saxena}},\ }\bibfield  {title} {\bibinfo {title} {Efficient computation of the structural phase behavior of block copolymers},\ }\href {https://doi.org/10.1103/PhysRevE.65.041806} {\bibfield  {journal} {\bibinfo  {journal} {Phys. Rev. E}\ }\textbf {\bibinfo {volume} {65}},\ \bibinfo {pages} {041806} (\bibinfo {year} {2002})}\BibitemShut {NoStop}%
\bibitem [{\citenamefont {Vigil}\ \emph {et~al.}(2022)\citenamefont {Vigil}, \citenamefont {Quah}, \citenamefont {Sun}, \citenamefont {Delaney},\ and\ \citenamefont {Fredrickson}}]{Vigil_2022}%
  \BibitemOpen
  \bibfield  {author} {\bibinfo {author} {\bibfnamefont {D.~L.}\ \bibnamefont {Vigil}}, \bibinfo {author} {\bibfnamefont {T.}~\bibnamefont {Quah}}, \bibinfo {author} {\bibfnamefont {D.}~\bibnamefont {Sun}}, \bibinfo {author} {\bibfnamefont {K.~T.}\ \bibnamefont {Delaney}},\ and\ \bibinfo {author} {\bibfnamefont {G.~H.}\ \bibnamefont {Fredrickson}},\ }\bibfield  {title} {\bibinfo {title} {Self-consistent field theory predicts universal phase behavior for linear, comb, and bottlebrush diblock copolymers},\ }\href {https://doi.org/10.1021/acs.macromol.2c00192} {\bibfield  {journal} {\bibinfo  {journal} {Macromolecules}\ }\textbf {\bibinfo {volume} {55}},\ \bibinfo {pages} {4237} (\bibinfo {year} {2022})}\BibitemShut {NoStop}%
\bibitem [{\citenamefont {Frigo}(1999)}]{FFTW}%
  \BibitemOpen
  \bibfield  {author} {\bibinfo {author} {\bibfnamefont {M.}~\bibnamefont {Frigo}},\ }\href@noop {} {\emph {\bibinfo {title} {A Fast Fourier Transform Compiler}}},\ Proceedings of the 1999 ACM SIGPLAN Conference on Programming Language Design and Implementation (PLDI '99)\ (\bibinfo {address} {Atlanta, Georgia},\ \bibinfo {year} {1999})\BibitemShut {NoStop}%
\bibitem [{\citenamefont {Hoffman}\ and\ \citenamefont {Frankel}(2018)}]{Hoffman_2018}%
  \BibitemOpen
  \bibfield  {author} {\bibinfo {author} {\bibfnamefont {J.~D.}\ \bibnamefont {Hoffman}}\ and\ \bibinfo {author} {\bibfnamefont {S.}~\bibnamefont {Frankel}},\ }\href {https://doi.org/10.1201/9781315274508} {\emph {\bibinfo {title} {Numerical Methods for Engineers and Scientists}}}\ (\bibinfo  {publisher} {CRC Press},\ \bibinfo {year} {2018})\BibitemShut {NoStop}%
\bibitem [{\citenamefont {Qiang}\ and\ \citenamefont {Li}(2020)}]{Qiang_2020b}%
  \BibitemOpen
  \bibfield  {author} {\bibinfo {author} {\bibfnamefont {Y.}~\bibnamefont {Qiang}}\ and\ \bibinfo {author} {\bibfnamefont {W.}~\bibnamefont {Li}},\ }\bibfield  {title} {\bibinfo {title} {Accelerated pseudo-spectral method of self-consistent field theory via crystallographic fast fourier transform},\ }\href {https://doi.org/10.1021/acs.macromol.0c01974} {\bibfield  {journal} {\bibinfo  {journal} {Macromolecules}\ }\textbf {\bibinfo {volume} {53}},\ \bibinfo {pages} {9943} (\bibinfo {year} {2020})}\BibitemShut {NoStop}%
\end{thebibliography}%

\end{document}